\documentclass[longauth]{aa}
\usepackage{natbib}
\usepackage[varg]{txfonts}
\usepackage{graphicx}

\usepackage{booktabs}
\usepackage{multirow}
\usepackage{tabularx}
\usepackage{lscape}
\usepackage{longtable}

\newcolumntype{L}[1]{>{\raggedright\arraybackslash}p{#1}}
\newcolumntype{C}[1]{>{\centering\arraybackslash}p{#1}}
\newcolumntype{R}[1]{>{\raggedleft\arraybackslash}p{#1}}

\usepackage{color} 

\def\Msun{\ifmmode{\mathrm M_\odot}\else{$M_\odot$}\fi}

\newcommand{\SigStar}{\ensuremath{\Sigma_\mathrm{\star}}}
\newcommand{\SigMol}{\ensuremath{\Sigma_\mathrm{mol}}}
\newcommand{\SigSFR}{\ensuremath{\Sigma_\mathrm{SFR}}}
\newcommand{\tdep}{\ensuremath{\tau_\mathrm{dep}}}
\newcommand{\hi}{\ion{H}{I}}
\newcommand{\hii}{\ion{H}{II}}

\def\mod#1{#1}

\begin{document}

\defcitealias{2012MNRAS.421.3127N}{N12}
\defcitealias{2013ARA&A..51..207B}{B13}
\defcitealias{2015A&A...582A..86H}{HE15}

\title{Do spiral arms enhance star formation efficiency?}

%\author{M.~Querejeta\inst{1}, A.~K.~Leroy\inst{4}, S.~Meidt\inst{3}, E.~Schinnerer\inst{2},  E.~Emsellem\inst{5,6},  R.~S.~Klessen\inst{7,8}, J.~Sun\inst{4}, M.~Sormani\inst{4} \& PHANGS}

% AFFILIATIONS (add or correct if necessary):
% --------------------------------------------

\newcommand{\OSU}{\label{OSU} Department of Astronomy, The Ohio State University, 140 West 18th Avenue, Columbus, Ohio 43210, USA}

\newcommand{\Alberta}{\label{Alberta} Department of Physics, University of Alberta, Edmonton, AB T6G 2E1, Canada}

\newcommand{\ANU}{\label{ANU} Research School of Astronomy and Astrophysics, Australian National University, Canberra, ACT 2611, Australia}

\newcommand{\IPAC}{\label{IPAC} Caltech-IPAC, 1200 E. California Blvd. Pasadena, CA 91125, USA}

\newcommand{\Carnegie}{\label{Carnegi} Observatories of the Carnegie Institution for Science, 813 Santa Barbara Street, Pasadena, CA 91101, USA}

\newcommand{\CCAPP}{\label{CCAPP} Center for Cosmology and Astroparticle Physics, 191 West Woodruff Avenue, Columbus, OH 43210, USA}

\newcommand{\CfA}{\label{CfA} Harvard-Smithsonian Center for Astrophysics, 60 Garden Street, Cambridge, MA 02138, USA}

\newcommand{\CITEVA}{\label{CITEVA} Centro de Astronomía (CITEVA), Universidad de Antofagasta, Avenida Angamos 601, Antofagasta, Chile}

\newcommand{\CNRS}{\label{CNRS} CNRS, IRAP, 9 Av. du Colonel Roche, BP 44346, F-31028 Toulouse cedex 4, France}

\newcommand{\ESO}{\label{ESO} European Southern Observatory, Karl-Schwarzschild Stra{\ss}e 2, D-85748 Garching bei M\"{u}nchen, Germany}

\newcommand{\Heidelberg}{\label{Heidelberg} Astronomisches Rechen-Institut, Zentrum f\"{u}r Astronomie der Universit\"{a}t Heidelberg, M\"{o}nchhofstra\ss e 12-14, D-69120 Heidelberg, Germany}

\newcommand{\COOL}{\label{COOL} Cosmic Origins Of Life (COOL) Research DAO, coolresearch.io}

\newcommand{\ICRAR}{\label{ICRAR} International Centre for Radio Astronomy Research, University of Western Australia, 35 Stirling Highway, Crawley, WA 6009, Australia}

\newcommand{\IRAM}{\label{IRAM} Institut de Radioastronomie Millim\'{e}trique (IRAM), 300 Rue de la Piscine, F-38406 Saint Martin d'H\`{e}res, France}

\newcommand{\ITA}{\label{ITA} Universit\"{a}t Heidelberg, Zentrum f\"{u}r Astronomie, Institut f\"{u}r Theoretische Astrophysik, Albert-Ueberle-Str 2, D-69120 Heidelberg, Germany}

\newcommand{\IWR}{\label{IWR} Universit\"{a}t Heidelberg, Interdisziplin\"{a}res Zentrum f\"{u}r Wissenschaftliches Rechnen, Im Neuenheimer Feld 205, D-69120 Heidelberg, Germany}

\newcommand{\JHU}{\label{JHU} Department of Physics and Astronomy, The Johns Hopkins University, Baltimore, MD 21218, USA}

\newcommand{\Leiden}{\label{Leiden} Leiden Observatory, Leiden University, P.O. Box 9513, 2300 RA Leiden, The Netherlands}

\newcommand{\Maryland}{\label{Maryland} Department of Astronomy, University of Maryland, College Park, MD 20742, USA}

\newcommand{\MPE}{\label{MPE} Max-Planck-Institut f\"{u}r extraterrestrische Physik, Giessenbachstra{\ss}e 1, D-85748 Garching, Germany}

\newcommand{\MPIA}{\label{MPIA} Max-Planck-Institut f\"{u}r Astronomie, K\"{o}nigstuhl 17, D-69117, Heidelberg, Germany}

\newcommand{\Nagoya}{\label{Nagoya} Department of Physics, Nagoya University, Furo-cho, Chikusa-ku, Nagoya, Aichi 464-8602, Japan}

\newcommand{\NRAO}{\label{NRAO} National Radio Astronomy Observatory, 520 Edgemont Road, Charlottesville, VA 22903-2475, USA}

\newcommand{\OAN}{\label{OAN} Observatorio Astron\'{o}mico Nacional (IGN), C/Alfonso XII, 3, E-28014 Madrid, Spain}

\newcommand{\ObsParis}{\label{ObsParis} Sorbonne Universit\'{e}, Observatoire de Paris, Universit\'{e} PSL, CNRS, LERMA, F-75014, Paris, France}

\newcommand{\Princeton}{\label{Princeton} Department of Astrophysical Sciences, Princeton University, 4 Ivy Ln., Princeton, NJ 08544 USA}

\newcommand{\UToledo}{\label{UToledo} University of Toledo, 2801 W. Bancroft St., Mail Stop 111, Toledo, OH, 43606}

\newcommand{\Toulouse}{\label{Toulouse} Universit\'{e} de Toulouse, UPS-OMP, IRAP, F-31028 Toulouse cedex 4, France}

\newcommand{\UBonn}{\label{UBonn} Argelander-Institut f\"ur Astronomie, Universit\"at Bonn, Auf dem H\"ugel 71, 53121 Bonn, Germany}

\newcommand{\UChile}{\label{UChile} Departamento de Astronom\'{i}a, Universidad de Chile, Camino del Observatorio 1515, Las Condes, Santiago, Chile}

\newcommand{\UConn}{\label{UConn} Department of Physics, University of Connecticut, Storrs, CT, 06269, USA}

\newcommand{\UCSD}{\label{UCSD} Center for Astrophysics and Space Sciences, Department of Physics,  University of California, San Diego, 9500 Gilman Drive, La Jolla, CA 92093, USA}

\newcommand{\UCSDAA}{\label{UCSDAA} Department of Astronomy \& Astrophysics,  University of California, San Diego, 9500 Gilman Drive, La Jolla, CA 92093, USA}

\newcommand{\UGent}{\label{UGent} Sterrenkundig Observatorium, Universiteit Gent, Krijgslaan 281 S9, B-9000 Gent, Belgium}

\newcommand{\ULyon}{\label{ULyon} Univ Lyon, Univ Lyon 1, ENS de Lyon, CNRS, Centre de Recherche Astrophysique de Lyon UMR5574,\\ F-69230 Saint-Genis-Laval, France}

\newcommand{\UMass}{\label{UMass} University of Massachusetts—Amherst, 710 N. Pleasant Street, Amherst, MA 01003, USA}

\newcommand{\UWyoming}{\label{UWyoming} Department of Physics and Astronomy, University of Wyoming, Laramie, WY 82071, USA}

\newcommand{\LAM}{\label{LAM} Aix Marseille Univ, CNRS, CNES, LAM (Laboratoire d’Astrophysique de Marseille), Marseille, France}

\newcommand{\UHawaii}{\label{UHawaii} Institute for Astronomy, University of Hawaii, 2680 Woodlawn Drive, Honolulu, HI 96822, USA}

\newcommand{\UCM}{\label{UCM} Departamento de F\'{\i}sica de la Tierra y Astrof\'{\i}sica, Universidad Complutense de Madrid, E-28040, Spain}

\newcommand{\IPARC}{\label{IPARC} Instituto de F\'{\i}sica de Part\'{\i}culas y del Cosmos IPARCOS, Facultad de Ciencias F\'{\i}sicas, Universidad Complutense de Madrid, E-28040, Spain}

\newcommand{\STScI}{\label{STScI} Space Telescope Science Institute, 3700 San Martin Drive, Baltimore, MD 21218, USA}

\newcommand{\McMaster}{\label{McMaster} Department of Physics and Astronomy, McMaster University, 1280 Main Street West, Hamilton, ON L8S 4M1, Canada}

\newcommand{\INAF}{\label{INAF} INAF -- Osservatorio Astrofisico di Arcetri, Largo E. Fermi 5, I-50157, Firenze, Italy}

\newcommand{\Sydney}{\label{Sydney} Sydney Institute for Astronomy, School of Physics A28, The University of Sydney, NSW 2006, Australia}

\newcommand{\UA}{\label{UA} Centro de Astronomía (CITEVA), Universidad de Antofagasta, Avenida Angamos 601, Antofagasta, Chile}

\newcommand{\CITA}{\label{CITA} Canadian Institute for Theoretical Astrophysics (CITA), University of Toronto, 60 St George St, Toronto, ON M5S 3H8, Canada}

\newcommand{\ASIAA}{\label{ASIAA} Institute of Astronomy and Astrophysics, Academia Sinica, No. 1, Sec. 4, Roosevelt Road, Taipei 10617, Taiwan}

\newcommand{\TKU}{\label{TKU} Department of Physics, Tamkang University, No.151, Yingzhuan Rd., Tamsui Dist., New Taipei City 251301, Taiwan}

\newcommand{\PSMA}{\label{PSMA} Penn State Mont Alto, 1 Campus Drive, Mont Alto, PA  17237, USA}

\newcommand{\ILL}{\label{ILL} Institut Laue-Langevin, 71 avenue des Martyrs, F-38042 Grenoble, France}

\newcommand{\TUM}{\label{TUM} Technical University of Munich, School of Engineering and Design, Department of Aerospace and Geodesy, Chair of Remote Sensing Technology, Arcisstr. 21, 80333 Munich, Germany}

\newcommand{\Surrey}{\label{Surrey} Department of Physics, University of Surrey, Guildford GU2 7XH, UK}

\newcommand{\Oxford}{\label{Oxford} Sub-department of Astrophysics, Department of Physics, University of Oxford, Keble Road, Oxford OX1 3RH, UK}

\newcommand{\AIP}{\label{AIP} Leibniz-Institut for Astrophysik Potsdam (AIP), An der Sternwarte 16, 14482 Potsdam, Germany}

\newcommand{\StAndrews}{\label{StAndrews} School of Physics and Astronomy, University of St Andrews, North Haugh, St Andrews, KY16 9SS}

\newcommand{\IAC}{\label{IAC}{Instituto de Astrof\'isica de Canarias, C/ V\'ia L\'actea s/n, E-38205, La Laguna, Spain}}

\newcommand{\ULL}{\label{ULL}{Departamento de Astrof\'isica, Universidad de La Laguna, Av. del Astrof\'isico Francisco S\'anchez s/n, E-38206, La Laguna, Spain}}

\newcommand{\insubria}{ \label{insubria} Universit{\`a} dell’Insubria, via Valleggio 11, 22100 Como, Italy}
% =========================================================

% PAPER TEAM:
% ------------
\author{%
Miguel~Querejeta\inst{\ref{OAN}}              % 0000-0002-0472-1011
\and Adam~K.~Leroy\inst{\ref{OSU}}            % 0000-0002-2545-1700
\and Sharon~E.~Meidt\inst{\ref{UGent}}        % 0000-0002-6118-4048
\and Eva~Schinnerer\inst{\ref{MPIA}}          %
\and Francesco~Belfiore\inst{\ref{INAF}} %
\and Eric~Emsellem\inst{\ref{ESO},\ref{ULyon}} %
\and Ralf~S.~Klessen\inst{\ref{ITA},\ref{IWR}}% 0000-0002-0560-3172
\and Jiayi~Sun\inst{\ref{Princeton}}  % 0000-0003-0378-4667
\and Mattia~Sormani\inst{\ref{insubria},\ref{Surrey}}
%
%
% PLEASE ADD YOUR NAME AND AFFILIATION HERE (same format as above)
% PHANGS team in alphabetical order:
% -----------------------------------
%
\and Ivana~Be\v{s}li\'c\inst{\ref{ObsParis}}
\and Yixian~Cao\inst{\ref{MPE}}%0000-0001-5301-1326
\and M\'elanie~Chevance\inst{\ref{ITA},\ref{COOL}}
\and Dario~Colombo\inst{\ref{UBonn}} % 0000-0001-9793-6400
\and Daniel~A.~Dale\inst{\ref{UWyoming}} %0000-0002-5782-9093
\and Santiago Garc\'ia-Burillo\inst{\ref{OAN}} %
\and Simon~C.~O.~Glover\inst{\ref{ITA}} %0000-0001-6708-1317
\and Kathryn~Grasha\inst{\ref{ANU}}  
\and Brent~Groves\inst{\ref{ICRAR}}   % \orcidlink{0000-0002-9768-0246}
\and Eric.~W.~Koch\inst{\ref{CfA}} % 0000-0001-9605-780X
\and Lukas~Neumann\inst{\ref{UBonn}} % 0000-0001-9793-6400
\and Hsi-An~Pan \inst{\ref{TKU}} %0000-0002-1370-6964
\and Ismael~Pessa\inst{\ref{AIP}} %0000-0002-0873-5744
\and Jérôme~Pety\inst{\ref{IRAM},\ref{ObsParis}} %0000-0002-0873-5744
\and Francesca Pinna\inst{\ref{IAC},\ref{ULL}}
%0000-0001-5965-3530
\and Lise~Ramambason\inst{\ref{ITA}} %0000-0002-9190-9986
\and Alessandro Razza\inst{\ref{UChile}} %0000-0001-7876-1713
\and Andrea~Romanelli\inst{\ref{ITA}}
\and Erik~Rosolowsky\inst{\ref{Alberta}}
\and Marina Ruiz-Garc\'ia\inst{\ref{OAN}} %
\and Patricia S\'anchez-Bl\'azquez\inst{\ref{UCM}} %
\and Rowan~Smith\inst{\ref{StAndrews}} %0000-0002-0820-1814
\and Sophia Stuber\inst{\ref{MPIA}} %
\and Leonardo~Ubeda\inst{\ref{STScI}}  
\and Antonio Usero\inst{\ref{OAN}} %
\and Thomas~G.~Williams\inst{\ref{Oxford}}
}

\institute{\OAN{} \and \OSU{} \and \UGent{} \and \MPIA{} \and \INAF{} \and \ESO{} \and \ULyon{} \and \ITA{} \and \IWR{} \and \Princeton{} \and \insubria{} \and \Surrey{}  \and \ObsParis{}  \and \MPE{} \and \COOL{} \and \UBonn{} \and \UWyoming \and \ANU{} \and \ICRAR{} \and \CfA{} \and \TKU{} \and \AIP{} \and \IRAM{} \and \IAC{} \and \ULL{}  \and \UChile{} \and \Alberta{} \and \UCM{} \and \StAndrews{} \and \STScI{} \and \Oxford{}}

%\institute{Observatorio Astron{\'o}mico Nacional (IGN), C/ Alfonso XII 3, E-28014 Madrid, Spain, \email{m.querejeta@oan.es}
%\and Department of Astronomy, The Ohio State University, 140 West 18th Ave, Columbus, OH 43210, USA 
%\and Sterrenkundig Observatorium, Universiteit Gent, Krijgslaan 281 S9, B-9000 Gent, Belgium 
%\and  Max-Planck-Institut f\"{u}r Astronomie, K\"{o}nigstuhl 17, D-69117 Heidelberg, Germany 
%\and European Southern Observatory, Karl-Schwarzschild-Stra{\ss}e 2, D-85748 Garching, Germany  
%\and Univ Lyon, Univ Lyon1, ENS de Lyon, CNRS, Centre de Recherche Astrophysique de Lyon UMR5574, F-69230 Saint-Genis-Laval, France 
%\and Universit\"{a}t Heidelberg, Zentrum f\"{u}r Astronomie, Albert-Ueberle-Stra{\ss}e 2, D-69120 Heidelberg, Germany  
%\and Universit\"{a}t Heidelberg, Interdisziplin\"{a}res Zentrum f\"{u}r Wissenschaftliches Rechnen, INF 205, D-69120 Heidelberg, Germany }

\date{Received ..... / Accepted .....}

\abstract {Spiral arms, as those of our own Milky Way, are some of the most spectacular features in disc galaxies. It has been argued that star formation should proceed more efficiently in spiral arms as a result of gas compression. Yet, observational studies have so far yielded contradictory results. Here, we examine arm/interarm surface density contrasts at $\sim$100\,pc resolution in 28 spiral galaxies from the PHANGS survey. We find that the arm AND interarm contrast in stellar mass surface density ($\Sigma_\star$) is very modest, typically a few tens of percent. This is much smaller than the contrasts measured for molecular gas ($\Sigma_\mathrm{mol}$) or star formation rate ($\Sigma_\mathrm{SFR}$) surface density, which typically reach a factor of ${\sim}2{-}3$. However, $\Sigma_\mathrm{mol}$ and $\Sigma_\mathrm{SFR}$ contrasts show a significant correlation with the enhancement in $\Sigma_\star$, suggesting that the small stellar contrast largely dictates the stronger accumulation of gas and star formation. All these contrasts increase for grand-design spirals compared to multi-armed and flocculent systems (and for galaxies with high stellar mass). The median star formation efficiency (SFE) of the molecular gas is $16$\% higher in spiral arms than in interarm regions, with a large scatter, and the contrast increases significantly (median SFE contrast $2.34$) for  regions of particularly enhanced stellar contrast ($\Sigma_\star$ contrast $>1.97$).
The molecular-to-atomic gas ratio ($\Sigma_\mathrm{mol}$/$\Sigma_\mathrm{atom}$) is higher in spiral arms, pointing to a transformation of atomic to molecular gas. 
As a consequence, the total gas contrast ($\Sigma_\mathrm{mol} + \Sigma_\mathrm{atom}$) slightly drops compared to $\Sigma_\mathrm{mol}$ (median $4$\% lower, working at $\sim$kpc resolution), while the SFE contrast increases when we include atomic gas (median $8$\% higher than for $\Sigma_\mathrm{mol}$).
The contrasts show important fluctuations with galactocentric radius. We confirm that our results are robust against a number of effects, such as spiral mask width, tracers, resolution, and binning. In conclusion, the boost in the SFE of molecular gas in spiral arms is generally modest or absent, except for locations with \mod{exceptionally} large stellar contrasts.
}


\titlerunning{Do spiral arms enhance star formation efficiency?}
\authorrunning{M.~Querejeta et al.}

\maketitle 
\section{Introduction} 
\label{Sec:introduction}

Spiral arms stand out as some of the most iconic features of galaxies, including our own Milky Way. Approximately two out of three galaxies in the local Universe display some kind of spiral structure \citep[e.g.][]{2010ApJS..186..427N,2013MNRAS.435.2835W,2015ApJS..217...32B}, including well-delineated grand-design spirals (like M51 or NGC\,1365) or more sparse multi-armed (e.g. M100 or NGC\,4254) and flocculent spiral components (e.g. NGC\,2775 or NGC\,4298). While spiral arms are recognised as active sites of star formation, it is still unclear if this results from the mere accumulation of large amounts of gas or if it is due to a more efficient conversion of gas into stars in spiral arms than in interarm regions.

There are a few reasons we would expect spiral arms to lead to enhancements in the star formation process. By collecting gas, spirals elevate star formation rates (SFRs) by locally increasing the gas density on scales of several hundreds of pcs. They may also more actively `trigger' star formation by boosting the star formation efficiency (SFE), such that more stars are formed per unit mass of gas than in regions of comparable density elsewhere in the galaxy. This `triggering' of star formation is believed to either reflect a reduction in shear in spiral arms, which promotes gravitational collapse and the formation of stars \citep[e.g.][]{1987ApJ...312..626E,1993prpl.conf...97E,2002ApJ...570..132K,2017MNRAS.470.4261D}, or the development of shocks in the gas in response to spiral density waves \citep{1969ApJ...158..123R,1975ApJ...196..381R,2004MNRAS.349..909G}.

Several observational studies have quantified the increase of gas and star formation rate surface density in spiral arms \citep[e.g.][]{1988Natur.334..402V,1993A&A...274..123G,2003PASJ...55..191N,2009A&A...495..795H,2020ApJ...901L...8S,2021A&A...656A.133Q}.
Leveraging the high resolution of PHANGS-ALMA maps \citep{2021ApJS..257...43L} that probe CO emission on ${\sim} 100$\,pc scales, \citet{2021ApJ...913..113M} found evidence that gas density contrasts scale non-linearly with contrasts in the underlying stellar spiral density, consistent with shocks actively elevating the density in the arm. 
Even so, observational evidence suggests that the existence of a spiral shock may not immediately lead to an enhancement in the SFE. Studies comparing SFE in spiral arms and the interarm region lead to ambiguous, inconsistent results, with some pointing to enhanced SFE in the arms 
\citep[e.g.][]{1987PhDT........11L,1988Natur.334..402V,1990ApJ...349..497C,1990ApJ...356..135L,1996MNRAS.283..251K} and some suggesting overall similar SFE between spiral and interarm regions, despite local fluctuations \citep{2003AJ....126.2831H,2010ApJ...725..534F,2012MNRAS.426..701M,2012ApJ...757..155R,2015MNRAS.452..289E,2016ApJ...827..103K,2018MNRAS.479.2361R,2021A&A...656A.133Q}.
This might reflect additional mechanisms that act in opposition to a potential boost in SFE. For example, cloud-cloud collisions, sometimes seen as triggers of star formation, can also lead to elevated gas velocity dispersions that shift gas out of a self-gravitating state \citep{2007MNRAS.374.1115D,2014PASA...31...35D}. Shear and differential gas flows in the arm can be another source of stability against gravitational collapse that suppresses star formation \citep{2013ApJ...779...45M}.

To unambiguously determine the efficiency of star formation in spiral arms, it is crucial to first control for as many independent factors as possible. The measured SFE can critically depend on the employed tracer, conversion factors, observational resolution, and the definition of spiral arms. Yet, even for a consistent definition of arms, different galaxies seem to show a divergent behaviour regarding SFE in spiral arms, as we recently showed for the PHANGS sample of galaxies \citep{2021A&A...656A.133Q}. Here, we aim to examine this problem in greater detail by assembling a comprehensive set of homogeneous measurements of spiral arm-to-interarm contrasts in stellar mass, gas, SFR, and SFE, applying a single consistent strategy to all spiral arms and employing state-of-the art calibrations for the gas density and star formation rate from PHANGS.

PHANGS\footnote{Physics at High Angular resolution in Nearby GalaxieS; \url{http://www.phangs.org}} is a multi-wavelength effort to map nearby galaxies at ${\lesssim} 1''$ resolution to study the star formation cycle. Spanning large programmes on ALMA \citep{2021ApJS..257...43L}, MUSE \citep{emsellem21}, HST \citep{2022ApJS..258...10L}, and JWST \citep{2023ApJ...944L..17L,2024arXiv240115142W} and further complemented by a large number of ancillary observations (including atomic gas), PHANGS provides a unique view on the distribution of molecular clouds in galaxies, their collapse to form stars, and the feedback associated with the star formation process. 

This paper is a natural follow-up companion to \citet{2021ApJ...913..113M}, where a study of the relation between azimuthal contrasts in stellar mass and molecular gas was presented, without explicitly identifying spiral arms. The present paper builds on \citet{2021A&A...656A.133Q}, where we presented environmental masks of morphological features for PHANGS galaxies (including spirals) and discussed how molecular gas and star formation are distributed across environments. In \citet{2021A&A...656A.133Q}, the distribution of surface densities, efficiency, and contrasts was studied at kpc resolution. Here, we examine the effect of higher resolution (${\sim}100$\,pc instead of ${\sim}1$\,kpc) on arm and interarm contrasts. We also consider how contrasts change if we employ thinner arm masks, with an alternative arm definition that we introduce here. While in \citet{2021A&A...656A.133Q} we measured a single arm/interarm contrast per galaxy, here we study the contrast in individual spiral segments, and we also analyse variations with galactocentric radius. Another novelty of this paper is that we account for the spiral arm enhancement in stellar mass surface density and its relation to the contrast in molecular gas and star formation rate surface densities. We also consider the effect of including atomic gas.

The structure of this paper is the following. In Sect.~\ref{Sec:data}, we describe our tracers of molecular and atomic gas, star formation, and stellar mass, as well as the environmental masks and binning approach. The main results of the paper are presented in Sect.~\ref{Sec:results} and discussed in Sect.~\ref{Sec:discussion}. We conclude with a summary in Sect.~\ref{Sec:concl}.

\section{Data and methods} 
\label{Sec:data}

\begin{figure*}[t]
\begin{center}
\includegraphics[trim=0 0 0 0, clip,width=0.95\textwidth]{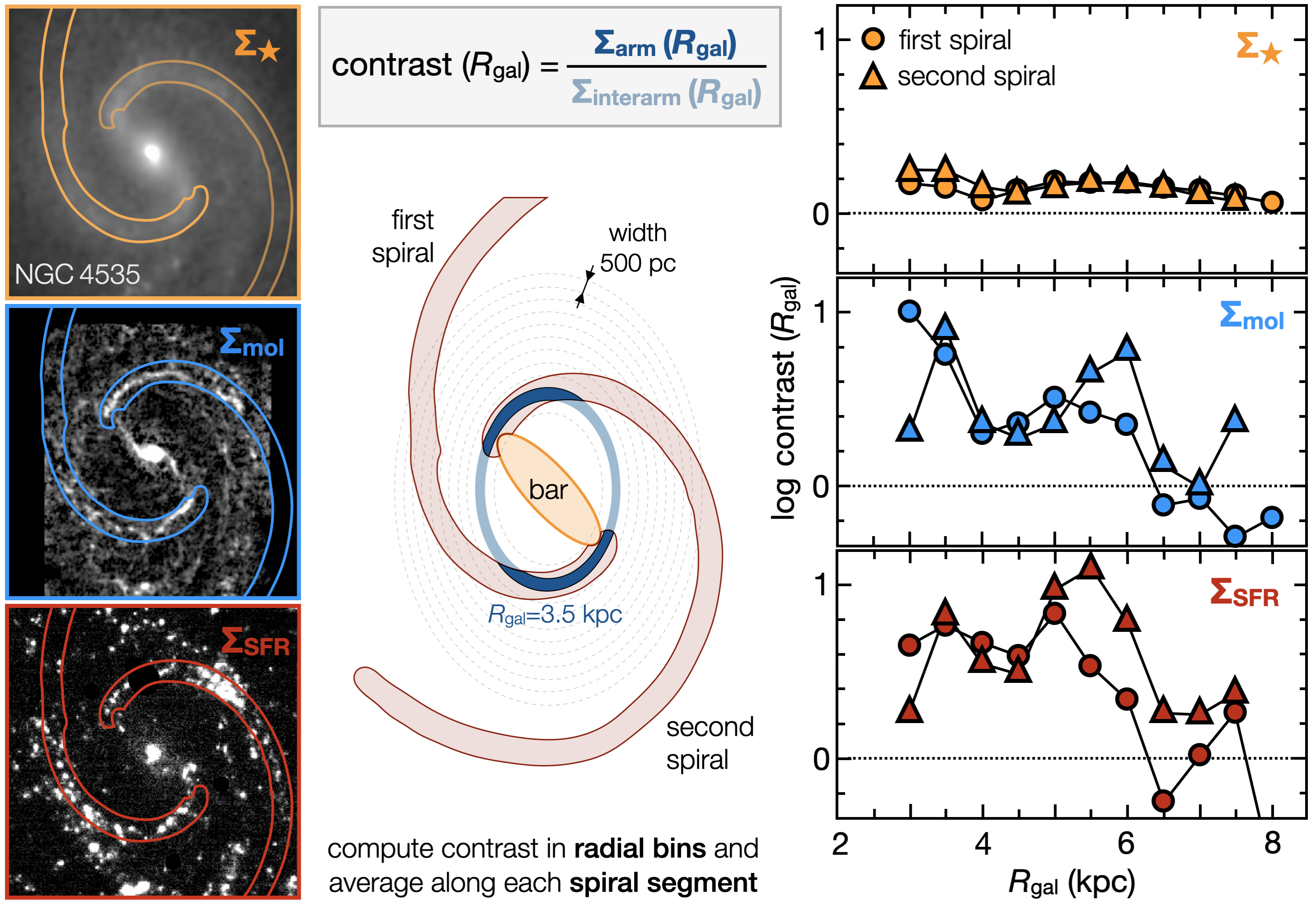}
\end{center}
\caption{Cartoon illustrating the region covered by the spiral mask on the \SigStar, \SigMol, and \SigSFR\ maps of NGC\,4535, and the nominal binning approach used in this paper. We consider uniform bins in galactocentric radius (assuming the inclination and PA from \citealt{2020ApJ...897..122L}); the blue-shaded ring illustrates one of these bins at $R_{\rm gal}=3.5$\,kpc. The right panels show the radial profiles of the arm/interarm contrast (in log scale) for \SigStar, \SigMol, and \SigSFR.}
\label{fig:method}
\end{figure*}

We focus on the 28 spiral galaxies from PHANGS--ALMA where a spiral mask was explicitly defined in \citet{2021A&A...656A.133Q}. This choice limits our study to galaxies with long and continuous spiral segments, which can be traced over most of the star-forming disc (the remaining galaxies do not have a spiral mask defined in \citealt{2021A&A...656A.133Q}, as the identification of arms becomes increasingly subjective and less meaningful). This results in a sample made up predominantly of grand-design spirals, excluding a larger fraction of multi-armed and flocculent spirals, where the distinction between arm and interarm becomes fuzzier. 
{We perform this study at a cloud-scale resolution of ${\sim}100$\,pc, working with the highest-resolution maps available to us in each case\footnote{We note that matching to the best common CO physical resolution ($180$\,pc) has a minimal effect on the contrasts (scatter of ${\sim}3$\%, without systematic offsets), because this is still small compared to the width of the spiral masks (${\sim}1{-}2$\,kpc).}, which typically corresponds to ${\sim} 1\arcsec$ for CO and H$\alpha$.}
In Sect.~\ref{Sec:sample}, we explain how the PHANGS--ALMA sample was selected. We measure molecular gas surface densities based on the ALMA observations of \mbox{CO(2--1)} as explained in Sect.\,\ref{Sec:PHANGS--ALMA}. In Sect.\,\ref{Sec:SFRs} we outline our measurement of star formation rates. We describe the \hi{} data in Sect.\,\ref{Sec:HIdata}. We determine stellar mass surface densities based on near-infrared (NIR) observations as explained in Sect.~\ref{Sec:stellarmass}. The masks that we use are presented in Sect.~\ref{Sec:masks} including a new, thinner alternative for spiral arms. In Sect.\,\ref{Sec:method}, we introduce our nominal approach to measure contrasts in radial bins.

\subsection{Sample}
\label{Sec:sample}

The galaxies studied in this paper are part of the nominal PHANGS--ALMA sample of 74 galaxies, excluding extensions \citep{2021ApJS..257...43L}. PHANGS--ALMA targeted virtually all the massive, star-forming galaxies out to $D \approx 17$\,Mpc, which are not highly inclined and which are visible from the ALMA site. The galaxies are representative of the $z=0$ star formation `main sequence' \citep[e.g.][]{2007ApJ...660L..43N}. We refer the interested reader to \citet{2021ApJS..257...43L} for a detailed description of the sample selection and properties.

This sub-sample of 28 galaxies with well-defined spiral arms spans stellar masses $9.5 \lesssim \log(M_\star / {\rm M}_\odot) \lesssim 11.1$, with a median $\log(M_\star / {\rm M}_\odot) = 10.5$, which is very similar to the median stellar mass in the whole PHANGS--ALMA sample ($10.4$). In terms of molecular gas mass ($8.6 \lesssim \log(M_{\rm mol} / {\rm M}_\odot) \lesssim 10.2$), the median of this spiral sub-sample ($\log(M_{\rm mol} / {\rm M}_\odot) = 9.5$) is slightly higher than the median in PHANGS--ALMA ($9.2$); for star formation rates ($-0.44 \lesssim \log({\rm SFR} / [{\rm M}_\odot\,{\rm yr}^{-1}]) \lesssim 1.23$), the median is moderately higher ($\log({\rm SFR} / [{\rm M}_\odot\,{\rm yr}^{-1}]) = 0.36$) than across PHANGS--ALMA ($0.00$). We list the galaxies studied in this paper in Table~\ref{table:sample} along with some basic properties.

\subsection{Molecular gas surface densities} 
\label{Sec:PHANGS--ALMA}

We measure molecular gas surface densities from the $^{12}$\mbox{CO(2--1)} transition using the public data from the PHANGS--ALMA survey. The observations cover the star-forming disc (typically close to the ${\sim}0.5 R_{25}$ radius\footnote{$R_{25}$ is the radius where the average surface brightness is $\mu_{\rm B} = 25$\,mag\,arcsec$^{-2}$.}) and accounting for ${\sim} 70\%$ of all the CO emission \citep{2021ApJS..255...19L}. To determine the CO integrated intensity, we employ the zeroth-order moment maps based on the `broad' masks presented in \citet{2021ApJS..255...19L} and publicly released in July 2022 (PHANGS--ALMA version~4.0). These masks feature high CO flux completeness at the expense of somewhat higher noise and thus are more suitable for our purpose.

The conversion of the \mbox{CO(2--1)} integrated intensity (in K\,km\,s$^{-1}$) to molecular gas surface density (in M$_\odot$\,pc$^{-2}$) requires adopting both a conversion factor, $\alpha_{\rm CO}$, and a line ratio, $R_{21}= \mbox{CO(2--1)}/\mbox{CO(1--0)}$. 
We employed the $\alpha_{\rm CO}$ prescription adopted in many previous PHANGS studies, which includes a radial trend according to the metallicity gradient in galaxies (\citealt{2020ApJ...901L...8S}, based on \citealt{2017MNRAS.470.4750A}).
% Our nominal PHANGS choice for $\alpha_{\rm CO}$ relies on a prescription which varies only radially (to track the predominant radial gradient in metallicity; Sun+20).
We calculated azimuthal arm/interarm contrasts in radial bins (Sect.~\ref{Sec:method} and Fig.~\ref{fig:method}); thus, in practice, any radial dependence of $\alpha_{\rm CO}$ is effectively cancelled out (other than very small variations within each radial bin). This means that we would measure nearly identical contrasts if we assumed a constant Galactic  of $\alpha_{\rm CO} = 4.35$\,M$_\odot$\,pc$^{-2}$\,(K\,km\,s$^{-1}$)$^{-1}$ or any other prescription that varies only with radius. However, azimuthal variations in the real $\alpha_{\rm CO}$ are not completely negligible and next we estimate the uncertainty introduced by assuming a purely radial conversion factor.

Under the assumption that $\alpha_{\rm CO}$ primarily tracks metallicity variations within galaxies, we indeed expect a predominantly radial trend with much more limited azimuthal effects. Systematic arm-to-interarm gas metallicity variations have been found in some galaxies \citep[e.g.][]{2017ApJ...846...39H,2018A&A...618A..64H,2020MNRAS.492.4149S}, typically of the order of $\sim$0.05\,dex in $12+\log({\rm O/H})$, while in other cases such arm-to-interarm variations were found to be even more limited or absent \citep[e.g.][]{2019ApJ...887...80K,2022MNRAS.509.1303W}. According to the metallicity-dependent PHANGS $\alpha_{\rm CO}$ prescription \citep{2020ApJ...901L...8S}, a change of 0.05\,dex in metallicity would imply a change of 0.07-0.09\,dex in $\alpha_{\rm CO}$. Therefore, we conservatively assume a  0.1\,dex uncertainty on the molecular surface density contrasts as a consequence of neglecting azimuthal changes in $\alpha_{\rm CO}$.

\citet{2024ApJ...961...42T}  recently put forward an $\alpha_{\rm CO}$ prescription that depends on CO velocity dispersion, tracing the CO optical depth (while, for near-solar metallicities, the correlation of $\alpha_{\rm CO}$ with metallicity is interpreted as coincidental, as both tend to drop with increasing radius). The CO velocity dispersion measured on 150\,pc scales and averaged (intensity-weighted) over 1.5\,kpc-sized apertures, $\langle \Delta v \rangle_{\rm 150\,pc}$, does not vary too much between arm and interarm regions. For our sample, the median $\langle \Delta v \rangle_{\rm 150\,pc}$ is only $13$\% higher in spiral arms (with a scatter of ${\sim}25$\%), which has an impact of only ${\sim}10$\% on $\alpha_{\rm CO}$. This is well within the 0.1\,dex uncertainty that we assume (but could introduce a systematic effect on \SigMol\ and SFE contrasts at the ${\sim}10$\% level).

For simplicity, we adopted a constant line ratio of $R_{21}=0.65$ \citep{2013AJ....146...19L,denBrok21}. The latest studies examining $R_{21}$ variations in galaxies, albeit at coarser resolution, measure a typical azimuthal scatter of $\sim$20\% in $R_{21}$ as long as we exclude galaxy outskirts \citep{denBrok21,2022ApJ...927..149L}. This roughly agrees with previous studies, which typically found standard deviations of around 0.1\,dex in $R_{21}$ within a given galaxy \citep{2013AJ....146...19L,2021PASJ...73..257Y}.
The \mbox{CO(2--1)} line ratio is known to vary with local conditions, typically increasing towards galaxy centres and, more generally, mildly increasing with $\Sigma_{\rm SFR}$. \citet{2022ApJ...927..149L} suggest a power-law dependence, $R_{21} \propto \Sigma_{\rm SFR}^{0.13}$, comparing measurements across different galaxies. As a corollary, this could suggest slightly higher $R_{21}$ in spiral arms than {in} matched interarm regions, since $\Sigma_{\rm SFR}$ is typically higher in spirals. However, \citet{denBrok21} found very limited differences in $R_{21}$ between arm and interarm within galaxies (if any, paradoxically larger $R_{21}$ in interarm regions, as seen in M51; \citealt{2022A&A...662A..89D}). Therefore, there is no clear indication for a systematically higher $R_{21}$ in spiral arms, and the observed variations are well within the 0.1\,dex azimuthal scatter that we assume.
Thus, we take 0.1\,dex as a reasonable uncertainty on the molecular contrast due to arm-to-interarm variations in $R_{21}$. Adding this in quadrature with the uncertainty in $\alpha_{\rm CO}$, we consider a global uncertainty of 0.14\,dex in our estimate of molecular gas surface density contrast.

\subsection{Star formation rate measurements} 
\label{Sec:SFRs}

Razza et al.\ (in prep.) obtained ground-based narrow-band H$\alpha$ images for a total of 65 galaxies from the PHANGS--ALMA parent sample. These were observed with the Wide Field Imager (WFI) at the MPG \mbox{2.2-metre} telescope in La Silla or with the DirectCCD camera at the du~Pont \mbox{2.5-metre} telescope in Las Campanas between 2016 and 2019 (these maps have appeared previously, e.g.\ in \citealt{2022ApJ...927....9P}). Both broad-band and narrow-band images were obtained, which allowed us to derive H$\alpha$ continuum-subtracted images. The resolution, limited by seeing, is typically ${\sim} 1\arcsec$ at a full width at half maximum (FWHM) ranging from $0.6\arcsec$ to $1.3\arcsec$, ${\sim}100$\,pc. The H$\alpha$ maps were corrected for filter transmission as well as [{\sc N\,ii}] contamination (assuming $[\textsc{N\,ii}]/\textrm{H}\alpha = 0.3$; see \citealt{2019ApJ...887...49S}).

A robust conversion from H$\alpha$ fluxes to SFR should also account for obscured star formation; otherwise, extinction will make us underestimate SFRs in regions that are strongly obscured by dust. 
We use narrow-band H$\alpha$ observations to trace star formation, and calibrate the impact of extinction using Balmer-corrected H$\alpha$ maps.
This is possible for a subset of 13 galaxies, and for a smaller field of view than the narrow-band maps, where we have extinction-corrected H$\alpha$ SFR maps from PHANGS--MUSE \citep{2022A&A...659A.191E,2023A&A...670A..67B} at ${\sim} 1\arcsec$ resolution. 
These rely on the Balmer decrement (see e.g. \citealt{2012MNRAS.419.1402G}), which allows us to measure extinction based on the $L_{\rm H\alpha}/L_{\rm H\beta}$ ratio (assuming case B recombination, temperature $T = 10^4$\,K, and density $n_{\rm e} = 10^2$\,cm$^{-3}$; see \citealt{2023A&A...670A..67B} for details).

As shown in Appendix~\ref{sec:AppendixMUSE}, there is a tight correlation between the MUSE SFR contrasts with and without extinction (Spearman rank coefficient 0.97), with a systematic amplification of contrasts when extinction is accounted for. We find that the following power-law captures this systematic amplification very well:

\begin{equation}
    \log\left({\rm SFR_{contrast}^{ext-corr}}\right)=-0.0013 + 1.1799\, \log\left({\rm H\alpha_{contrast}^{no-ext}}\right),
\end{equation}

\noindent where ${\rm SFR_{contrast}^{ext-corr}}$ is the arm/interarm contrast of SFR from MUSE, including the extinction correction based on the Balmer decrement, and ${\rm H\alpha_{contrast}^{no-ext}}$ is the arm/interarm contrast of H$\alpha$ from MUSE, switching off the extinction correction. The scatter around this relation (${\sim}0.1$\,dex) gives us an idea of the uncertainty associated with this empirical approach.

Additionally, in Appendix~\ref{sec:AppendixMUSE} we also consider the uncertainty due to narrow-band H$\alpha$ calibration. We estimate this by comparing the narrow-band H$\alpha$ contrasts to the MUSE-based H$\alpha$ contrasts (ignoring extinction). The scatter in this case is slightly smaller than the effect of extinction, but not negligible (0.059\,dex). Therefore, we add both in quadrature to obtain a representative uncertainty on the SFR contrasts of 0.122\,dex. In Appendix~\ref{Sec:DIG} we discuss that diffuse ionised gas (DIG) is unlikely to have a major impact on SFR contrasts according to our observational strategy.

\subsection{\hi{} data} 
\label{Sec:HIdata}

In Section~\ref{Sec:effect_HI}, we consider the effect of including neutral atomic gas in the contrast measurements. We employ 21\,cm interferometric maps from a number of surveys from the Karl G. Jansky Very Large Array (VLA). Most of them come from new VLA observations for PHANGS (data that have previously appeared in \citealt{2022AJ....164...43S} and \citealt{2023arXiv231100407C}), while others come from THINGS \citep{2008AJ....136.2563W}, VIVA \citep{2009AJ....138.1741C}, HERACLES \citep{2009AJ....137.4670L}, or specific programs on individual galaxies.
The resolution of the \hi{} maps ranges from ${\sim}10{-}60''$ (FWHM), depending on the target, which translates into a physical resolution of ${\sim}0.5{-}4$\,kpc for our spiral galaxy sample.

We apply the following equation to transform \hi{} observations to atomic gas surface density, which assumes an optically thin 21~cm emission:

\begin{equation}
    \frac{\Sigma_{\rm atom}}{\rm M_\odot \, pc^{-2}} = 2.0\times10^{-2} \left(\frac{I_{\rm HI}}{\rm K\,km\,s^{-1}}\right) \, \cos{i}~,
    \label{eq:SigHI}
\end{equation}

\noindent where $\Sigma_{\rm atom}$ includes the 1.36 factor to account for helium and heavier elements. The $\cos{i}$ term corrects for galaxy inclination (adopted from the PHANGS Sample Table 1.6, following \citealt{2020ApJ...897..122L}).

\subsection{Stellar mass surface densities} 
\label{Sec:stellarmass}

For most galaxies, we use a set of stellar mass maps based on {\it Spitzer} $3.6$\,$\mu$m imaging obtained from the {\it Spitzer} Survey of Stellar Structure in Galaxies (S$^4$G; \citealt{2010PASP..122.1397S}), with a PSF of $1.7''$. The mass maps rely on an Independent Component Analysis (ICA) correction for dust emission, present at $3.6$\,$\mu$m on top of the photospheric emission from old stars. The method, introduced in \citet{2012ApJ...744...17M}, was applied to the entire S$^4$G sample in \citet{2015ApJS..219....5Q}. It makes use of the neighbouring 4.5\,$\mu$m band and exploits the fact that stellar and dust emission are expected to have very different [3.6]-[4.5] colours. Then, ICA (similar to principal component analysis, PCA) identifies the global stellar and dust [3.6]-[4.5] colours that best describe the two underlying components. Since dust emission at $3.6$\,$\mu$m is strongest around star-forming regions, the arm/interarm contrast nearly always decreases when shifting from the original IRAC $3.6$\,$\mu$m  to the ICA-based dust-corrected stellar mass map. For the galaxies which are not in S$^4$G, we use the original IRAC $3.6$\,$\mu$m maps (not corrected with ICA, see \citealt{2021A&A...656A.133Q} for details).

The conversion from IRAC fluxes to stellar mass surface densities requires adopting a mass-to-light (M/L) ratio. We assume a constant M/L on the ICA-corrected $3.6$\,$\mu$m maps following \citet{2014ApJ...788..144M}, which is expected to provide a reliable conversion to stellar masses within 0.1\,dex (accounting for unconstrained age and metallicity differences, \citealt{2014ApJ...788..144M}). For our purposes, we do not care about the exact normalisation of M/L, or about possible radial variations, as they will cancel out in the arm/interarm contrasts. We still conservatively assume an uncertainty of 0.1\,dex on the arm/interarm stellar surface density contrasts to account for possible arm/interarm variations in M/L (the range of M/L expected at $3.6$\,$\mu$m after correcting with ICA; \citealt{2014ApJ...788..144M}). To test the robustness of our choice for the M/L and our adopted error bar, for 13 galaxies we compare the IRAC contrasts with contrasts measured from the MUSE stellar mass map and we find excellent agreement (see
Appendix~\ref{Sec:stellarcon}). 
We also consider the typical change of 0.04\,dex on the arm/interarm contrast due to the ICA correction as a conservative error bar. Adding both in quadrature, we obtain a global uncertainty of 0.11\,dex for the stellar arm/interarm contrasts. Figure~\ref{fig:MUSE_SigStar_vs_IRAC-ICA} confirms that the stellar contrasts based on the ICA-corrected maps agree well with independent estimates from stellar population fitting in PHANGS--MUSE, and that the uncertainty of 0.11\,dex assumed here is reasonable (see Appendix~\ref{Sec:stellarcon} for details).

\subsection{Spiral masks}
\label{Sec:masks}

\begin{figure}[t]
\begin{center}
\includegraphics[trim=0 0 0 0, clip,width=0.48\textwidth]{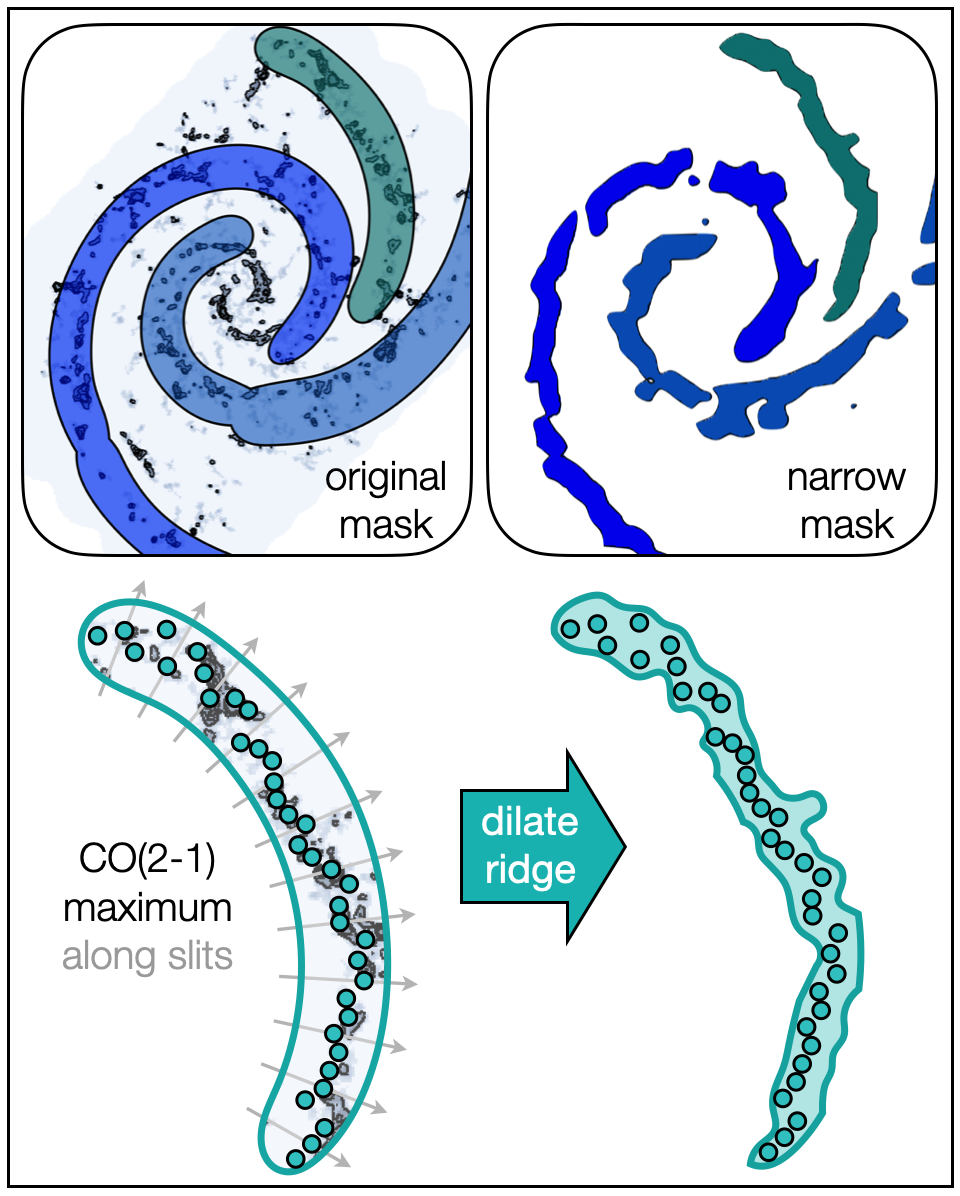}
\end{center}
\caption{Construction of narrow spiral masks, illustrated with NGC\,628. The top panels show the original (left) and new narrow mask (right). The bottom panel illustrates the construction of the narrow mask on the north-western spiral segment of NGC\,628 (green colour). The ridge of peak emission for each tracer (CO, H$\alpha$) is identified along successive cuts perpendicular to the analytic log-spiral function (grey arrows), and limited to the original spiral mask footprint. The ridge is dilated to provide a narrower mask, which is by construction always a subset of the original mask. The final mask is the union of the narrow masks based on the CO and  H$\alpha$ peaks.
}
\label{fig:thin_masks}
\end{figure}

\subsubsection{Original spiral masks} 
\label{Sec:origsp}

Throughout this paper, we use the spiral component of the environmental masks defined in \citet{2021A&A...656A.133Q}. These masks delimit morphological features visually identified on NIR images which, in addition to spiral arms, include centres, bars, rings, lenses, bulges, and discs.

Specifically, spiral masks were constructed following three steps. Firstly, regions of bright $3.6$\,$\mu$m emission were identified along each spiral arm (using an unsharp-mask approach) and fitted with an analytic log-spiral function in the plane of the galaxy. Secondly, analytic log-spiral curves were assigned a width determined empirically based on CO emission. Finally, the resulting masks were visually inspected and the starting and ending point of some segments were extended in order to enforce continuity. For most galaxies, the near-infrared (NIR) images and the analytic log-spiral fits come from S$^4$G \citep{2015A&A...582A..86H}, but we also relied on archival and newly obtained {\it Spitzer} $3.6$\,$\mu$m observations in some cases, where we also performed the log-spiral fitting, as explained in \citet{2021A&A...656A.133Q}.

The spiral masks from \citet{2021A&A...656A.133Q} consist of perfectly smooth, dilated log-spiral segments, with a typical width of ${\sim}1{-}2$\,kpc in order to accommodate most $3.6$\,$\mu$m, CO, and H$\alpha$ emission along the arm. Locally, the distribution of molecular gas or star formation in the arms often looks thinner, but the presence of kinks and irregularities requires this width in order to warrant full coverage along the entire arm (Fig.\,\ref{fig:method}). In this paper, we also consider a more restrictive strategy that results in thinner (and more irregular) spiral masks, as explained next. In Appendix~\ref{Sec:narrow}, we show that contrasts are not strongly affected {by} the width of the spiral masks, and thus, for simplicity, across this paper we use the original masks presented in \citet{2021A&A...656A.133Q}.

\subsubsection{A set of thinner spiral maskss} 
\label{Sec:thinnersp}

To test how the width of the adopted spiral regions affects the arm/interarm contrast, we construct a set of narrower spiral masks\footnote{\url{https://www.canfar.net/storage/vault/list/phangs/RELEASES/Querejeta_etal_2024}}. These masks are not used for the main measurements in this paper, which rely on the public masks from \citet{2021A&A...656A.133Q}. The alternative narrow masks introduced here are used in Appendix~\ref{Sec:narrow} to quantify the effect of mask width, concluding that it should not significantly affect our main conclusions.

The strategy to build narrower masks is illustrated in Fig.\,\ref{fig:thin_masks}. Along the log-spiral curve that defines the backbone of the spiral masks from \citet{2021A&A...656A.133Q}, we determine the line perpendicular to the log-spiral function at each point, and identify the position of the maximum CO or H$\alpha$ intensity along this perpendicular line within the boundaries of the original spiral mask. This process is repeated for all the pixels along the log-spiral curve, resulting in a distribution of pixels of peak intensity from each of these cuts perpendicular to the spiral arm. The corresponding set of pixels are assigned a value of 1, while the rest is set to zero, and the image is smoothed with an empirically chosen Gaussian kernel of ${\rm FWHM}=7.5''$, on which we impose a threshold of 0.01. This is essentially equivalent to dilating the ridge of peak emission, ignoring isolated outliers.

This process is applied independently to the CO and H$\alpha$ maps, resulting in two narrow masks that are very similar, but not identical. We take the union of the two masks to yield a final narrow mask. By construction, the narrow mask is always a subset of the original spiral mask, as we force the dilated mask to be within the original mask. The narrow mask typically covers ${\sim}50$\% of the area of the original mask.
In Appendix~\ref{Sec:narrow} we show how this choice of spiral mask affects the arm/interarm contrast measurements.

\subsection{Strategies to measure arm/interarm contrasts}
\label{Sec:method}

There are different approaches to derive arm/interarm contrasts in nearby galaxies. As illustrated in Fig.\,\ref{fig:method}, our nominal method splits each galaxy into a number of radial bins in the plane of the galaxy; for each radial bin, we consider the mean\footnote{Using the median instead of the mean does not qualitatively affect our conclusions. We prefer the mean as we find that it is more robust, particularly for the interarm region, where the median often falls below the noise level (as already pointed out by \citealt{2021ApJ...913..113M}).} surface density within the footprint of each spiral segment at that radius (arm value), and divide it by the mean surface density within the ring outside the spiral mask (interarm value). We considered the intensity maps without clipping and, therefore, we should not be biased by non-detections. At the outermost radial bins, we require at least 50\% of the spiral and interarm pixels in the ring to lie within the PHANGS--ALMA field of view for the measurement to be considered.
The width of the radial bins is chosen to be 500\,pc in the plane of the galaxy to mitigate stochastic sampling effects\footnote{This is just slightly above the maximum characteristic separation length between independent star-forming regions measured for PHANGS galaxies by \citet{2022MNRAS.516.3006K}; we consider the effect of varying this bin width in Appendix~\ref{sec:sanity_checks}.}, which results in elliptical annuli in the plane of the sky ({according to} the disc inclination and position angle from the PHANGS sample table 1.6, which follow \citealt{2020ApJ...897..122L}).
This allows us to plot each radial bin independently and examine trends with galactocentric radius. 
Each galaxy typically contributes around a dozen measurements in 500\,pc-wide radial bins. Additionally, we also plot with a different symbol the average contrast across each spiral segment (typically two or three datapoints per galaxy, depending on how many spiral arms the galaxy has).

This radial approach has the advantage of simplicity, being easily reproducible, and assigning each pixel uniquely to one bin. To zeroth-order, gas in a spiral galaxy moves on approximately circular orbits, and stellar or gas surface density follow roughly exponential radial profiles; this motivates the interest of splitting bins by radius. However, since the pitch angle of spiral arms varies considerably from galaxy to galaxy, this method implies that, sometimes, a 500\,pc radial bin will encompass a very large area of a spiral, spanning a large interval in azimuth (see Fig.\,1), while in other cases, when the spiral is nearly perpendicular to the elliptical annulus, the resulting spiral bin will have a much smaller area. Thus, despite providing a uniform sampling in galactocentric radius, this method yields an inhomogeneous sampling in terms of spiral arm area.

In Appendix~\ref{sec:sanity_checks}, we consider two alternatives for the sampling of the arm/interarm contrasts. They both consider regularly spaced `boxes' along each spiral segment, whose boundaries cut the log-spiral spine perpendicularly at regular intervals of 500\,pc. 
The first alternative {defines} the interarm as all non-spiral pixels at matched galactocentric radii for each bin.
In the second alternative approach, the spiral bins are defined exactly the same way, but the interarm is measured as adjacent boxes, on both sides immediately next to each spiral bin, to provide a more local reference for the interarm region. Even though the sampling method affects the actual values of the contrasts, our results remain qualitatively the same independently of these different choices.

We note that these methods differ from the arm/interarm contrasts presented in \citet{2021A&A...656A.133Q}. In that paper, the arm-to-interarm contrast was calculated as the ratio of the mean surface density within the spiral mask (considering all spiral segments simultaneously) to the mean surface density in the area outside the spiral mask at matched galactocentric radius. This results in a single number for each contrast per galaxy, and therefore differs from the present approach, which examines each spiral segment {separately} and considers the radial variation of the contrast. Our approach here is also different from the one in \citet{2021ApJ...913..113M}, who examined contrasts based on percentiles of stellar and molecular gas surface brightness as a function of radius, without explicit reference to the spiral morphology. Since we want to specifically study the role of spiral arms in this paper, we instead use the spiral masks from \citet{2021A&A...656A.133Q}. Our results yield a consistent picture with \citet{2021ApJ...913..113M} regarding the relation between stellar and gas contrasts.

\section{Results} 
\label{Sec:results}

We start by presenting our measurements of arm/interarm contrasts at ${\sim}100$\,pc scales
%\footnote{We use the maps at the best resolution available to us in each case, but we note that matching to the best common CO physical resolution ($180$\,pc) has a minimal effect on the contrasts (scatter of ${\sim}3$\%, without systematic offsets), because this is still small compared to the width of the spiral masks (${\sim}1{-}2$\,kpc).}
in stellar mass, molecular gas, and star formation surface density, as well as the SFE (in Sect.~\ref{Sec:highres}). In Sect~\ref{Sec:GD_vs_rest}, we demonstrate that grand-design spirals tend to have larger contrasts. In Sect.~\ref{Sec:correl_stellar}, we examine how contrasts strongly correlate with each other. 
In Sect.~\ref{Sec:SFE_in_spirals} we address the question of whether spiral arms systematically boost star formation efficiency, and we consider the role of atomic gas in Sect.~\ref{Sec:effect_HI}.
We examine the radial variation of contrasts in Sect.~\ref{Sec:radial} {and symmetry between opposite spiral segments in Sect.~\ref{Sec:results_symmetry}. Finally, we assess the effect of resolution in Sect.~\ref{Sec:effect_res}.}
More technical details are presented in the Appendices, including an empirical calibration to account for extinction on SFR contrasts (Appendix~\ref{sec:AppendixMUSE}), a comparison with alternative stellar mass tracers (Appendix~\ref{Sec:stellarcon}), assessing the impact of DIG (Appendix~\ref{Sec:DIG}), and mask width (Appendix~\ref{Sec:narrow}), as well as additional sanity checks (Appendix~\ref{sec:sanity_checks}).

\subsection{Range of arm/interarm contrasts in \SigStar, \SigMol, \SigSFR, and SFE} 
\label{Sec:highres}

\begin{figure*}[t]
\begin{center}
\includegraphics[trim=0 0 0 0, clip,width=0.95\textwidth]{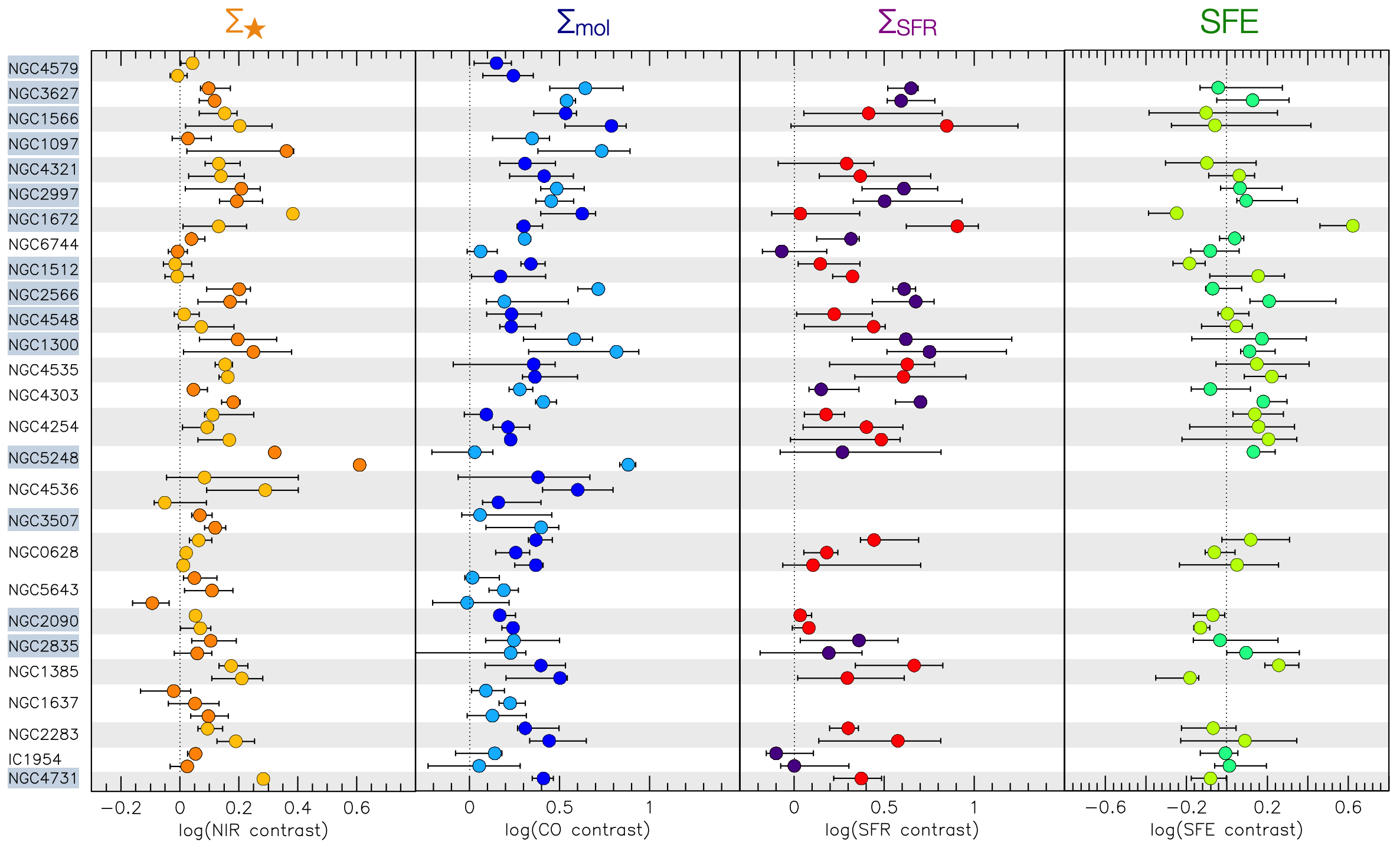}
\end{center}
\caption{Range of contrasts for each spiral segment per galaxy, ordered by decreasing stellar mass of the host galaxy from top to bottom. The circles indicate the median value along each segment and the error bars extend from the 25$^{\rm th}$ to the 75$^{\rm th}$ percentile. For easier visual differentiation, all spiral segments within a given galaxy have the same colour hue. Grand-design spirals have their names highlighted in pale blue.
}
\label{fig:contrast_ranges}
\end{figure*}

\begin{table*}[t!]
\begin{center}
\caption[h!]{arm/interarm contrast measurements.}
\begin{tabular}{lcccc}
\hline\hline
& $\Sigma_\star$  & $\Sigma_\mathrm{mol}$  & $\Sigma_\mathrm{SFR}$  & SFE \\
\hline
   \noalign{\smallskip}
Nominal contrasts & $1.28_{-0.26}^{+0.49}$ & $2.22_{-0.96}^{+2.19}$ & $2.56_{-1.61}^{+4.59}$ & $1.16_{-0.58}^{+1.23}$ \\
   \noalign{\smallskip}
\hline
   \noalign{\smallskip}
Contrasts from MUSE & $1.21_{-0.29}^{+0.46}$ & $2.28_{-0.96}^{+2.12}$ & $2.72_{-1.57}^{+4.55}$ & $1.21_{-0.54}^{+0.93}$ \\
   \noalign{\smallskip}
\hline
   \noalign{\smallskip}
Low resolution (1.5\,kpc) & $1.15_{-0.21}^{+0.48}$ & $1.49_{-0.49}^{+1.35}$ & $1.50_{-0.57}^{+2.20}$ & $1.04_{-0.31}^{+0.56}$ \\
   \noalign{\smallskip}
\hline
   \noalign{\smallskip}
Narrow masks & $1.28_{-0.24}^{+0.53}$ & $2.53_{-1.02}^{+2.83}$ & $3.11_{-1.97}^{+5.45}$ & $1.21_{-0.59}^{+1.23}$ \\
   \noalign{\smallskip}
\hline
   \noalign{\smallskip}
Thinner radial bins (250 pc) & $1.27_{-0.25}^{+0.47}$ & $2.25_{-1.10}^{+2.48}$ & $2.54_{-1.56}^{+4.92}$ & $1.15_{-0.61}^{+1.33}$ \\
   \noalign{\smallskip}
\hline
   \noalign{\smallskip}
Wider radial bins (1000 pc) & $1.27_{-0.29}^{+0.49}$ & $2.14_{-0.81}^{+2.48}$ & $2.34_{-1.34}^{+4.56}$ & $1.14_{-0.51}^{+1.09}$ \\
   \noalign{\smallskip}
\hline
   \noalign{\smallskip}
Perpendicular bins along spiral arms & $1.34_{-0.32}^{+0.68}$ & $2.31_{-1.13}^{+2.94}$ & $2.64_{-1.74}^{+7.30}$ & $1.16_{-0.63}^{+1.62}$ \\
   \noalign{\smallskip}
\hline
   \noalign{\smallskip}
   Perpendicular bins with adjacent interarm & $1.14_{-0.18}^{+0.27}$ & $2.19_{-1.24}^{+3.37}$ & $2.76_{-1.78}^{+6.13}$ & $1.17_{-0.73}^{+1.91}$ \\
   \noalign{\smallskip}
\hline
   \noalign{\smallskip}
Nominal - Grand design spirals & $1.34_{-0.33}^{+0.57}$ & $2.73_{-1.36}^{+3.06}$ & $3.12_{-2.08}^{+6.74}$ & $1.16_{-0.57}^{+1.56}$ \\
   \noalign{\smallskip}
Nominal - Rest of spirals & $1.23_{-0.19}^{+0.30}$ & $1.90_{-0.78}^{+1.04}$ & $2.00_{-1.14}^{+3.15}$ & $1.17_{-0.58}^{+1.06}$ \\
   \noalign{\smallskip}
\hline
   \noalign{\smallskip}
%   Nominal - Barred (23 galaxies) & $1.276_{-0.266}^{+0.451}$ & $2.277_{-1.073}^{+3.067}$ & $2.658_{-1.703}^{+5.780}$ & $1.153_{-0.578}^{+1.534}$ \\
%   \noalign{\smallskip}
%\hline
%   \noalign{\smallskip}
%Nominal - Non-barred (5 galaxies) & $1.275_{-0.236}^{+0.543}$ & $2.086_{-0.782}^{+1.174}$ & $2.246_{-1.274}^{+3.581}$ & $1.171_{-0.567}^{+1.168}$ \\
%   \noalign{\smallskip}
%\hline
%   \noalign{\smallskip}
Nominal - High M$_\star$ & $1.34_{-0.34}^{+0.49}$ & $2.70_{-1.34}^{+3.09}$ & $3.14_{-2.14}^{+6.91}$ & $1.17_{-0.61}^{+1.51}$ \\
   \noalign{\smallskip}
Nominal - Low M$_\star$ & $1.24_{-0.19}^{+0.39}$ & $1.80_{-0.68}^{+1.24}$ & $1.85_{-0.95}^{+3.18}$ & $1.13_{-0.54}^{+1.21}$ \\
   \noalign{\smallskip}
\hline
   \noalign{\smallskip}
Nominal - High SFR & $1.33_{-0.30}^{+0.51}$ & $2.46_{-1.15}^{+2.73}$ & $3.07_{-2.17}^{+5.28}$ & $1.18_{-0.60}^{+1.54}$ \\
   \noalign{\smallskip}
Nominal - Low SFR & $1.19_{-0.20}^{+0.44}$ & $2.02_{-0.90}^{+2.00}$ & $2.13_{-1.17}^{+3.99}$ & $1.11_{-0.53}^{+1.00}$ \\
   \noalign{\smallskip}
\hline
   \noalign{\smallskip}
Nominal - High $\Delta$MS & $1.41_{-0.34}^{+0.48}$ & $2.32_{-1.02}^{+2.92}$ & $3.05_{-2.10}^{+5.33}$ & $1.23_{-0.64}^{+1.57}$ \\
   \noalign{\smallskip}
Nominal - Low $\Delta$MS & $1.17_{-0.19}^{+0.39}$ & $2.06_{-0.93}^{+1.99}$ & $2.16_{-1.25}^{+3.84}$ & $1.09_{-0.53}^{+0.85}$ \\
%   \noalign{\smallskip}
%\hline
%   \noalign{\smallskip}
%Nominal - High T & $1.302_{-0.261}^{+0.523}$ & $2.237_{-0.952}^{+2.288}$ & $2.396_{-1.445}^{+6.026}$ & $1.172_{-0.587}^{+1.187}$ \\
%   \noalign{\smallskip}
%\hline
%   \noalign{\smallskip}
%Nominal - Low T & $1.178_{-0.231}^{+0.416}$ & $2.016_{-0.821}^{+2.376}$ & $2.761_{-1.759}^{+3.540}$ & $1.089_{-0.514}^{+1.615}$ \\
\noalign{\smallskip}
\hline
\hline
\end{tabular}
\label{table:stats}
\end{center}
\tablefoot{arm/interarm contrasts in the surface density of stellar mass, molecular gas, star formation rates, and SFE (median and 16$^{\rm th}$--84$^{\rm th}$ percentile range, measured on galactocentric rings of 500\,pc width). The nominal contrasts rely on the masks from \citet{2021A&A...656A.133Q}, while the narrow masks refer to those introduced here in Sect.~\ref{Sec:thinnersp}. The two alternative sampling schemes based on perpendicular bins are explained in Appendix~\ref{sec:sanity_checks} (Fig.~\ref{fig:binning_alternatives}). Low and high M$_\star$ refer to the bottom or top 50\% of stellar masses in the sample (threshold $3.4 \times 10^{10}$\,M$_\odot$); analogously for SFRs (threshold $2.3$\,M$_\odot$\,yr$^{-1}$) and $\Delta$MS (threshold $0.234$\,dex).}
\end{table*}

For each spiral segment, we measure the arm/interarm contrast in the surface density of stellar mass (\SigStar), molecular gas (\SigMol), star formation rate (\SigSFR), and SFE of the molecular gas (which we define as the inverse of the depletion time, ${\rm SFE} = 1/ \tdep = \SigSFR / \SigMol$). The measurements are performed on radial bins following elliptical annuli of matched galactocentric radii with a width of 500\,pc, as explained in Sect.\,\ref{Sec:method} and illustrated in Fig.\,\ref{fig:method}. We only consider measurements if at least 50\% of the pixels in a given spiral bin (and 50\% of the corresponding interarm region) fall within the PHANGS--ALMA footprint, which is typically rectangular.
This condition excludes NGC\,1365 completely, as the CO coverage beyond the bar is very limited. When performing this analysis, we also lacked narrow-band H$\alpha$ data for six galaxies\footnote{The galaxies without high-resolution PHANGS SFR maps are: NGC1097, NGC1637, NGC3507, NGC4536, NGC4579, and NGC5643.}, for which we could only consider $\Sigma_\star$ and $\Sigma_\mathrm{mol}$. This results in a total of 27 spiral galaxies (excluding NGC\,1365, as explained above) and 59 spiral segments with \SigStar\ and \SigMol\ measurements, out of which 21 galaxies (44 spiral segments) also have \SigSFR\ information. Table~\ref{table:stats} summarises the contrast distributions that we obtain for different galaxy subsets and approaches. To make measurements more directly comparable, in Table~\ref{table:stats} and the following figures involving SFRs, we only include stellar and CO contrasts for galaxies where SFR is available (i.e.\ the \SigStar\ and \SigMol\ ranges correspond to the same set of radial bins as the SFR contrasts).

Figure~\ref{fig:contrast_ranges} shows the median and dispersion of arm/interarm contrasts for each spiral segment across PHANGS galaxies {ordered by decreasing stellar mass}. The stellar contrast tends to be smaller and with more limited fluctuations, whereas the CO or SFR contrasts tend to span large ranges (their $x$ axis covers different ranges), implying that, even within a given galaxy and along a single spiral segment, there are significant fluctuations with galactocentric radius. 
In some cases, these fluctuations even lead to negative contrasts in the logarithmic scale, which is not necessarily unphysical, and simply points to radial bins where the mean interarm value is slightly higher than the mean arm value.
Looking at the distribution of SFE contrast, we see that it is not systematically positive; while some galaxies do seem to show a slight preference for either enhanced or suppressed SFE in spirals, in most cases, the SFE contrast shows values both above and below unity within a galaxy. This already highlights the diversity of physical conditions in spiral arms; we examine SFE in more detail below in Sect.~\ref{Sec:SFE_in_spirals}.

\begin{figure*}[t]
\begin{center}
\includegraphics[trim=0 0 0 0, clip,width=0.95\textwidth]{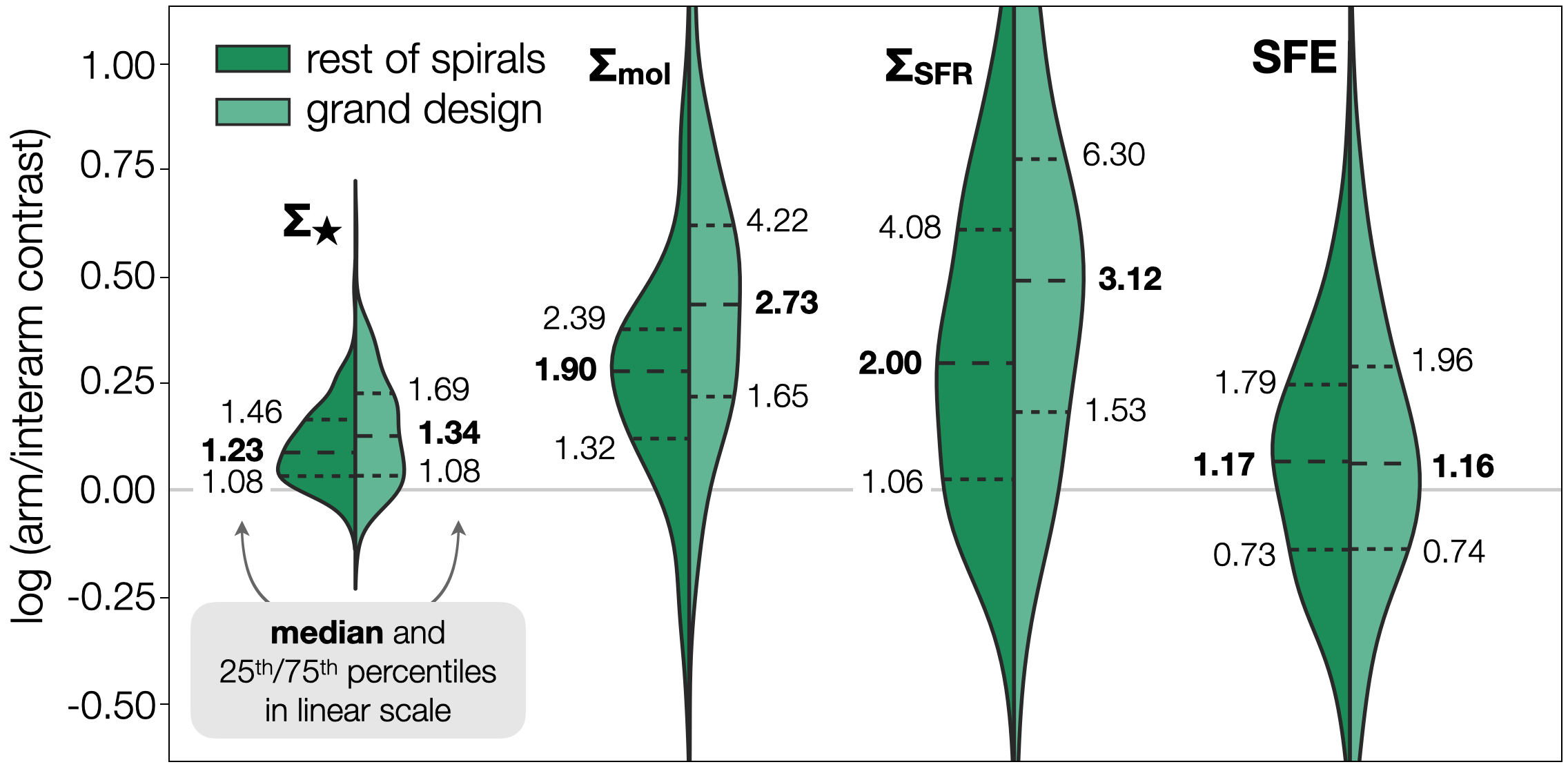}
\end{center}
\caption{Violin plot showing the distribution of stellar mass, molecular gas, star formation rate surface density, and SFE ($1/\tau_{\rm dep}$) arm/interarm contrasts in logarithmic scale for all radial bins across the PHANGS targets. The violin plots are split into grand-design galaxies (right) and the rest of spirals (left). The long dashed line shows the median of the distribution in each case, while the short dashed lines display the 25$^{\rm th}$ and 75$^{\rm th}$ percentiles of the data, with labels indicating for reference the corresponding values on a linear scale.
}
\label{fig:violin_plot_GD-vs-rest}
\end{figure*}

Figure~\ref{fig:violin_plot_GD-vs-rest} shows the aggregated distribution of contrasts across all the PHANGS radial bins in our sample at ${\sim}100$\,pc resolution. For a more fair comparison, here we only include bins for which \SigStar, \SigMol, and \SigSFR\ measurements exist simultaneously at a given radius. It immediately stands out that the range of stellar mass contrasts is fairly modest compared to the contrast in molecular gas or star formation. The stellar contrast is typically between a few percent and around 50\%, much smaller than $\Sigma_\mathrm{mol}$ or $\Sigma_\mathrm{SFR}$ contrasts, which often reach factors of ${\sim}2{-}3$ (see Table\,\ref{table:stats}). This is because even a moderate spiral potential can trigger a strong response in the gas (as shown by \citealt{1969ApJ...158..123R}). \citet{2021ApJ...913..113M} also found much smaller stellar than molecular gas contrasts in the PHANGS sample (based on flux percentiles in galactocentric radial bins, without explicitly defining spiral masks). 

The \SigMol\ and \SigSFR\ contrasts follow a wide distribution, from just a few percent up to ${\sim}10$ in the most extreme cases. The median contrast is slightly higher for \SigSFR\ than \SigMol, but the scatter is also particularly pronounced, exceeding the 16$^{\rm th}$--84$^{\rm th}$ percentile range of \SigMol\ contrasts by as much as a factor of two (Table\,\ref{table:stats}). This implies that the fluctuations in  arm/interarm contrast are larger for star formation than for molecular gas. 
This could reflect the fact that spiral arms act directly on the gas, and indirectly on star formation, involving additional physics that controls the collapse of gas to form stars and which introduces scatter. We discuss this further in Sect.\,\ref{Sec:discusscontr}.

One of the main questions that motivate this paper is quantifying whether star formation proceeds more efficiently in spiral arms than in interarm regions when examined at 100\,pc scales. The right panel of Fig.\,\ref{fig:violin_plot_GD-vs-rest} provides an answer to this: in some cases SFE is indeed enhanced in spiral arms, but in (many) other cases the opposite happens. We examine this problem in greater detail below in Sect.~\ref{Sec:SFE_in_spirals}.

The magnitude of the molecular and SFR contrasts that we measure broadly agrees with previous results from the literature (most of them at lower spatial resolution), where typical gas surface densities were found to be a few times higher in spiral arms than in interarm regions \citep[e.g.][]{1988Natur.334..402V,1993A&A...274..123G,1996MNRAS.283..251K,2003PASJ...55..191N,2009A&A...495..795H,2020ApJ...901L...8S}. In \citet{2021A&A...656A.133Q} we also measured for PHANGS galaxies, using the same nominal masks albeit at lower resolution, that spiral arms have typically ${\sim}2$~times higher molecular gas and SFR surface densities than the interarm, with a range from less than a factor of~$2$ up to a factor of ${\sim}10$ on a linear scale. 
Our results are also in qualitative agreement with the analysis presented in \citet{2021A&A...650A.134P} and \citet{2022A&A...663A..61P}, who examined star formation scaling relations involving \SigSFR, \SigMol, and \SigStar\ in the PHANGS--MUSE sample (18 galaxies). Their main conclusion is that the relations involving stellar surface density vary strongly with galactic environment, with a higher normalisation in \SigSFR\ and \SigMol\ as a function of \SigStar\ for spiral arms than for `disc' (which includes interarm regions). This implies that, while globally there is a nearly linear correlation between stellar mass and molecular gas mass in galaxies, in detail, the same relation cannot hold everywhere within discs, because \SigMol\ arm/interarm contrasts (or \SigSFR\ contrasts) are much larger than the contrasts in \SigStar (as explicitly shown by \citealt{2021ApJ...913..113M}). This results in a higher normalisation in the scaling relations of \SigMol\ and \SigSFR\ as a function of \SigStar\ for spiral arms \citep{2021A&A...650A.134P,2022A&A...663A..61P}.

%Equivalently, the stellar $C_\star$ coefficient in the \SigSFR-\SigMol-\SigStar\ plane is higher in spiral arms. This can be understood as an indirect manifestation of the higher \SigMol\ and \SigSFR\ arm/interarm contrasts compared to the \SigStar\ contrasts, which agrees with what we have measured in this paper. Indeed, since the molecular gas and SFR arm/interarm contrasts exceed the stellar contrast, this means that, at fixed \SigStar, we expect a higher \SigMol\ and \SigSFR\ in spiral arms than in the interarm, as implied by the higher normalisation in the scaling relations of \SigMol\ and \SigSFR\ as a function of \SigStar\ \citep{2021A&A...650A.134P,2022A&A...663A..61P}.

In detail, the measured contrast in a given galaxy depends on resolution and, to a lesser degree, on methodology (which we discuss in Sect.~\ref{Sec:discuss_methodology}). For example, for NGC\,4321, \citet{1996A&A...308...27K} obtained an average molecular contrast of ${\sim}1.7$ and ${\sim}1.8$ for the north and south spiral, respectively, at a resolution of $15''$ (${\sim}1$\,kpc). They targeted individual pointings with the Nobeyama 45\,m telescope along the arms and immediate interarm regions (similar to the second binning alternative that we introduce in Appendix\,\ref{sec:sanity_checks}). Despite the methodological differences, these molecular contrasts are very similar to the ones we obtain at 1.5\,kpc resolution, $1.8$ and $1.6$ for each arm, respectively. However, when we shift to high resolution, the molecular contrasts become 2.7 and 2.4 for the north and south arm, respectively. This warns against direct comparison of contrasts derived at different resolutions.

\subsection{Grand-design spirals show larger contrasts (and effect of $M_\star$, SFR, $\Delta$MS)} 
\label{Sec:GD_vs_rest}

The violin plots in Fig.~\ref{fig:violin_plot_GD-vs-rest} are split into grand-design spirals (17 galaxies) and the rest (multi-armed and flocculent, 11 galaxies). We follow the NIR visual classification from \citet{2015ApJS..217...32B} for galaxies in the S$^4$G survey, and the definitions adopted in \citet{2021ApJ...913..113M} otherwise.
We find a significantly different distribution between the contrasts in grand-design galaxies and the rest for \SigStar, \SigMol, and \SigSFR. For each tracer, the contrast is higher for grand-design spirals: the median {surface density} enhancement is 50\% higher for stellar mass, and almost a factor of two higher for molecular gas and star formation. 
\citet{2021ApJ...913..113M} also found that, at fixed 3.6\,$\mu$m contrast, CO contrasts are slightly larger in grand-design spirals than in multi-arm and flocculent galaxies. 
We ran a Kolmogorov-Smirnov test for the two populations, finding a vanishingly small probability ($p<0.1\%$) that the grand-design contrasts are drawn from the same population as the rest of spirals for \SigStar, \SigMol, and \SigSFR. This is not the case for SFE contrasts, though ($p=0.49$), which do not show a significant difference for grand-design spirals. A Jackknife approach also confirms this: if we remove one galaxy at a time and recompute medians, in all cases the \SigStar, \SigMol, and \SigSFR\ contrast remains higher in grand-design spirals than in the rest. If we consider the \SigStar\ contrast for an extended field of view (all \SigStar\ measurements across {entire} spiral arms, no longer limited to the PHANGS--ALMA field of view), the difference between grand-design spirals and the rest becomes even more acute, with a median \SigStar\ enhancement that is three times larger for grand-design (61\%) than for the other spiral morphologies (21\%), as shown in Fig.\,\ref{fig:violin_plot_GD-vs-rest-extended}.

The difference in contrasts for grand-design spirals is also mirrored by a difference if we focus on the most massive galaxies. If we split the sample into the 50\% of galaxies with the higher stellar masses (above the median $3.4 \times 10^{10}$\,M$_\odot$) and the lower stellar masses (below that median), we find higher contrasts for the higher-mass sub-sample, by a similar amount as for grand-design spirals (Table\,\ref{table:stats}).
The effect is very similar to the split between grand design galaxies and the rest; indeed, grand-design spirals are known to occur preferentially on galaxies with higher stellar masses \citep[e.g.][]{2017MNRAS.471.1070B,2023MNRAS.518.1022S}, and this is also the case for PHANGS galaxies \citep{2023A&A...676A.113S}. Out of the 14 galaxies with the highest stellar masses, 12 galaxies (86\%) are grand design, so both sub-samples are closely related. Yet, the difference is not driven exclusively by morphology, because even within grand-design spirals, we also {found}  larger contrasts for the galaxies with higher stellar masses. We find moderate but positive rank correlation coefficients between the \SigStar, \SigMol, and \SigSFR\ contrasts and total stellar mass of each galaxy (Spearman $\rho \sim 0.2$ for all datapoints, $p$ value $<1$\%), but not between SFE contrast and the stellar mass of the galaxy. The correlations are similarly strong if we instead consider total SFR, and we also find a similar change in the distribution of contrasts when splitting the {spiral} sample into two halves based on total SFR (above and below the median of 2.29\,M$_\odot$\,yr$^{-1}$; also related to the split by stellar mass or morphology). However, we do not find any remarkable differences when splitting the sample based on being barred or unbarred, or based on early {or} late Hubble type.

In Table\,\ref{table:stats}, we also consider the offset from the star-forming main sequence of galaxies ($\Delta$MS, based on \citealt{2019ApJS..244...24L}), i.e.\ the vertical offset of each galaxy with respect to the best fit to the relation between SFR and $M_\star$. We find a clear trend for higher contrasts in the galaxies that lie preferentially above the main sequence ($\Delta$MS above $0.234$\,dex, the median across the sample), comparable to the one for high $M_\star$. Interestingly, in this case we also find a difference for SFE contrast, which increases for galaxies in the upper half of the sample in terms of $\Delta$MS (median $1.23$ as opposed to $1.09$). Compared to the separation into grand design spirals and the rest, this slightly different SFE behaviour arises from similarly high SFR contrasts but somewhat more limited molecular contrasts for galaxies with a high $\Delta$MS. This is perhaps not too surprising; by imposing a high $\Delta$MS, we are preferentially selecting galaxies with comparatively high SFR (high SFR/$M_\star$), but at the same time those with a more limited $M_\star$, which results statistically in less extreme \SigMol\ contrasts. In any case, this demonstrates that the position of a galaxy on the star-forming main sequence matters in terms of the type of spiral response (even though, of course, there is significant scatter, as attested by the large range of 16$^{\rm th}$--84$^{\rm th}$ percentile range).

\subsection{Correlations among contrasts} 
\label{Sec:correl_stellar}

\begin{figure*}[t]
\begin{center}
\includegraphics[trim=0 0 0 0, clip,width=0.95\textwidth]{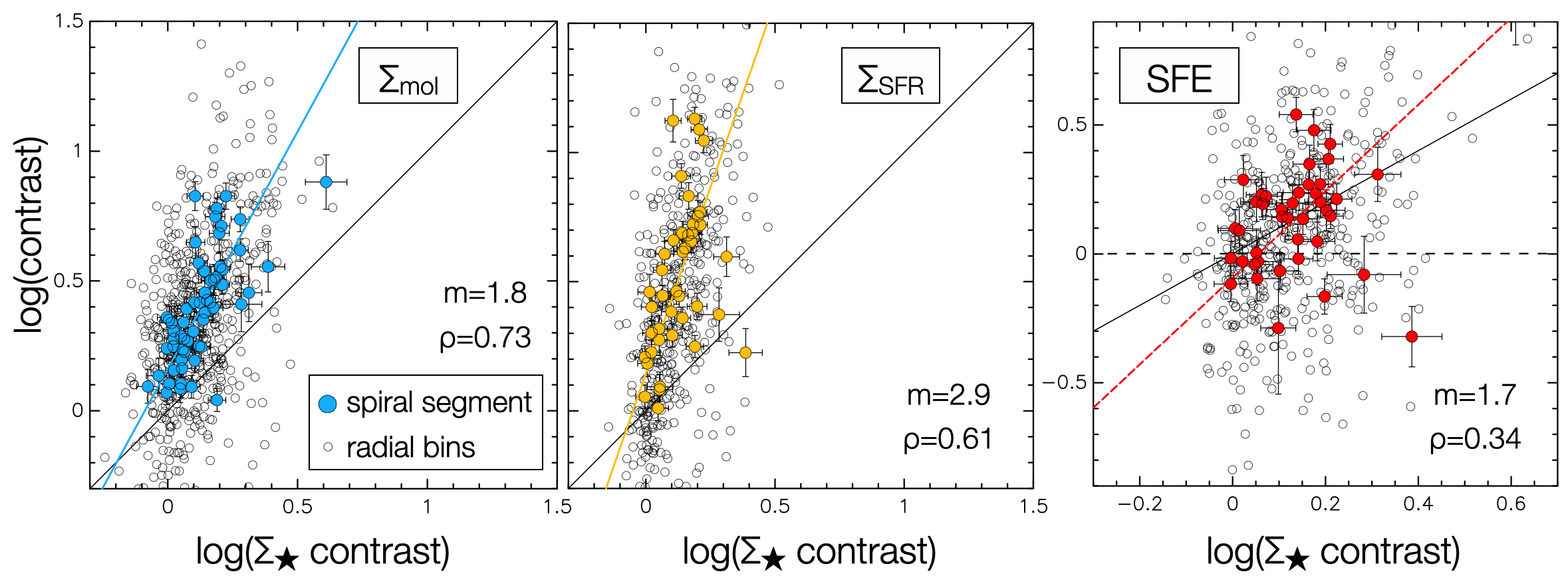}
\end{center}
\caption{arm/interarm molecular gas, SFR, and SFE contrast as a function of stellar mass contrast. The solid black line marks the 1:1 relationship. All radial datapoints are shown as open circles in the background. The coloured solid line indicates the best bisector fit to the coloured circles, which represent the average per spiral segment. The fits are provided in Table\,\ref{table:stats_appendix}, and we list the slope as $m$ inside the plot. The strength of the correlation for the coloured circles is indicated by $\rho$.).
}
\label{fig:CO_SFR_SFE_vs_NIR}
\end{figure*}

\begin{figure}[t]
\begin{center}
\includegraphics[trim=0 0 0 0, clip,width=0.43\textwidth]{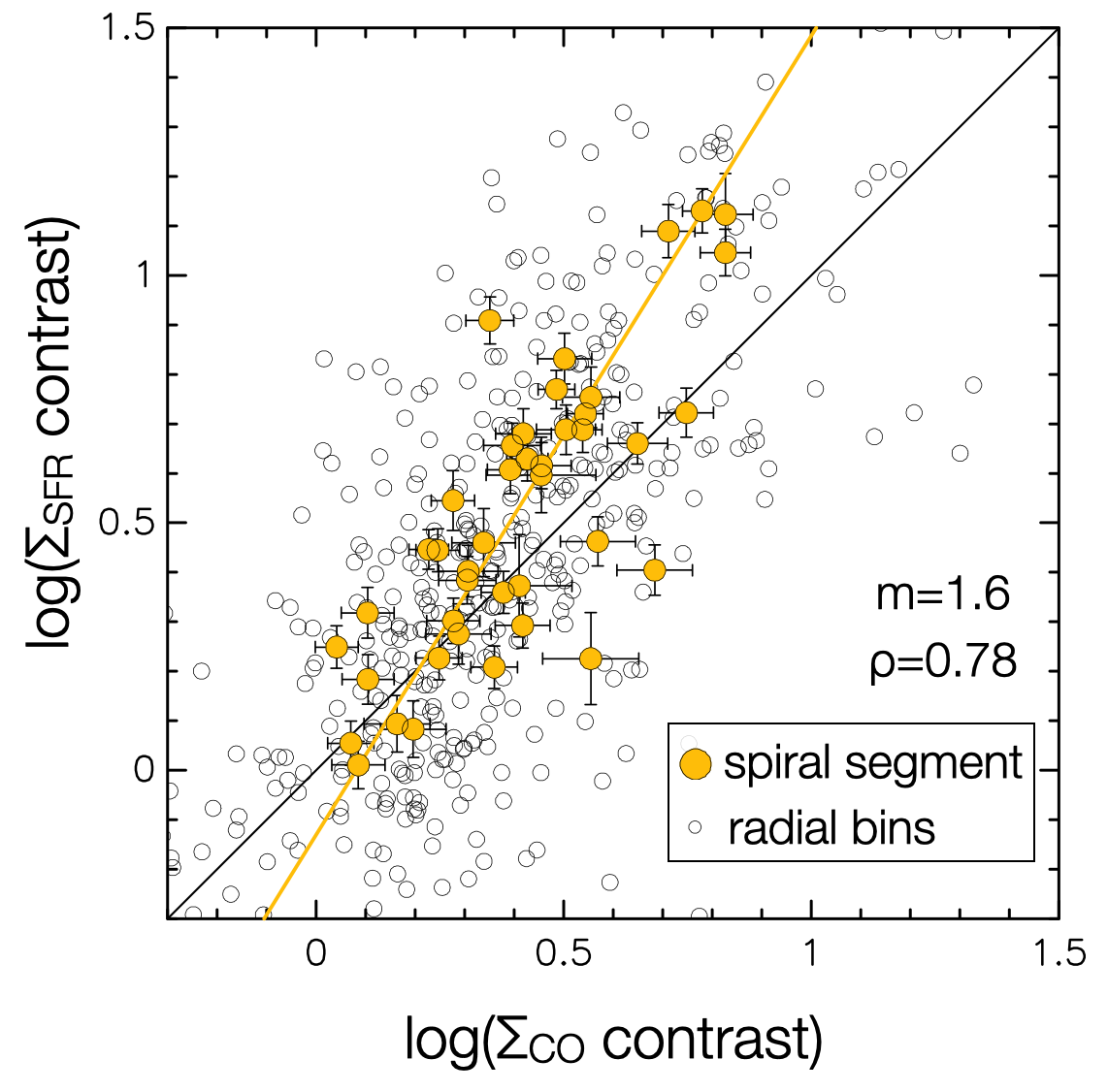}
\end{center}
\caption{arm/interarm molecular gas versus SFR contrast. Symbols as in Fig.\,\ref{fig:CO_SFR_SFE_vs_NIR}.
}
\label{fig:CO_vs_SFR}
\end{figure}

Figure~\ref{fig:CO_SFR_SFE_vs_NIR} shows a clear correlation between the contrast in \SigMol\ and \SigStar, and a similarly strong correlation arises between \SigSFR\ and \SigStar. This suggests that the stellar contrast largely dictates how strongly molecular gas and star formation pile up in a given arm. This is perhaps not too surprising, since the stellar contrast sets the depth of the spiral gravitational potential well. The Spearman rank correlation coefficients are $0.73$ and $0.61$, respectively. The trend is superlinear, in the sense that the response of the gas (and star formation) is amplified relative to the contrast in the stellar surface density. In the case of CO, the slope from a bisector fit is $m=1.8$, while for SFR it is even steeper, $m=2.9$. The rank coefficients and linear fits refer to the mean contrasts over entire spiral segments (more details, including all radial bins and uncertainties for the fits, in Table\,\ref{table:stats_appendix}).

\citet{2021ApJ...913..113M} measured CO contrasts at 150\,pc resolution as a function of 3.6\,$\mu$m contrast, and found that they follow approximately a 2:1 relation on a log–log scale, with some vertical offset above the reference 2:1 line, and with a Pearson correlation coefficient $R \sim 0.5$.
Our measured slope of $m=1.8$ is thus in good agreement with \cite{2021ApJ...913..113M}, but we do not find a vertical offset here, and we find a tighter correlation with Spearman $\rho \sim 0.73$, which could arise from the different method followed (percentiles versus environmental masks). This superlinear relation is consistent with the expectation if gas is compressed as it flows supersonically in response to non-axisymmetric stellar structures \citep{1969ApJ...158..123R,2021ApJ...913..113M}.

Despite the strong correlation between stellar contrast and the contrast in molecular gas and SFR, the correlation between the contrasts in SFE and stellar surface density is much weaker ($\rho = 0.34$). If we perturb the data points using a Gaussian distribution according to the error bars, the weak positive correlation vanishes (with a bootstrap of $N=10\,000$ runs, the average rank coefficient drops to $\rho=-0.09$). We comment {on} the impact on SFE  in more detail below in Sect.~\ref{Sec:SFE_in_spirals}. 

We also examine the contrast in \SigSFR\ as a function of \SigMol\ contrast, as shown in Fig.~\ref{fig:CO_vs_SFR}. If the efficiency remains on average similar in spiral arms and interarm regions, we expect the \SigSFR\ as a function of \SigMol\ contrast to be connected by the Kennicutt-Schmidt relation \citep{1959ApJ...129..243S,1998ApJ...498..541K}.
Indeed, we find an even stronger correlation ($\rho=0.78$) than {between} \SigMol\ or \SigSFR\ and \SigStar. The best-fit regression relating the \SigSFR\ contrast to the \SigMol\ contrast is slightly superlinear (slope $m=1.6$), implying that the star formation contrast exceeds on average the molecular contrast slightly for the cases where the \SigMol\ contrast is highest. In other words, the spiral segments or bins with the highest contrasts typically show slightly more efficient star formation in the arms than in interarm regions. This is in agreement with the finding that there is a slightly increased SFE contrast for the highest \SigStar\ contrast (which at the same time corresponds to the highest \SigMol\ contrasts). We discuss this in more detail in Sect.~\ref{Sec:SFE_in_spirals}.

\subsection{Star formation efficiency in spiral arms} 
\label{Sec:SFE_in_spirals}

%(4) SFE contrast shows scatter, but median close to unity

Here we analyse whether spiral arms preferentially trigger the collapse of molecular gas to form stars, systematically boosting its SFE. We emphasise that by default we consider the efficiency associated with molecular gas only, ${\rm SFE} = \SigSFR / \SigMol = 1/ \tdep$ (sometimes labelled ${\rm SFE_{mol}}$). This is different from the SFE of the total gas (\hi{}$+$H$_2$); we will examine the effect of including atomic gas below in Sect.~\ref{Sec:effect_HI}.

The right panel of Fig.\,\ref{fig:violin_plot_GD-vs-rest} demonstrates that SFE is not always enhanced in spiral arms. 
From a total of 422 radial bins, we find 244 bins in which SFE is higher in spiral arms (while in 178 bins it is higher in interarm regions). Thus, if we pick up random locations along spiral arms, in roughly six out of ten cases we will find a higher SFE in spirals.

The median SFE contrast across all radial bins is $1.16$. If we bootstrap the SFE values within their error bars $10\,000$ times, we confirm that the enhancement of the median with respect to unity is statistically significant (at the $4\sigma$ level). 
With a Jackknife approach, if we randomly remove half of the datapoints and repeat the process $N=10\,000$ times, the mean remains $1.16$ with a standard deviation of $0.03$ (in fact, only in ${\sim}5$ out of $10\,000$ random cases does removing half of the bins result in a mean SFE contrast below unity). This confirms that the result is robust. Yet, this enhancement of ${\sim}10{-}20$\% in SFE is modest compared to the actual enhancement in \SigMol\ (median enhancement 120\%), \SigSFR\ (160\%), and is even below the limited enhancement of \SigStar (28\%).

Our SFE measurements rely on H$\alpha$ fluxes corrected for extinction following the empirical calibration from Appendix\,\ref{sec:AppendixMUSE}. If we were to take SFR as directly proportional to H$\alpha$, then the SFE enhancement would completely vanish, as shown in Fig.~\ref{fig:violin_plot_GD-vs-rest-Ha-only}. This emphasises the importance of the extinction correction for SFR and SFE contrasts. For the subsample of 13 galaxies in PHANGS--MUSE, we find a median SFE contrast of $1.21$ from extinction-corrected H$\alpha$ (Balmer decrement). For the same galaxies and field of view as PHANGS--MUSE, our nominal SFE contrast has a median of $1.19$, which reassuringly demonstrates the success of our strategy to account for extinction on arm/interarm contrasts (Appendix\,\ref{sec:AppendixMUSE}). Conversely, applying the velocity dispersion-dependent $\alpha_{\rm CO}$ from \citet{2024ApJ...961...42T} leads to an increase of \SigMol\ contrasts at the ${\sim}10$\% level, which results in SFE contrasts which are ${\sim}10$\% higher.

In Figure~\ref{fig:CO_SFR_SFE_vs_NIR} we found a strong correlation between \SigStar\ and \SigMol\ or \SigSFR\ contrasts, but a weaker correlation for SFE contrasts (right panel of Fig.~\ref{fig:CO_SFR_SFE_vs_NIR}, $\rho = 0.34$). This suggests that, while the stellar mass contrast is a clear driver of the accumulation of gas and, indirectly, star formation, it is not as directly linked, in general, with a boost in SFE. Yet, if we focus on the largest \SigStar\ contrasts, the median SFE contrast increases significantly. For example, for the top 50\% in  \SigStar\ contrasts ($>1.23$, or $>0.09$ in log), the median SFE contrast rises to $1.40$, and goes up to $2.34$ if we focus on the top 10\% of \SigStar\ contrasts ($>1.97$, or $>0.29$ in log). Therefore, the largest stellar contrasts do result, on average, in a noticeable increase of SFE. When splitting the sample into grand-design spirals and the rest, a difference in SFE contrasts is not apparent.

Our findings agree with some previous studies which did not find significant differences among spiral types in terms of SFE, either observationally \citep[e.g.][]{2008AJ....136.2846B,2008AJ....136.2782L,2016ApJ...827..103K,2010ApJ...725..534F} or using simulations \citep[e.g.][]{2011MNRAS.417.1318D}. Our results also roughly agree with the analysis of PHANGS--MUSE galaxies presented in \citet{2021A&A...650A.134P,2022A&A...663A..61P}. The arm versus disc variation in the resolved Kennicutt-Schmidt relation (\SigSFR\ vs \SigMol) was found to be very limited, the same as the corresponding coefficients in the \SigSFR-\SigMol-\SigStar\ plane, in agreement with the SFE arm/interarm contrast oscillating around unity that we have found in this paper.  In particular, \citet{2021A&A...650A.134P} found that spirals show slightly higher \SigSFR\ at fixed \SigMol\ compared to disc values, but only below $\Sigma_\mathrm{mol} \lesssim 100$\,M$_\odot$/pc$^2$ (at higher $\Sigma_\mathrm{mol}$, discs are actually slightly above spirals in the molecular Kennicutt-Schmidt relation; see their Fig.~9). We emphasise that the binned data in \citet{2021A&A...650A.134P} aggregate measurements from different galaxies and the disc environment includes both interarm regions and discs without spiral masks. Therefore, this is not necessarily representative of the arm/interarm contrasts measured within individual galaxies, and the fact that the offset is small agrees with our overall findings. Our results also agree with Romanelli et al.\ (in prep.), who do not find significant differences between arm and interarm regions in terms of molecular cloud SFE (inferred cloud lifetime divided by depletion time).

\subsection{Atomic gas contrast and its effect on SFE} 
\label{Sec:effect_HI}

\begin{figure}[t]
\begin{center}
\includegraphics[trim=0 0 0 0, clip,width=0.45\textwidth]{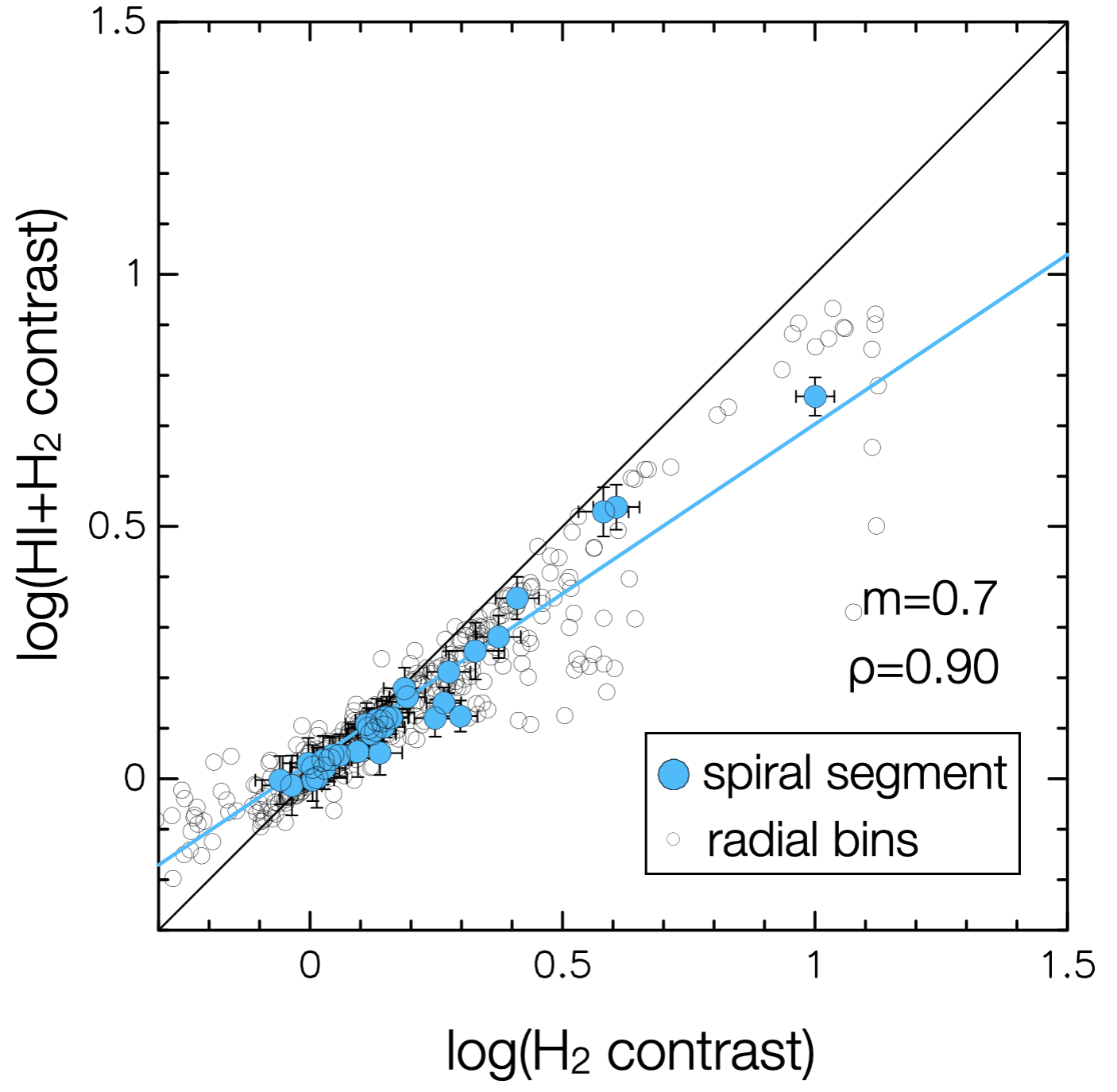}
\end{center}
\caption{Effect of \hi{} on the spiral arm/interarm contrasts. The plot shows the contrast of total gas (adding up both atomic and molecular surface densities, \hi{}$+$H$_2$) as a function of the molecular contrast (H$_2$). The measurement are performed matching the resolution of \hi{} for each galaxy and over the same field of view. Symbols as in Fig.\,\ref{fig:CO_SFR_SFE_vs_NIR}.
}
\label{fig:effect_HI}
\end{figure}

\begin{table}[t!]
\begin{center}
\caption[h!]{Effect of including atomic gas on arm/interarm contrasts.}
\begin{tabular}{lcc}
\hline\hline
\noalign{\smallskip}
 & $\Sigma_\mathrm{gas}$ contrast  &  SFE contrast \\
\noalign{\smallskip}
\hline
\noalign{\smallskip}
\hi{}  &  $1.22_{-0.21}^{+0.57}$ & $1.21_{-0.41}^{+1.05}$ \\
   \noalign{\smallskip}
\hline
   \noalign{\smallskip}
\hi{}+H$_2$  &  $1.26_{-0.27}^{+0.88}$ & $1.09_{-0.29}^{+0.49}$ \\
   \noalign{\smallskip}
\hline
   \noalign{\smallskip}
H$_2$  (\hi{} galaxies) & $1.31_{-0.35}^{+1.31}$ & $1.01_{-0.24}^{+0.36}$ \\
\noalign{\smallskip}
\hline
   \noalign{\smallskip}
H$_2$ high resol.\ (\hi{} galaxies) & $2.43_{-1.08}^{+3.24}$ & $1.17_{-0.55}^{+1.05}$ \\
   \noalign{\smallskip}
\hline
   \noalign{\smallskip}
$\Sigma_\mathrm{mol}$/$\Sigma_\mathrm{atom}$  & $1.13_{-0.27}^{+0.55}$ & --- \\
%& $1.218_{-0.376}^{+1.019}$ \multicolumn{2}{c}{xxxx}
\noalign{\smallskip}
\hline
\hline
\end{tabular}
\label{table:includingHI}
\end{center}
\tablefoot{Gas and SFE contrasts (median and 16$^{\rm th}$--84$^{\rm th}$ percentile range, measured on galactocentric rings of 500\,pc width). Limited to 15 spiral galaxies where \hi{} observations achieve a spatial resolution better than 2.5\,kpc.}
\end{table}

So far, we have considered the contrasts based on molecular gas, without accounting for atomic gas. This seems like a reasonable choice, since molecular gas often dominates the inner parts of galaxies, and is more closely related to star formation \citep{2008AJ....136.2846B,2013AJ....146...19L}. Another reason to focus on the molecular phase is the higher resolution of our CO observations compared to \hi{}. Yet, the arm/interarm contrast in atomic gas is interesting, and it can tell us not only about the accumulation of gas, but also the transformation of atomic to molecular gas driven by spiral compression.

In this Section, we consider the impact of including atomic gas in arm/interarm contrasts, which we can unfortunately only examine at lower resolution (a few tens of arcsec for \hi{}, i.e.\ $\sim$kpc scales, instead of ${\sim} 100$\,pc scales for CO). We limit our analysis to the cases where \hi{} achieves a physical resolution of 2.5\,kpc or better (15 galaxies, keeping in each case the original \hi{} resolution, matching CO to this resolution), restricted to the extent of the PHANGS--ALMA field of view. Table~\ref{table:includingHI} lists the distribution of arm/interarm contrasts for this subsample of \hi{} galaxies. 
The H$_2$ contrast drops very significantly from high to low resolution (matched to \hi{}), and the SFE contrast also drops slightly.
We examine the effect of resolution more closely in Sect.~\ref{Sec:effect_res}. Even at matched resolution, the \hi{} contrast is typically ${\sim}10$\% lower than the H$_2$ contrast (while \hi{}+H$_2$ shows intermediate contrasts). This is also confirmed if we explicitly look at the molecular-to-atomic ratio ($\Sigma_\mathrm{mol}$/$\Sigma_\mathrm{atom}$), which shows a $13$\% higher median in spiral arms.

Figure~\ref{fig:effect_HI} compares point-by-point the contrast of total gas (\hi{}+H$_2$) as a function of molecular contrast (H$_2$) matched to the resolution of \hi{}. The correlation is very strong (Spearman $\rho=0.90$). Most datapoints cluster around relatively low contrast values (mostly ${\lesssim} 0.5$ in log scale), and in those cases the agreement between \hi{}+H$_2$ and H$_2$ contrast is good (following the 1:1 line). Overall, the drop in the contrasts when including \hi{} is small (median 6\% lower for the average over segments, only 3\% lower for all radial bins). For exceptionally high contrasts, however, the H$_2$ contrast exceeds the one for total gas by $\sim$0.1{-}0.2\,dex ($30{-}60$\%). The relation between \hi{}+H$_2$ and H$_2$ contrast can be fitted by a power-law with slope $0.67$ (but we note that the fit tends to overestimate the difference with respect to the 1:1 line for the largest contrasts).

The boost in molecular-to-atomic ratio ($\Sigma_\mathrm{mol}$/$\Sigma_\mathrm{atom}$) in spirals points to some conversion of atomic to molecular gas due to the compression associated with spiral arms. At the same time, photodissociation of H$_2$ downstream of the arm could play a role \citep[e.g.][]{2013ApJ...779...42S}. In any case, we find a higher H$_2$ than \hi{} contrast, which suggests that spiral arms might trigger the formation of molecules and do not simply gather preexisting molecular gas; in addition, spiral arms might enhance the density enough to form CO (and not only H$_2$), transforming some CO-dark molecular gas to molecular gas that efficiently emits in the CO line \citep[e.g.][]{2007MNRAS.376.1747D,2014MNRAS.441.1628S}. We caution that the limited resolution of the \hi{} observations might wash out some of these effects. Observations of \hi{} at higher physical resolution would be highly desirable to deepen our understanding of such phase transitions associated with spiral arms.

Including atomic gas also affects the SFE. Table~\ref{table:includingHI} shows that the median SFE contrast increases by $8$\% when considering the total gas (H$_2$+\hi{}) instead of only H$_2$. As expected, the change is even more extreme if we consider \hi{} only, which results in an SFE contrast typically ${\sim}20$\% higher than when considering only H$_2$. 
The difference is particularly strong for the largest molecular contrasts, where considering the SFE of the total gas (\hi{}$+$H$_2$) results in contrasts nearly a factor of ${\sim}2$ higher than focusing exclusively on the molecular phase. 
To make further progress, it becomes essential to obtain observations of atomic gas at higher physical resolution.
We discuss this further in Sect.~\ref{Sec:discussSFE}.

\subsection{Radial trends in arm/interarm contrast} 
\label{Sec:radial}

\begin{figure*}[t]
\begin{center}
\includegraphics[trim=0 0 0 0, clip,width=0.85\textwidth]{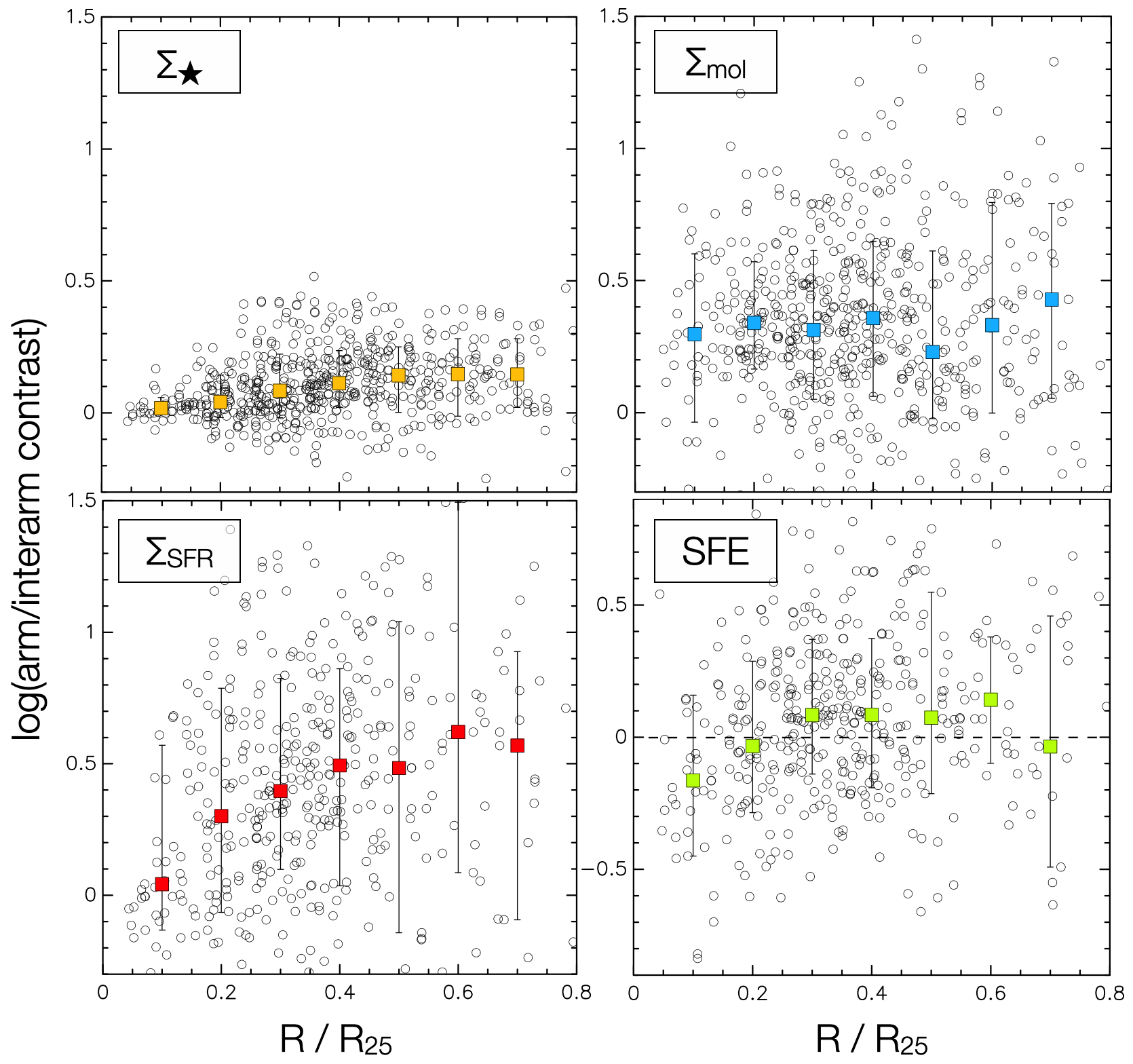}
\end{center}
\caption{Range of contrasts as a function of normalised radius ($R/R_{25}$). The squares denote the running medians in bins of 0.2\,dex and the vertical error bars indicate the 16$^{\rm th}$--84$^{\rm th}$ percentile range.
}
\label{fig:Rnorm}
\end{figure*}

Here we examine whether the arm/interarm contrast shows significant fluctuations with galactocentric radius. 
Figure\,\ref{fig:Rnorm} shows our measurements of the arm/interarm contrast in elliptical annuli (radial bins of 500\,pc) as a function of the galactocentric radius of each bin normalised by $R_{25}$. 
There is a very significant scatter in the contrasts at fixed $R/R_{25}$, but the running medians also show that there are some (mild) trends with radius. The enhancement of the stellar contrast above unity becomes on average a factor of ${\sim}8$ larger from the innermost to the outer radii that we sample; the median contrast in stellar mass surface density monotonically rises from just 5\% in the innermost bin, up to 38\% at $0.5\,R_{25}$, and then plateaus around this value. The trend is confirmed if we use the MUSE-based stellar mass maps instead (Appendix~\ref{Sec:stellarcon}). Thus, spiral arm stellar contrasts are damped in inner galaxies relative to middle radii; stellar surface densities become much larger as we approach galaxy centres, while the interarm values become comparatively higher with respect to the spiral density enhancement.

The molecular gas contrast shows a more stable behaviour with radius, with running medians fluctuating around a factor $2$-$2.5$. Compared to the molecular contrast, the SFR contrast shows a strong decline towards the innermost normalised radial bins. A visual inspection of the maps suggests that this is mostly due to efficient star formation in inner rings or around bar ends, beyond the actual log-spiral mask. In other words, precisely around the point where log-spiral arms end in galactic centres or connect to a bar, there is plenty of scattered star formation, which falls inside the interarm footprint. As a consequence, the SFE contrast also shows its lowest value for the innermost bin (factor of ${\sim}0.7$), largely associated with the surroundings of bar ends and star-forming rings beyond spirals, while over most radii the running medians remain around unity or slightly above.

\subsection{Symmetry of contrasts in two-armed spirals} 
\label{Sec:results_symmetry}

The large majority of the spiral galaxies in our sample are dominated by two spiral arms (22 out of 28, or 79\%).
Here, for two-armed spirals, we examine whether the radial variations in contrasts that we observe are symmetric or largely uncorrelated for opposite spiral arms.

If the contrasts were exactly symmetric, we would expect a perfect {correlation between} opposite spiral arms (rank coefficient approaching unity), with scatter only due to measurement uncertainties. On the other hand, if the radial fluctuations along a given arm were totally independent from the other arm, we would expect a lack of correlation ($\rho$ close to zero). If we calculate the Spearman rank correlation coefficient for contrasts in opposite arms, we find a relatively strong correlation for the stellar component ($\rho = 0.61$), a weaker correlation for the molecular and SFR contrast ($\rho =0.36$ and $0.43$, respectively), and an even smaller correlation for the SFE contrast ($\rho =0.21$). Therefore, there seems to be some degree of connection between both arms (the arms somehow `know about each other'), but the large scatter implies that fluctuations along one arm are not always perfectly mirrored by changes in the contrast of the other spiral arm.
In Sect.\,\ref{Sec:symmetry} we discuss the stronger arm-to-arm correlation in NIR than in CO or SFR. 

\subsection{{Effect of resolution}} 
\label{Sec:effect_res}

\begin{figure*}[t]
\begin{center}
\includegraphics[trim=0 0 0 0, clip,width=0.85\textwidth]{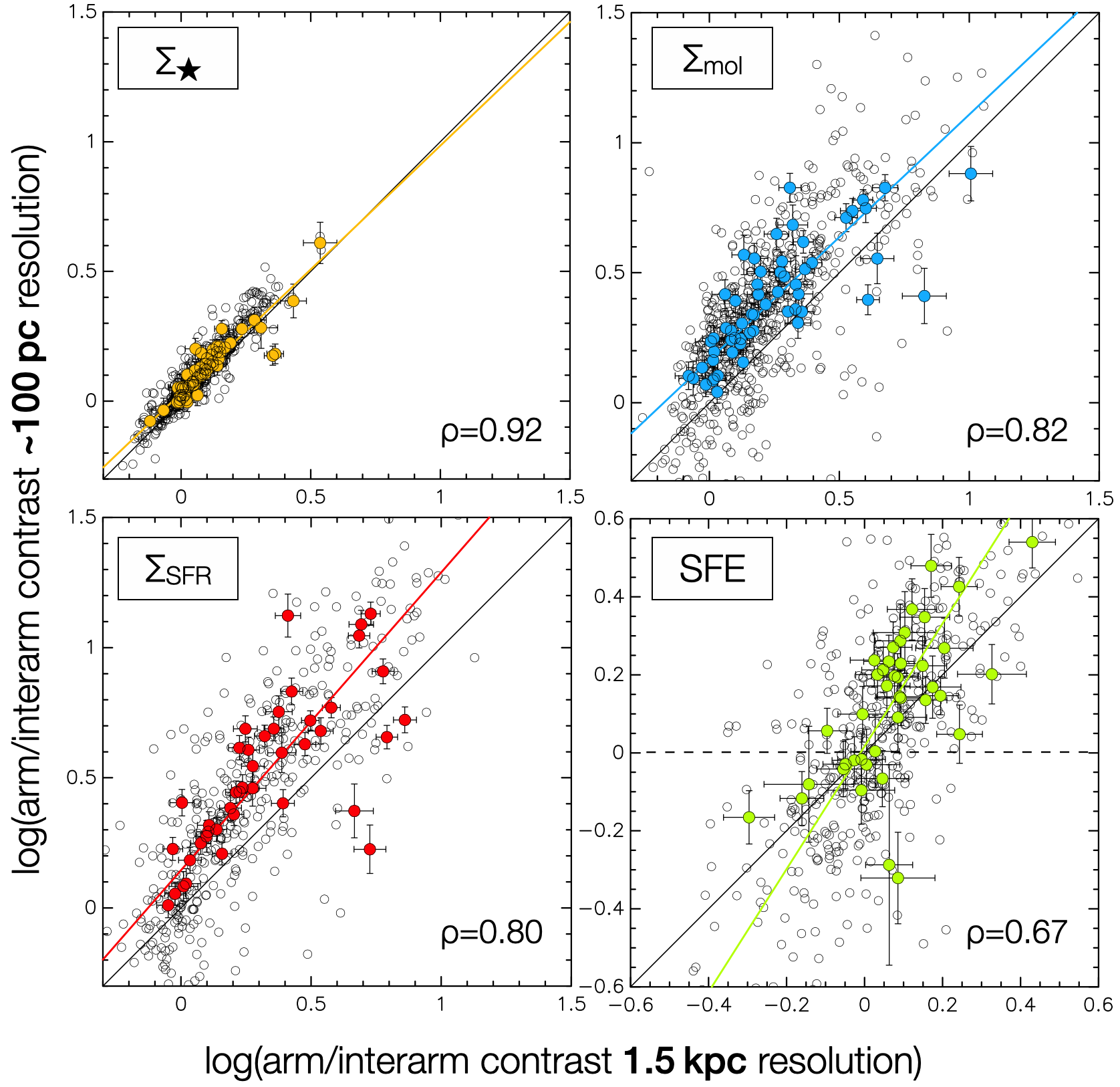}
\end{center}
\caption{Comparison of contrasts at our high working resolution (${\sim}$100\,pc) versus contrasts for the maps degraded to 1.5\,kpc resolution. 
}
\label{fig:resol}
\end{figure*}

The main difference between the arm/interarm contrasts presented in this paper and those from \citet{2021A&A...656A.133Q} and other previous works is the resolution of the data. Here  we consider contrasts measured from data at ${\sim}$100\,pc resolution as opposed to previous studies working on kpc-resolutions. In this Section, we directly assess the effect of resolution on the contrasts by comparing our ${\sim}$100\,pc resolution results with the results if we convolve the maps to 1.5\,kpc resolution. This is the physical resolution used in the contrasts presented in \citet{2021A&A...656A.133Q}, as it is the best common resolution possible across all PHANGS galaxies limited by the SFR tracer (WISE 22\,$\mu$m). In order to convolve the maps to 1.5\,kpc resolution, we employ the kernels from \citet{2011PASP..123.1218A}.

Figure~\ref{fig:resol} shows that there is a reasonably good correlation between the contrast measured at 1.5\,kpc and at 100\,pc resolution (ranks $\rho \sim 0.8{-}0.9$ for \SigStar, \SigMol, and \SigSFR, $\rho=0.67$ for SFE). In any case, there is some scatter, especially for SFE, which is a more indirect quantity. More importantly, there is a systematic offset towards higher contrasts, especially for \SigMol\ and \SigSFR, when we shift to high resolution. This is well expected because using data at 1.5\,kpc will somewhat dilute the signal. The contrasts measured at 100\,pc are typically 8\%, 38\% and 44\% higher than at 1.5\,kpc resolution for \SigStar, \SigMol, and \SigSFR, respectively (median {difference}). The lower change for \SigStar\  is probably because the stellar mass distribution is intrinsically smoother than \SigMol\ or \SigSFR. For SFE, if we focus on the contrasts above unity, the median increase is 9\% (up to 23\% if we focus on contrasts above 1.3). Indeed, at high contrast values, the amplification at high resolution is slightly larger for SFR than for CO (for contrasts above 2, the median increase for SFR is 80\%, as opposed to 36\% median amplification for CO); this is why the highest SFE contrasts also get slightly boosted at high resolution compared to low resolution.

\section{Discussion}
\label{Sec:discussion}

We start by discussing the magnitude of our measured \SigStar, \SigMol, and \SigSFR\ contrasts and their variation among and within galaxies in Sect.\,\ref{Sec:discusscontr}. Then
we discuss their radial variations and symmetry in Sect.\,\ref{Sec:symmetry}.
We address the question of whether spiral arms systematically boost SFE in Sect.\,\ref{Sec:discussSFE}.

\subsection{Magnitude of \SigStar, \SigMol, \SigSFR\ contrasts and variation with galaxy properties} 
\label{Sec:discusscontr}

Spiral arms are often regarded as locations where most gas and star formation accumulates in galaxies. Yet, as shown by \citet{2021A&A...656A.133Q} for PHANGS, the overall contribution of arm and interarm to total molecular gas and SFR is fairly similar, while the interarm region spans a larger area (in agreement with findings from previous studies such as \citealt{2010ApJ...725..534F}).
To examine this problem further, we have performed detailed measurements of the arm/interarm contrast of stellar mass, molecular gas, and star formation rate surface density, as well as the resulting efficiency, across PHANGS galaxies.

One of the main results of this paper is that the stellar surface density arm/interarm contrast is very small (about a few tens of percent), but, as it determines the gravitational potential, it is sufficient and critical to trigger a non-linear response in the gas which results in the accumulation of molecular gas and star formation in spiral arms (typical contrast of a factor ${\sim}2{-}3$). Using a morphology-independent approach to quantify gas and stellar density contrasts in these same galaxies, \citet{2021ApJ...913..113M} {found}  similar results. 
In this work, we have examined the robustness of this result {against}  different methodological choices (e.g.\ spiral mask width, tracers, binning; see Appendix\,\ref{sec:sanity_checks}).

We {found}  that \SigSFR\ contrasts show a larger scatter than \SigMol\ contrasts. This suggests that there are more fluctuations in the contrast of \SigSFR, which is probably a result of the intrinsic stochasticity of the star formation process, where a small fraction of the molecular gas is transformed into stars \citep[e.g.][]{2010ApJ...722.1699S,2014MNRAS.439.3239K,2018ApJ...861L..18U,2019Natur.569..519K,2020MNRAS.493.2872C}. While spiral arms act directly on the gas, the effect on star formation is indirect, and modulated by changes in SFE (e.g.\ due to local conditions in the ISM such as pressure or magnetic fields; \citealt{2021ApJ...911..128K}). Our choice of 500\,pc-wide radial bins somewhat mitigates stochastic sampling effects (which would be stronger in smaller apertures; the characteristic separation length between independent star-forming regions measured by \citealt{2022MNRAS.516.3006K} for PHANGS galaxies is ${\sim}200{-}400$\,pc). However, local variations in SFE can still result in a wider distribution of SFR contrasts compared to CO contrasts. In the spiral galaxy M100, \citet{2018ApJ...863...59E} found regularly spaced NIR clumps along filaments (with typical separations of ${\sim}400$\,pc), which they interpret as the result of large-scale gravitational instabilities in shocked gas \citep[see also][]{2020NatAs...4.1064H}. All these fluctuations can explain the occasional appearance of bins where the interarm surface density is slightly higher than in the arm, leading to negative contrasts in the logarithmic scale. 

Our results point to a tight relationship between stellar mass, galaxy dynamics, morphology, and the reorganisation of the ISM in spiral galaxies. 
We found that grand-design spirals show higher stellar contrasts (up to three times higher median for an extended field of view), and molecular and SFR contrasts are also higher (${\sim}2$ higher median). This agrees with findings from \citet{2021ApJ...913..113M} on larger CO contrasts in grand-design galaxies. Regarding the stellar contrast, \citet{2017MNRAS.471.1070B} found significantly lower arm/interarm contrasts in flocculent galaxies, but similar ones in multi-armed and grand-design spirals. Only two galaxies in our sample are classified as flocculent (NGC\,1385 and NGC\,2283), as opposed to nine multi-armed and 17 grand-design. Therefore, the difference between grand-design and the rest in our case is mostly driven by multi-armed spirals (and not flocculent).

Since the most massive spirals in our sample tend to be grand-design, we also find higher contrasts for the galaxies with largest stellar mass. The increased grand-design contrasts could have a dynamical origin, as both internally and externally excited density waves  (e.g.\ due to bars or interactions) can manifest themselves as grand-design spirals. In turn, this results in an accumulation of stellar mass in a symmetric, two-armed spiral pattern, which can drive shocks that accumulate gas and trigger star formation (following the picture from \citealt{1969ApJ...158..123R}).

Another strong piece of evidence supporting the link between dynamics and \SigMol\ and \SigSFR\ contrasts is the strong correlation between these quantities and \SigStar\ contrasts. Specifically, we find superlinear relations (slope $m=1.8$ for \SigMol, $m=2.9$ for \SigSFR), which agrees with similar findings by \citet{2021ApJ...913..113M}, consistent with gas compressed as it flows supersonically as it enters spiral arms. Specifically, the slope of $m=2$ between gas and stellar contrast can be expected for hydrodynamic shocks with a Mach number that is set by the underlying stellar density contrast \citep{2021ApJ...913..113M}. Scatter is expected around this relation due to a number of additional effects, including gas self-gravity and magnetic fields.

\subsection{Radial variation and symmetry in arm/interarm contrasts} 
\label{Sec:symmetry}

Our interpretation of the strong correlation between \SigMol\ or \SigSFR\ and \SigStar\ contrasts, and the larger contrasts found in grand-design spirals, is that galaxy dynamics largely dictates, through perturbations in the gravitational potential, the accumulation of gas and, subsequently, star formation in spiral arms. Yet, this is a relation that operates `on average', while the individual radial bins, and even mean values over entire spiral segments, show significant scatter around this relation. That means that on top of the stellar contrast there must be local effects at play, potentially also related to spiral arms (e.g.\ magnetic fields, local flows, local shear, gravitational instabilities leading to regularly spaced star-forming clumps), in addition to observational uncertainties. This is even more acute in SFR contrasts, as it is an indirect quantity which is also sensitive to many other factors (stability of molecular clouds, and specific snapshot in which we capture the star formation cycle). 

We found a clear decrease in the stellar contrast as we approach galaxy centres. The median contrast in stellar mass drops from ${\sim}40$\% at $0.5\,R_{25}$ to just 5\% at $0.1\,R_{25}$. Overall, stellar surface densities quickly increase towards galaxy centres, which can even harbour bulges on top of a roughly exponential stellar disc. This means that the baseline interarm value is {more} elevated for the smallest radii, and a significant increase in spiral surface density can still appear as proportionally more limited due to this stellar bulge dilution. Furthermore, it becomes comparatively harder to excite spiral arms when stellar populations have a larger velocity dispersion (dynamically `hotter', i.e.\ higher $\sigma_\star / v_{\rm rot}$), which is typically the case when we move towards the centres of galaxies.

We also found that \SigSFR\ contrast drops by more than a factor of ten over the same radial range. As commented on in Sect.~\ref{Sec:radial}, this drop is associated with inner rings or the surroundings of bar ends, which are actively star-forming and fall within the interarm footprint.
More generally, the contact points of spirals and bars, as well as galaxy centres, show complex effects where the dichotomy between spiral and interarm star formation becomes less clear. In addition to the spiral perturbation on the gaseous disc, those inner radial bins reflect other dynamical effects, such as rings connected to Lindblad resonances \citep[e.g.][]{1996FCPh...17...95B}.

Two-fold rotational symmetry is expected in standard models of grand-design spiral structure, where an $m=2$ Fourier mode dominates. By focusing on the two-armed spirals in our sample, we found that there is a positive and significant degree of correlation between opposite spiral arms with $\rho \sim 0.6$ for the stellar component and $\rho \sim 0.4$ for the molecular and SFR component. This is consistent with an interpretation where dynamical processes drive a largely symmetric response in the stellar gravitational potential of the disc, and this leads to the accumulation of gas (and indirectly star formation), but on top of other mechanisms that ultimately set the details of the local gas response. 

The effect of arm symmetry on SFE is more subtle. If radial fluctuations in SFE were driven by a (radial) dynamical process, we could expect a significant correlation among arms. Yet, if the fluctuations in SFE were driven by local processes and/or time evolution (stochasticity and sampling issues), we would expect the measurements in opposite arms to be mostly uncorrelated. Indeed, what we find is that the correlation of SFE in opposite arms is much weaker ($\rho \sim 0.2$) than the correlation of \SigMol\ or \SigSFR\ contrasts ($\rho \sim 0.4$), suggesting that, while axisymmetric dynamical processes could play some modulating role, local effects and stochasticity largely wash out any symmetric trends.

\subsection{Considering whether the SFE is enhanced in spiral arms} 
\label{Sec:discussSFE}

While it is clear that spiral arms accumulate gas and star formation, the question remains as to what extent the conversion of gas into stars proceeds more efficiently in spiral arms. For example, shocks or cloud-cloud collisions could trigger the collapse of molecular clouds and induce more efficient star formation. 

Previous studies have addressed this long-standing problem using different tracers (e.g.\ gas including only \hi{}, only H$_2$, or both) at different physical resolutions (typically ${\sim}1$\,kpc), yielding contradictory results, with some papers claiming that spiral arms are sites of enhanced SFE \citep[e.g.][]{1990ApJ...356..135L,1996A&A...308...27K}, and others arguing that spirals pile up gas (and indirectly star formation) but do not boost the efficiency at which gas transforms into stars \citep[e.g.][]{2010ApJ...725..534F,2016ApJ...827..103K}. Galactic studies also point to little or no difference in the SFE between arm and interarm regions in the Milky Way \citep[e.g.][]{2012MNRAS.426..701M,2015MNRAS.452..289E,2018MNRAS.479.2361R,2021MNRAS.500.3050U}. To shed light on this problem, here we have considered the largest sample of spiral arm/interarm contrasts examined so far (28 nearby spirals) using homogeneous observations at $\sim$100\,pc resolution.

Our observations do not support the simple, naive picture where spiral arms always increase the efficiency by which molecular gas forms stars. 
If we pick up a random location, in roughly 6 out of 10 cases we will find higher SFE in spiral arms, but in as many as 4 out of 10 cases, the opposite will hold: more efficient star formation in interarm regions (Fig.\,\ref{fig:violin_plot_GD-vs-rest}, right panel). The median SFE enhancement across the whole sample is 16\%, which, albeit statistically significant, is limited compared to the spiral increase in \SigMol\ (median enhancement 120\%) or \SigSFR\ (160\%). We note that systematic variations in the $\alpha_{\rm CO}$ conversion factor could result in an additional $10$\% increase of SFE in arms compared to interarm regions (if the $\alpha_{\rm CO}$ prescription from \citealt{2024ApJ...961...42T} can be extrapolated to arm/interarm differences; see Sect.~\ref{Sec:PHANGS--ALMA}).

While we found clearly enhanced stellar contrasts in grand-design spirals, we do not find a significant difference in the SFE contrast between grand-design spirals and the rest (Fig.\,\ref{fig:violin_plot_GD-vs-rest}). SFE and \SigStar\ contrasts show a positive correlation, but fairly weak ($\rho = 0.34$), as opposed to the stronger correlation between \SigMol\ or \SigSFR\ and \SigStar\ contrasts ($\rho = 0.73$ and $0.61$, respectively). Therefore, the impact of global dynamics and morphology on the regulation of SFE seems more subtle. Overall, local or stochastic effects seem to play a stronger role in setting the spiral SFE.

However, if we focus on the largest stellar contrasts, we do find a more acute boost in spiral SFE. For the top 10\% of \SigStar\ contrasts ($>1.97$) the median SFE contrast rises to as much as $2.34$. As we have seen, even a weak stellar spiral arm (contrast ${\sim}10{-}20$\%) results in a significant, non-linear increase of molecular gas (and, indirectly, SFR) surface densities. Thus, to zeroth order, spiral arms simply pile up gas and star formation in a very efficient way. However, we need a very strong stellar contrast in the spiral (approaching a factor ${\sim}2$ or more) to produce a significant boost in the molecular SFE.

Our \hi{} observations, albeit at lower resolution, confirm that the molecular-to-atomic ratio ($\Sigma_\mathrm{mol}$/$\Sigma_\mathrm{atom}$) increases in spiral arms. Thus, to some degree, spiral arms promote a phase transition from \hi{} to H$_2$. As a consequence, the contrasts associated with the total gas are weaker than molecular contrasts. By promoting the formation of cold, potentially star-forming clouds, spiral arms make more gas susceptible to form stars, and could indirectly stimulate the formation of molecules. In that sense, they do result in enhanced star formation beyond just gathering gas, even if the molecular gas that is eventually available forms stars with a similar efficiency. As a corollary, the SFE contrast of the total gas (\hi{}+H$_2$) is higher than referred exclusively to the molecular phase. To better understand the role of atomic gas in spiral arms, it is imperative to advance towards high-resolution \hi{} mapping of spiral galaxies with matched molecular gas observations; this is already possible for very nearby spirals such as IC\,342 \citep{2023A&A...680A...4Q}.

The inclusion of atomic gas can explain some of the discrepancies within the literature. For example, early studies such as \citet{1987PhDT........11L}, \citet{1988Natur.334..402V} or \citet{1990ApJ...349..497C} relied exclusively on \hi{} to estimate gas and SFE contrasts. This results in a lower gas contrast and a boost of the SFE contrast (because part of the atomic gas transforms to molecular gas in spiral arms). Other studies referred to the SFE of the total (\hi{}+H$_2$) gas, such as \citet{1996A&A...308...27K}, which can also lead to higher observed SFE, especially for the largest contrasts. On the other hand, studies limited to molecular gas typically found similar SFE values in spiral arms and interarm regions \citep[e.g.][]{2010ApJ...725..534F,2016ApJ...827..103K,2021A&A...650A.134P,2021A&A...656A.133Q}.

Regarding simulations, there is no clear support for a picture where star formation proceeds more efficiently in spiral arms. Hydrodynamical galaxy simulations indeed suggest that there is an accumulation of dense (molecular) gas in spiral arms, but the star formation rate per unit gas mass remains similar in arms and interarm \citep{2006MNRAS.365...37B,2009MNRAS.396.1579D,2011MNRAS.417.1318D,2020ApJ...898...35K,2020MNRAS.492.2973T}. According to these studies, spiral arms are sites where competing effects meet, %on the one hand leading to the formation of larger molecular clouds, but on the other hand increasing velocity dispersion, which hinders collapse \citep{2017ApJ...845..133S}.
on the one hand building up higher gas densities and larger molecular clouds, and on the other hand involving shear, flows, and elevated velocity dispersions that hinder collapse \citep{2007MNRAS.374.1115D,2013ApJ...779...45M,2017ApJ...845..133S,2018ApJ...854..100M}.

Summing up, our results suggest that the impact of spiral arms on SFE is, to first order, a matter of definition. If we ask whether spiral arms simply `gather' or also `trigger' star formation, the answer is that they certainly passively gather, but to some extent also actively enhance star formation, because some \hi{} transforms to H$_2$ and, thus, more gas is susceptible to form stars. On the other hand, if the question is whether the available molecular gas forms stars more efficiently in spiral arms, the answer would be on average negative; the triggering effect is at least very limited, except in the cases where the stellar contrast is very large. Here we have not examined effects such as detailed gas flows, shear patterns, or magnetic fields within spirals, but some of these effects might be able to explain why we find more or less efficient conversion of molecular gas into stars in different locations along spirals.

\subsection{Methodology and measurement caveats} 
\label{Sec:discuss_methodology}

Table\,\ref{table:stats} summarises the distribution of contrasts for different methodological choices. {This includes using lower resolution maps (Sect.~\ref{Sec:effect_res}), a narrow version of the spiral masks (Appendix~ \ref{Sec:narrow}), radial bins of different width, and two alternative sampling approaches (Appendix~\ref{sec:sanity_checks}).}

The most relevant effect is resolution. Indeed, contrasts get strongly diluted when shifting from ${\sim} 100$\,pc to 1.5\,kpc resolution (Fig.~\ref{fig:resol}). The median stellar contrast drops from $1.28$ to $1.15$, while the median molecular and SFR contrasts drop from $2.22$ to $1.49$ and from $2.56$ to $1.50$, respectively. Importantly, the dilution effect is not totally symmetric for molecular gas and star formation, such that the SFE is also affected when degrading resolution (its median contrast drops from $1.16$ to $1.04$; even more if we degrade the resolution further to match \hi{}, see Table\,\ref{table:includingHI}).

We also test how results are affected by the precise definition of the spiral mask. Using the narrower spiral masks introduced in Sect.~\ref{Sec:thinnersp}, Fig.~\ref{fig:orig_vs_narrow} in Appendix~\ref{Sec:narrow} shows that, despite covering a significantly smaller area (${\sim} 1/2$), the contrasts for the narrower masks are very similar to the nominal ones or just slightly higher. The stellar contrast remains virtually unchanged, while the median molecular and SFR contrasts are just ${\sim}10{-}20$\% higher for the narrow masks. The resulting SFE is just $4$\% higher for the narrow masks, so the difference is fairly limited, and much smaller than implied by the change of resolution.

A quick visual inspection of CO or SFR maps could suggest that the narrow masks better capture the enhanced densities in spirals. Yet, the relatively limited change in contrasts when using the narrow masks {demonstrates} that spirals are not as concentrated as one could perhaps expect. Indeed, the immediate outskirts of the spiral ridge must also have elevated surface densities compared to interarm regions, as otherwise we would expect a more acute dilution of the contrast associated with the broader masks (which typically span a twice larger area). In other words, the narrow masks better capture the ridge of molecular emission and star formation (useful for some applications), but the effect of the spiral perturbation seems to extend further out, which justifies the use of the more regular and broader nominal masks across this paper.

% effect of binning approach

As shown in Table\,\ref{table:stats}, the distributions of contrasts for \SigStar, \SigMol, \SigSFR, and SFE are very similar if we choose either narrower radial bins (250\,pc) or wider bins (1000\,pc), instead of the nominal 500\,pc-wide bins. 
In Appendix~\ref{sec:sanity_checks}, instead of elliptical annuli of a fixed width we also consider a different binning strategy, splitting the spiral masks into `boxes' which follow each arm, with perpendicular cuts at regular lengths along the arm, and two alternative definitions of interarm. The impact of these alternative binning approaches on the contrasts is slightly larger that just changing the width of the radial bins, but does not lead to very strong systematic differences, as shown in Table\,\ref{table:stats} and Appendix~\ref{sec:sanity_checks}.

As already discussed in Sect.~\ref{Sec:effect_HI} and Sect.~\ref{Sec:discussSFE}, including atomic gas does have a very significant impact on the gas and SFE contrasts. 
Table~\ref{table:includingHI} quantifies these differences.

% Caveats:
In the Appendix, we also consider some caveats associated with our measurements. In Appendix~\ref{Sec:stellarcon} we confirm that our nominal stellar mass tracer ({\it Spitzer} $3.6$\,$\mu$m corrected for dust emission with ICA) is robust when compared to MUSE-based \SigStar\ maps (from stellar population fitting).
Appendix~\ref{sec:AppendixMUSE} considers the correction to narrow-band H$\alpha$ to account for extinction; for the same subset of galaxies and field of view, the contrasts derived this way agree with MUSE SFRs (extiction-corrected H$\alpha$ using the Balmer decrement).
In Appendix~\ref{Sec:DIG} we argue that diffuse ionised gas (DIG) is unlikely to have a major impact on SFR contrasts following our observational approach. Based on geometrical arguments, we expect our broad spiral masks to capture most leaked ionising photons associated with spiral arms. Furthermore, our tests using \hii{} region masks suggest that most DIG in the interarm regions is associated with local \hii{} regions, and is not due to leaked photons from the spiral arms.

In addition to the distributions of contrasts presented in Table\,\ref{table:stats} for different alternatives, in Appendix~\ref{sec:sanity_checks}, and specifically in Table~\ref{table:stats_appendix}, we examine how the main correlations among contrasts are affected by these methodological choices. Based on this analysis, our main conclusions seem qualitatively robust against these details.

\section{Summary and conclusions} 
\label{Sec:concl}

Disc galaxies in the local universe tend to develop different kinds of spiral structures, where approximately logarithmic spiral segments appear punctuated by luminous regions associated with young stars. 
In this paper, we have examined the increase of surface densities present in spiral arms compared to the remaining disc positions at matched galactocentric radii. We have measured these arm/interarm contrasts in the 28 nearby galaxies from PHANGS that show a clear spiral structure (according to the environmental masks from \citealt{2021A&A...656A.133Q}), split into radial bins of 500\,pc width.

The main novelty here with respect to \citet{2021A&A...656A.133Q} is that these measurements were performed at $\sim$100\,pc resolution instead of $\sim$kpc resolution, and that we considered radial bins and not only averages over full spiral segments. The limiting factor in \citet{2021A&A...656A.133Q} was the requirement to convolve the data to the resolution of the MIR map that was used in the hybridisation process to account for obscured star formation (WISE 22\,$\mu$m at $15''$ resolution). In this paper, we have circumvented this limitation by employing PHANGS--MUSE observations to empirically calibrate the effect of extinction on SFR contrasts, starting directly from the H$\alpha$ contrast (see Appendix~\ref{sec:AppendixMUSE} for details). This allowed us to robustly measure the SFR contrast at  $\sim$100\,pc resolution and therefore (indirectly) the SFE contrasts, within some moderate error bars.  
Compared to \citet{2021A&A...656A.133Q}, in this paper we also considered the stellar contrast and correlations among contrasts, as well as radial trends and symmetry among spiral arms in a given galaxy. Finally, we also examined a number of additional effects, including the width of the spiral masks, resolution, binning, and the inclusion of atomic gas.

Our main findings are:

\begin{enumerate}

\item The arm/interarm {enhancement} in stellar mass surface density ($\Sigma_\star$) is very modest, typically a few $\sim$10\%  (median 28\%). On the other hand, the molecular gas ($\Sigma_\mathrm{mol}$) and star formation rate ($\Sigma_\mathrm{SFR}$) {enhancements} are much higher, spanning from a few tens of percent up to a factor of a few (median {contrast} 2.2 and 2.6, respectively; Fig.\,\ref{fig:violin_plot_GD-vs-rest} and Table~\ref{table:stats}).

\item The {spiral enhancement is}  larger in grand-design spirals compared to the rest (median 50\% higher for stellar mass, factor of ${\sim}2$ higher for molecular gas and star formation surface density; Fig.\,\ref{fig:violin_plot_GD-vs-rest}). This suggests that grand-design galaxies are particularly efficient in reorganising stellar mass, gas, and star formation into spiral structures.

\item The contrast in $\Sigma_\mathrm{mol}$ and $\Sigma_\mathrm{SFR}$ show a significant correlation with the $\Sigma_\star$ contrast, implying that a modest contrast in stellar mass surface density largely controls the accumulation of gas and star formation, which gets amplified compared to the perturbation in the gravitational potential (see Fig.\,\ref{fig:CO_SFR_SFE_vs_NIR}).

\item The contrasts show large fluctuations with galactocentric radius, with a larger scatter in \SigSFR\ than in \SigMol. There is a pronounced drop in \SigStar\ contrast towards galaxy centres. The \SigSFR\ contrasts also show a drop towards the innermost spiral bins (Fig.\,\ref{fig:Rnorm}). This appears to be associated with extended star formation outside spiral arms around bar ends and galaxy centres.

\item The star formation efficiency (SFE) of molecular gas is higher in spiral arms only in 60\% of the radial bins, with a median enhancement of 16\%. However, when focusing on the largest stellar contrasts ($\gtrsim 2$, top 10\%), the median SFE contrast of molecular gas increases to as much as $2.34$.

\item The molecular-to-atomic gas ratio ($\Sigma_\mathrm{mol}$/$\Sigma_\mathrm{atom}$) is higher in spiral arms, with lower atomic than molecular contrasts. When referred to the total gas (\hi{}+H$_2$) instead of molecular gas (H$_2$) the median SFE increases by $8$\% (at ${\sim}$kpc resolution available for \hi{}). In that sense, spiral arms induce a phase transition from atomic to molecular gas, making more gas available to form stars. Thus, to some degree, spiral arms do trigger star formation,  in addition to their role in gathering molecular gas.

\item We have considered a number of technical details, including the spiral mask width, tracers, resolution, and binning. There is a significant dilution of contrasts when we shift to lower resolution (from $\sim$100\,pc to 1.5\,kpc), and the details of binning, specific tracers used and conversion factors do have an impact on the precise numbers in each case. In any case, these choices do not change our conclusions qualitatively.

\end{enumerate} 

In summary, the modest stellar contrasts largely dictate the accumulation of molecular gas and star formation in spiral arms. However, only the largest stellar contrasts result in a substantial boost in how efficiently molecular gas transforms into stars. The efficiency associated with the total gas (including the atomic phase) is slightly higher, as spirals do end up triggering some transformation of atomic to molecular gas, making a larger fraction of the gas available for star formation.
In any case, the organisation of the ISM by spiral arms and the modulation of the SFE\ is a complex phenomenon that shows significant local fluctuations and is not fully mirrored by opposite arms {even} in grand-design spirals.

% Do not delete the next line
\small  % Do not delete
%
%%% Comment the following line if you do not have acknowledgments.
\begin{acknowledgements}   % Do not delete if you declare acknowledgments
This work was carried out as part of the PHANGS collaboration. 
{We would like to thank the anonymous referee for comments that helped us improve the manuscript.}

This work is based on observations and archival data obtained with the \textit{Spitzer} Space Telescope, which is operated by the Jet Propulsion Laboratory, California Institute of Technology under a contract with NASA.
This paper makes use of the following ALMA data:  ADS/JAO.ALMA\#2012.1.00650.S,  % (N628/M74) 
ADS/JAO.ALMA\#2013.1.00803.S,  % (N5128/CenA)
ADS/JAO.ALMA\#2013.1.01161.S,  % (N1365 + N5236/M83)
ADS/JAO.ALMA\#2015.1.00121.S,  % (N5236/M83) % 
ADS/JAO.ALMA\#2015.1.00782.S,  % (N1313 + N7793) 
ADS/JAO.ALMA\#2015.1.00925.S,  % (pilot low mass) 
ADS/JAO.ALMA\#2015.1.00956.S,  % (pilot high mass) 
ADS/JAO.ALMA\#2016.1.00386.S, % (N5236/M83) 
ADS/JAO.ALMA\#2017.1.00886.L,  % (large program) %ADS/JAO.ALMA\#2018.1.01321.S, \linebreak % (N253, N300, Circinus)
ADS/JAO.ALMA\#2018.1.01651.S.  % (main sample follow-up) % ADS/JAO.ALMA\#2018.A.00062.S. \linebreak % (ACA-only nearby) 
ALMA is a partnership of ESO (representing its member states), NSF (USA) and NINS (Japan), together with NRC (Canada), MOST and ASIAA (Taiwan), and KASI (Republic of Korea), in cooperation with the Republic of Chile. The Joint ALMA Observatory is operated by ESO, AUI/NRAO and NAOJ. The National Radio Astronomy Observatory is a facility of the National Science Foundation operated under cooperative agreement by Associated Universities, Inc.

MQ, SGB, MRG, and AU acknowledge support from the Spanish grant PID2022-138560NB-I00, funded by MCIN/AEI/10.13039/501100011033/FEDER, EU.
JS acknowledges support by the National Aeronautics and Space Administration (NASA) through the NASA Hubble Fellowship grant HST-HF2-51544 awarded by the Space Telescope Science Institute (STScI), which is operated by the Association of Universities for Research in Astronomy, Inc., under contract NAS~5-26555.
MCS acknowledges financial support from the European Research Council under the ERC Starting Grant `GalFlow' (grant 101116226) and from the Royal Society (URF\textbackslash R1\textbackslash 221118).
MC gratefully acknowledges funding from the DFG through an Emmy Noether Research Group (grant number CH2137/1-1). COOL Research DAO is a Decentralized Autonomous Organization supporting research in astrophysics aimed at uncovering our cosmic origins.
LN acknowledges funding from the Deutsche Forschungsgemeinschaft (DFG, German Research Foundation) - 516405419.
I.P. acknowledges funding by the European Research Council through ERC-AdG SPECMAP-CGM, GA 101020943.

\end{acknowledgements}

\bibliography{mq}{}

\begin{thebibliography}{108}
\expandafter\ifx\csname natexlab\endcsname\relax\def\natexlab#1{#1}\fi

\bibitem[{{Accurso} {et~al.}(2017){Accurso}, {Saintonge}, {Catinella}, {Cortese}, {Dav{\'e}}, {Dunsheath}, {Genzel}, {Gracia-Carpio}, {Heckman}, {Jimmy}, {Kramer}, {Li}, {Lutz}, {Schiminovich}, {Schuster}, {Sternberg}, {Sturm}, {Tacconi}, {Tran}, \& {Wang}}]{2017MNRAS.470.4750A}
{Accurso}, G., {Saintonge}, A., {Catinella}, B., {et~al.} 2017, \mnras, 470, 4750

\bibitem[{{Aniano} {et~al.}(2011){Aniano}, {Draine}, {Gordon}, \& {Sandstrom}}]{2011PASP..123.1218A}
{Aniano}, G., {Draine}, B.~T., {Gordon}, K.~D., \& {Sandstrom}, K. 2011, \pasp, 123, 1218

\bibitem[{{Belfiore} {et~al.}(2023){Belfiore}, {Leroy}, {Sun}, {Barnes}, {Boquien}, {Cao}, {Congiu}, {Dale}, {Egorov}, {Eibensteiner}, {Glover}, {Grasha}, {Groves}, {Klessen}, {Kreckel}, {Neumann}, {Querejeta}, {Sanchez-Blazquez}, {Schinnerer}, \& {Williams}}]{2023A&A...670A..67B}
{Belfiore}, F., {Leroy}, A.~K., {Sun}, J., {et~al.} 2023, \aap, 670, A67

\bibitem[{{Belfiore} {et~al.}(2022){Belfiore}, {Santoro}, {Groves}, {Schinnerer}, {Kreckel}, {Glover}, {Klessen}, {Emsellem}, {Blanc}, {Congiu}, {Barnes}, {Boquien}, {Chevance}, {Dale}, {Kruijssen}, {Leroy}, {Pan}, {Pessa}, {Schruba}, \& {Williams}}]{2022A&A...659A..26B}
{Belfiore}, F., {Santoro}, F., {Groves}, B., {et~al.} 2022, \aap, 659, A26

\bibitem[{{Bigiel} {et~al.}(2008){Bigiel}, {Leroy}, {Walter}, {Brinks}, {de Blok}, {Madore}, \& {Thornley}}]{2008AJ....136.2846B}
{Bigiel}, F., {Leroy}, A., {Walter}, F., {et~al.} 2008, \aj, 136, 2846

\bibitem[{{Bittner} {et~al.}(2017){Bittner}, {Gadotti}, {Elmegreen}, {Athanassoula}, {Elmegreen}, {Bosma}, \& {Mu{\~n}oz-Mateos}}]{2017MNRAS.471.1070B}
{Bittner}, A., {Gadotti}, D.~A., {Elmegreen}, B.~G., {et~al.} 2017, \mnras, 471, 1070

\bibitem[{{Bonnell} {et~al.}(2006){Bonnell}, {Dobbs}, {Robitaille}, \& {Pringle}}]{2006MNRAS.365...37B}
{Bonnell}, I.~A., {Dobbs}, C.~L., {Robitaille}, T.~P., \& {Pringle}, J.~E. 2006, \mnras, 365, 37

\bibitem[{{Buta} \& {Combes}(1996)}]{1996FCPh...17...95B}
{Buta}, R. \& {Combes}, F. 1996, \fcp, 17, 95

\bibitem[{{Buta} {et~al.}(2015){Buta}, {Sheth}, {Athanassoula}, {Bosma}, {Knapen}, {Laurikainen}, {Salo}, {Elmegreen}, {Ho}, {Zaritsky}, {Courtois}, {Hinz}, {Mu{\~n}oz-Mateos}, {Kim}, {Regan}, {Gadotti}, {Gil de Paz}, {Laine}, {Men{\'e}ndez-Delmestre}, {Comer{\'o}n}, {Erroz Ferrer}, {Seibert}, {Mizusawa}, {Holwerda}, \& {Madore}}]{2015ApJS..217...32B}
{Buta}, R.~J., {Sheth}, K., {Athanassoula}, E., {et~al.} 2015, \apjs, 217, 32

\bibitem[{{Cepa} \& {Beckman}(1990)}]{1990ApJ...349..497C}
{Cepa}, J. \& {Beckman}, J.~E. 1990, \apj, 349, 497

\bibitem[{{Chevance} {et~al.}(2020){Chevance}, {Kruijssen}, {Hygate}, {Schruba}, {Longmore}, {Groves}, {Henshaw}, {Herrera}, {Hughes}, {Jeffreson}, {Lang}, {Leroy}, {Meidt}, {Pety}, {Razza}, {Rosolowsky}, {Schinnerer}, {Bigiel}, {Blanc}, {Emsellem}, {Faesi}, {Glover}, {Haydon}, {Ho}, {Kreckel}, {Lee}, {Liu}, {Querejeta}, {Saito}, {Sun}, {Usero}, \& {Utomo}}]{2020MNRAS.493.2872C}
{Chevance}, M., {Kruijssen}, J.~M.~D., {Hygate}, A. P.~S., {et~al.} 2020, \mnras, 493, 2872

\bibitem[{{Chiang} {et~al.}(2023){Chiang}, {Sandstrom}, {Chastenet}, {Bolatto}, {Koch}, {Leroy}, {Sun}, {Teng}, \& {Williams}}]{2023arXiv231100407C}
{Chiang}, I.-D., {Sandstrom}, K.~M., {Chastenet}, J., {et~al.} 2023, arXiv e-prints, arXiv:2311.00407

\bibitem[{{Chung} {et~al.}(2009){Chung}, {van Gorkom}, {Kenney}, {Crowl}, \& {Vollmer}}]{2009AJ....138.1741C}
{Chung}, A., {van Gorkom}, J.~H., {Kenney}, J. D.~P., {Crowl}, H., \& {Vollmer}, B. 2009, \aj, 138, 1741

\bibitem[{{den Brok} {et~al.}(2022){den Brok}, {Bigiel}, {Sliwa}, {Saito}, {Usero}, {Schinnerer}, {Leroy}, {Jim{\'e}nez-Donaire}, {Rosolowsky}, {Barnes}, {Puschnig}, {Pety}, {Schruba}, {Be{\v{s}}li{\'c}}, {Cao}, {Eibensteiner}, {Glover}, {Klessen}, {Kruijssen}, {Meidt}, {Neumann}, {Tomi{\v{c}}i{\'c}}, {Pan}, {Querejeta}, {Watkins}, {Williams}, \& {Wilner}}]{2022A&A...662A..89D}
{den Brok}, J.~S., {Bigiel}, F., {Sliwa}, K., {et~al.} 2022, \aap, 662, A89

\bibitem[{{den Brok} {et~al.}(2021){den Brok}, {Chatzigiannakis}, {Bigiel}, {PHANGS}, {PHANGS}, \& {PHANGS}}]{denBrok21}
{den Brok}, J.~S., {Chatzigiannakis}, D., {Bigiel}, F., {et~al.} 2021, \mnras~submitted

\bibitem[{{Dobbs} \& {Baba}(2014)}]{2014PASA...31...35D}
{Dobbs}, C. \& {Baba}, J. 2014, \pasa, 31, e035

\bibitem[{{Dobbs} \& {Bonnell}(2007{\natexlab{a}})}]{2007MNRAS.374.1115D}
{Dobbs}, C.~L. \& {Bonnell}, I.~A. 2007{\natexlab{a}}, \mnras, 374, 1115

\bibitem[{{Dobbs} \& {Bonnell}(2007{\natexlab{b}})}]{2007MNRAS.376.1747D}
{Dobbs}, C.~L. \& {Bonnell}, I.~A. 2007{\natexlab{b}}, \mnras, 376, 1747

\bibitem[{{Dobbs} {et~al.}(2011){Dobbs}, {Burkert}, \& {Pringle}}]{2011MNRAS.417.1318D}
{Dobbs}, C.~L., {Burkert}, A., \& {Pringle}, J.~E. 2011, \mnras, 417, 1318

\bibitem[{{Dobbs} \& {Pringle}(2009)}]{2009MNRAS.396.1579D}
{Dobbs}, C.~L. \& {Pringle}, J.~E. 2009, \mnras, 396, 1579

\bibitem[{{Duarte-Cabral} \& {Dobbs}(2017)}]{2017MNRAS.470.4261D}
{Duarte-Cabral}, A. \& {Dobbs}, C.~L. 2017, \mnras, 470, 4261

\bibitem[{{Eden} {et~al.}(2015){Eden}, {Moore}, {Urquhart}, {Elia}, {Plume}, {Rigby}, \& {Thompson}}]{2015MNRAS.452..289E}
{Eden}, D.~J., {Moore}, T.~J.~T., {Urquhart}, J.~S., {et~al.} 2015, \mnras, 452, 289

\bibitem[{{Elmegreen}(1987)}]{1987ApJ...312..626E}
{Elmegreen}, B.~G. 1987, \apj, 312, 626

\bibitem[{{Elmegreen}(1993)}]{1993prpl.conf...97E}
{Elmegreen}, B.~G. 1993, in Protostars and Planets III, ed. E.~H. {Levy} \& J.~I. {Lunine}, 97

\bibitem[{{Elmegreen} {et~al.}(2018){Elmegreen}, {Elmegreen}, \& {Efremov}}]{2018ApJ...863...59E}
{Elmegreen}, B.~G., {Elmegreen}, D.~M., \& {Efremov}, Y.~N. 2018, \apj, 863, 59

\bibitem[{{Emsellem} {et~al.}(2021){Emsellem}, {Schinnerer}, {Santoro}, {Belfiore}, {Belfiore}, {Belfiore}, \& {Belfiore}}]{emsellem21}
{Emsellem}, E., {Schinnerer}, E., {Santoro}, F., {et~al.} 2021, \aap~submitted

\bibitem[{{Emsellem} {et~al.}(2022){Emsellem}, {Schinnerer}, {Santoro}, {Belfiore}, {Pessa}, {McElroy}, {Blanc}, {Congiu}, {Groves}, {Ho}, {Kreckel}, {Razza}, {Sanchez-Blazquez}, {Egorov}, {Faesi}, {Klessen}, {Leroy}, {Meidt}, {Querejeta}, {Rosolowsky}, {Scheuermann}, {Anand}, {Barnes}, {Be{\v{s}}li{\'c}}, {Bigiel}, {Boquien}, {Cao}, {Chevance}, {Dale}, {Eibensteiner}, {Glover}, {Grasha}, {Henshaw}, {Hughes}, {Koch}, {Kruijssen}, {Lee}, {Liu}, {Pan}, {Pety}, {Saito}, {Sandstrom}, {Schruba}, {Sun}, {Thilker}, {Usero}, {Watkins}, \& {Williams}}]{2022A&A...659A.191E}
{Emsellem}, E., {Schinnerer}, E., {Santoro}, F., {et~al.} 2022, \aap, 659, A191

\bibitem[{{Foyle} {et~al.}(2010){Foyle}, {Rix}, {Walter}, \& {Leroy}}]{2010ApJ...725..534F}
{Foyle}, K., {Rix}, H.~W., {Walter}, F., \& {Leroy}, A.~K. 2010, \apj, 725, 534

\bibitem[{{Garcia-Burillo} {et~al.}(1993){Garcia-Burillo}, {Guelin}, \& {Cernicharo}}]{1993A&A...274..123G}
{Garcia-Burillo}, S., {Guelin}, M., \& {Cernicharo}, J. 1993, \aap, 274, 123

\bibitem[{{Gittins} \& {Clarke}(2004)}]{2004MNRAS.349..909G}
{Gittins}, D.~M. \& {Clarke}, C.~J. 2004, \mnras, 349, 909

\bibitem[{{Groves} {et~al.}(2012){Groves}, {Brinchmann}, \& {Walcher}}]{2012MNRAS.419.1402G}
{Groves}, B., {Brinchmann}, J., \& {Walcher}, C.~J. 2012, \mnras, 419, 1402

\bibitem[{{Groves} {et~al.}(2023){Groves}, {Kreckel}, {Santoro}, {Belfiore}, {Zavodnik}, {Congiu}, {Egorov}, {Emsellem}, {Grasha}, {Leroy}, {Scheuermann}, {Schinnerer}, {Watkins}, {Barnes}, {Bigiel}, {Dale}, {Glover}, {Pessa}, {Sanchez-Blazquez}, \& {Williams}}]{2023MNRAS.520.4902G}
{Groves}, B., {Kreckel}, K., {Santoro}, F., {et~al.} 2023, \mnras, 520, 4902

\bibitem[{{Haffner} {et~al.}(2009){Haffner}, {Dettmar}, {Beckman}, {Wood}, {Slavin}, {Giammanco}, {Madsen}, {Zurita}, \& {Reynolds}}]{2009RvMP...81..969H}
{Haffner}, L.~M., {Dettmar}, R.~J., {Beckman}, J.~E., {et~al.} 2009, Reviews of Modern Physics, 81, 969

\bibitem[{{Henry} {et~al.}(2003){Henry}, {Quillen}, \& {Gutermuth}}]{2003AJ....126.2831H}
{Henry}, A.~L., {Quillen}, A.~C., \& {Gutermuth}, R. 2003, \aj, 126, 2831

\bibitem[{{Henshaw} {et~al.}(2020){Henshaw}, {Kruijssen}, {Longmore}, {Riener}, {Leroy}, {Rosolowsky}, {Ginsburg}, {Battersby}, {Chevance}, {Meidt}, {Glover}, {Hughes}, {Kainulainen}, {Klessen}, {Schinnerer}, {Schruba}, {Beuther}, {Bigiel}, {Blanc}, {Emsellem}, {Henning}, {Herrera}, {Koch}, {Pety}, {Ragan}, \& {Sun}}]{2020NatAs...4.1064H}
{Henshaw}, J.~D., {Kruijssen}, J.~M.~D., {Longmore}, S.~N., {et~al.} 2020, Nature Astronomy, 4, 1064

\bibitem[{{Herrera-Endoqui} {et~al.}(2015){Herrera-Endoqui}, {D{\'{\i}}az-Garc{\'{\i}}a}, {Laurikainen}, \& {Salo}}]{2015A&A...582A..86H}
{Herrera-Endoqui}, M., {D{\'{\i}}az-Garc{\'{\i}}a}, S., {Laurikainen}, E., \& {Salo}, H. 2015, \aap, 582, A86

\bibitem[{{Hitschfeld} {et~al.}(2009){Hitschfeld}, {Kramer}, {Schuster}, {Garcia-Burillo}, \& {Stutzki}}]{2009A&A...495..795H}
{Hitschfeld}, M., {Kramer}, C., {Schuster}, K.~F., {Garcia-Burillo}, S., \& {Stutzki}, J. 2009, \aap, 495, 795

\bibitem[{{Ho} {et~al.}(2018){Ho}, {Meidt}, {Kudritzki}, {Groves}, {Seibert}, {Madore}, {Schinnerer}, {Rich}, {Kobayashi}, \& {Kewley}}]{2018A&A...618A..64H}
{Ho}, I.~T., {Meidt}, S.~E., {Kudritzki}, R.-P., {et~al.} 2018, \aap, 618, A64

\bibitem[{{Ho} {et~al.}(2017){Ho}, {Seibert}, {Meidt}, {Kudritzki}, {Kobayashi}, {Groves}, {Kewley}, {Madore}, {Rich}, {Schinnerer}, {D'Agostino}, \& {Poetrodjojo}}]{2017ApJ...846...39H}
{Ho}, I.~T., {Seibert}, M., {Meidt}, S.~E., {et~al.} 2017, \apj, 846, 39

\bibitem[{{Kennicutt}(1998)}]{1998ApJ...498..541K}
{Kennicutt}, Jr., R.~C. 1998, \apj, 498, 541

\bibitem[{{Kim} {et~al.}(2022){Kim}, {Chevance}, {Kruijssen}, {Leroy}, {Schruba}, {Barnes}, {Bigiel}, {Blanc}, {Cao}, {Congiu}, {Dale}, {Faesi}, {Glover}, {Grasha}, {Groves}, {Hughes}, {Klessen}, {Kreckel}, {McElroy}, {Pan}, {Pety}, {Querejeta}, {Razza}, {Rosolowsky}, {Saito}, {Schinnerer}, {Sun}, {Tomi{\v{c}}i{\'c}}, {Usero}, \& {Williams}}]{2022MNRAS.516.3006K}
{Kim}, J., {Chevance}, M., {Kruijssen}, J.~M.~D., {et~al.} 2022, \mnras, 516, 3006

\bibitem[{{Kim} {et~al.}(2021){Kim}, {Ostriker}, \& {Filippova}}]{2021ApJ...911..128K}
{Kim}, J.-G., {Ostriker}, E.~C., \& {Filippova}, N. 2021, \apj, 911, 128

\bibitem[{{Kim} {et~al.}(2020){Kim}, {Kim}, \& {Ostriker}}]{2020ApJ...898...35K}
{Kim}, W.-T., {Kim}, C.-G., \& {Ostriker}, E.~C. 2020, \apj, 898, 35

\bibitem[{{Kim} \& {Ostriker}(2002)}]{2002ApJ...570..132K}
{Kim}, W.-T. \& {Ostriker}, E.~C. 2002, \apj, 570, 132

\bibitem[{{Knapen} \& {Beckman}(1996)}]{1996MNRAS.283..251K}
{Knapen}, J.~H. \& {Beckman}, J.~E. 1996, \mnras, 283, 251

\bibitem[{{Knapen} {et~al.}(1996){Knapen}, {Beckman}, {Cepa}, \& {Nakai}}]{1996A&A...308...27K}
{Knapen}, J.~H., {Beckman}, J.~E., {Cepa}, J., \& {Nakai}, N. 1996, \aap, 308, 27

\bibitem[{{Kreckel} {et~al.}(2016){Kreckel}, {Blanc}, {Schinnerer}, {Groves}, {Adamo}, {Hughes}, \& {Meidt}}]{2016ApJ...827..103K}
{Kreckel}, K., {Blanc}, G.~A., {Schinnerer}, E., {et~al.} 2016, \apj, 827, 103

\bibitem[{{Kreckel} {et~al.}(2019){Kreckel}, {Ho}, {Blanc}, {Groves}, {Santoro}, {Schinnerer}, {Bigiel}, {Chevance}, {Congiu}, {Emsellem}, {Faesi}, {Glover}, {Grasha}, {Kruijssen}, {Lang}, {Leroy}, {Meidt}, {McElroy}, {Pety}, {Rosolowsky}, {Saito}, {Sandstrom}, {Sanchez-Blazquez}, \& {Schruba}}]{2019ApJ...887...80K}
{Kreckel}, K., {Ho}, I.~T., {Blanc}, G.~A., {et~al.} 2019, \apj, 887, 80

\bibitem[{{Kruijssen} \& {Longmore}(2014)}]{2014MNRAS.439.3239K}
{Kruijssen}, J.~M.~D. \& {Longmore}, S.~N. 2014, \mnras, 439, 3239

\bibitem[{{Kruijssen} {et~al.}(2019){Kruijssen}, {Schruba}, {Chevance}, {Longmore}, {Hygate}, {Haydon}, {McLeod}, {Dalcanton}, {Tacconi}, \& {van Dishoeck}}]{2019Natur.569..519K}
{Kruijssen}, J.~M.~D., {Schruba}, A., {Chevance}, M., {et~al.} 2019, \nat, 569, 519

\bibitem[{{Lang} {et~al.}(2020){Lang}, {Meidt}, {Rosolowsky}, {Nofech}, {Schinnerer}, {Leroy}, {Emsellem}, {Pessa}, {Glover}, {Groves}, {Hughes}, {Kruijssen}, {Querejeta}, {Schruba}, {Bigiel}, {Blanc}, {Chevance}, {Colombo}, {Faesi}, {Henshaw}, {Herrera}, {Liu}, {Pety}, {Puschnig}, {Saito}, {Sun}, \& {Usero}}]{2020ApJ...897..122L}
{Lang}, P., {Meidt}, S.~E., {Rosolowsky}, E., {et~al.} 2020, \apj, 897, 122

\bibitem[{{Lee} {et~al.}(2023){Lee}, {Sandstrom}, {Leroy}, {Thilker}, {Schinnerer}, {Rosolowsky}, {Larson}, {Egorov}, {Williams}, {Schmidt}, {Emsellem}, {Anand}, {Barnes}, {Belfiore}, {Be{\v{s}}li{\'c}}, {Bigiel}, {Blanc}, {Bolatto}, {Boquien}, {den Brok}, {Cao}, {Chandar}, {Chastenet}, {Chevance}, {Chiang}, {Congiu}, {Dale}, {Deger}, {Eibensteiner}, {Faesi}, {Glover}, {Grasha}, {Groves}, {Hassani}, {Henny}, {Henshaw}, {Hoyer}, {Hughes}, {Jeffreson}, {Jim{\'e}nez-Donaire}, {Kim}, {Kim}, {Klessen}, {Koch}, {Kreckel}, {Kruijssen}, {Li}, {Liu}, {Lopez}, {Maschmann}, {Chen}, {Meidt}, {Murphy}, {Neumann}, {Neumayer}, {Pan}, {Pessa}, {Pety}, {Querejeta}, {Pinna}, {Rodr{\'\i}guez}, {Saito}, {S{\'a}nchez-Bl{\'a}zquez}, {Santoro}, {Sardone}, {Smith}, {Sormani}, {Scheuermann}, {Stuber}, {Sutter}, {Sun}, {Teng}, {Tre{\ss}}, {Usero}, {Watkins}, {Whitmore}, \& {Razza}}]{2023ApJ...944L..17L}
{Lee}, J.~C., {Sandstrom}, K.~M., {Leroy}, A.~K., {et~al.} 2023, \apjl, 944, L17

\bibitem[{{Lee} {et~al.}(2022){Lee}, {Whitmore}, {Thilker}, {Deger}, {Larson}, {Ubeda}, {Anand}, {Boquien}, {Chandar}, {Dale}, {Emsellem}, {Leroy}, {Rosolowsky}, {Schinnerer}, {Schmidt}, {Lilly}, {Turner}, {Van Dyk}, {White}, {Barnes}, {Belfiore}, {Bigiel}, {Blanc}, {Cao}, {Chevance}, {Congiu}, {Egorov}, {Glover}, {Grasha}, {Groves}, {Henshaw}, {Hughes}, {Klessen}, {Koch}, {Kreckel}, {Kruijssen}, {Liu}, {Lopez}, {Mayker}, {Meidt}, {Murphy}, {Pan}, {Pety}, {Querejeta}, {Razza}, {Saito}, {S{\'a}nchez-Bl{\'a}zquez}, {Santoro}, {Sardone}, {Scheuermann}, {Schruba}, {Sun}, {Usero}, {Watkins}, \& {Williams}}]{2022ApJS..258...10L}
{Lee}, J.~C., {Whitmore}, B.~C., {Thilker}, D.~A., {et~al.} 2022, \apjs, 258, 10

\bibitem[{{Leroy} {et~al.}(2021{\natexlab{a}}){Leroy}, {Hughes}, {Liu}, {Pety}, {Rosolowsky}, {Saito}, {Schinnerer}, {Schruba}, {Usero}, {Faesi}, {Herrera}, {Chevance}, {Hygate}, {Kepley}, {Koch}, {Querejeta}, {Sliwa}, {Will}, {Wilson}, {Anand}, {Barnes}, {Belfiore}, {Be{\v{s}}li{\'c}}, {Bigiel}, {Blanc}, {Bolatto}, {Boquien}, {Cao}, {Chandar}, {Chastenet}, {Chiang}, {Congiu}, {Dale}, {Deger}, {den Brok}, {Eibensteiner}, {Emsellem}, {Garc{\'\i}a-Rodr{\'\i}guez}, {Glover}, {Grasha}, {Groves}, {Henshaw}, {Jim{\'e}nez Donaire}, {Kim}, {Klessen}, {Kreckel}, {Kruijssen}, {Larson}, {Lee}, {Mayker}, {McElroy}, {Meidt}, {Mok}, {Pan}, {Puschnig}, {Razza}, {S{\'a}nchez-Bl'azquez}, {Sandstrom}, {Santoro}, {Sardone}, {Scheuermann}, {Sun}, {Thilker}, {Turner}, {Ubeda}, {Utomo}, {Watkins}, \& {Williams}}]{2021ApJS..255...19L}
{Leroy}, A.~K., {Hughes}, A., {Liu}, D., {et~al.} 2021{\natexlab{a}}, \apjs, 255, 19

\bibitem[{{Leroy} {et~al.}(2022){Leroy}, {Rosolowsky}, {Usero}, {Sandstrom}, {Schinnerer}, {Schruba}, {Bolatto}, {Sun}, {Barnes}, {Belfiore}, {Bigiel}, {den Brok}, {Cao}, {Chiang}, {Chevance}, {Dale}, {Eibensteiner}, {Faesi}, {Glover}, {Hughes}, {Jim{\'e}nez Donaire}, {Klessen}, {Koch}, {Kruijssen}, {Liu}, {Meidt}, {Pan}, {Pety}, {Puschnig}, {Querejeta}, {Saito}, {Sardone}, {Watkins}, {Weiss}, \& {Williams}}]{2022ApJ...927..149L}
{Leroy}, A.~K., {Rosolowsky}, E., {Usero}, A., {et~al.} 2022, \apj, 927, 149

\bibitem[{{Leroy} {et~al.}(2019){Leroy}, {Sandstrom}, {Lang}, {Lewis}, {Salim}, {Behrens}, {Chastenet}, {Chiang}, {Gallagher}, {Kessler}, \& {Utomo}}]{2019ApJS..244...24L}
{Leroy}, A.~K., {Sandstrom}, K.~M., {Lang}, D., {et~al.} 2019, \apjs, 244, 24

\bibitem[{{Leroy} {et~al.}(2021{\natexlab{b}}){Leroy}, {Schinnerer}, {Hughes}, {Rosolowsky}, {Pety}, {Schruba}, {Usero}, {Blanc}, {Chevance}, {Emsellem}, {Faesi}, {Herrera}, {Liu}, {Meidt}, {Querejeta}, {Saito}, {Sandstrom}, {Sun}, {Williams}, {Anand}, {Barnes}, {Behrens}, {Belfiore}, {Benincasa}, {Be{\v{s}}li{\'c}}, {Bigiel}, {Bolatto}, {den Brok}, {Cao}, {Chandar}, {Chastenet}, {Chiang}, {Congiu}, {Dale}, {Deger}, {Eibensteiner}, {Egorov}, {Garc{\'\i}a-Rodr{\'\i}guez}, {Glover}, {Grasha}, {Henshaw}, {Ho}, {Kepley}, {Kim}, {Klessen}, {Kreckel}, {Koch}, {Kruijssen}, {Larson}, {Lee}, {Lopez}, {Machado}, {Mayker}, {McElroy}, {Murphy}, {Ostriker}, {Pan}, {Pessa}, {Puschnig}, {Razza}, {S{\'a}nchez-Bl{\'a}zquez}, {Santoro}, {Sardone}, {Scheuermann}, {Sliwa}, {Sormani}, {Stuber}, {Thilker}, {Turner}, {Utomo}, {Watkins}, \& {Whitmore}}]{2021ApJS..257...43L}
{Leroy}, A.~K., {Schinnerer}, E., {Hughes}, A., {et~al.} 2021{\natexlab{b}}, \apjs, 257, 43

\bibitem[{{Leroy} {et~al.}(2009){Leroy}, {Walter}, {Bigiel}, {Usero}, {Weiss}, {Brinks}, {de Blok}, {Kennicutt}, {Schuster}, {Kramer}, {Wiesemeyer}, \& {Roussel}}]{2009AJ....137.4670L}
{Leroy}, A.~K., {Walter}, F., {Bigiel}, F., {et~al.} 2009, \aj, 137, 4670

\bibitem[{{Leroy} {et~al.}(2008){Leroy}, {Walter}, {Brinks}, {Bigiel}, {de Blok}, {Madore}, \& {Thornley}}]{2008AJ....136.2782L}
{Leroy}, A.~K., {Walter}, F., {Brinks}, E., {et~al.} 2008, \aj, 136, 2782

\bibitem[{{Leroy} {et~al.}(2013){Leroy}, {Walter}, {Sandstrom}, {Schruba}, {Munoz-Mateos}, {Bigiel}, {Bolatto}, {Brinks}, {de Blok}, {Meidt}, {Rix}, {Rosolowsky}, {Schinnerer}, {Schuster}, \& {Usero}}]{2013AJ....146...19L}
{Leroy}, A.~K., {Walter}, F., {Sandstrom}, K., {et~al.} 2013, \aj, 146, 19

\bibitem[{{Lord}(1987)}]{1987PhDT........11L}
{Lord}, S.~D. 1987, PhD thesis, Massachusetts Univ., Amherst.

\bibitem[{{Lord} \& {Young}(1990)}]{1990ApJ...356..135L}
{Lord}, S.~D. \& {Young}, J.~S. 1990, \apj, 356, 135

\bibitem[{{Mart{\'{\i}}nez-Garc{\'{\i}}a} {et~al.}(2017){Mart{\'{\i}}nez-Garc{\'{\i}}a}, {Gonz{\'a}lez-L{\'o}pezlira}, {Magris C.}, \& {Bruzual A.}}]{2017ApJ...835...93M}
{Mart{\'{\i}}nez-Garc{\'{\i}}a}, E.~E., {Gonz{\'a}lez-L{\'o}pezlira}, R.~A., {Magris C.}, G., \& {Bruzual A.}, G. 2017, \apj, 835, 93

\bibitem[{{Meidt} {et~al.}(2021){Meidt}, {Leroy}, {Querejeta}, {Schinnerer}, {Sun}, {van der Wel}, {Emsellem}, {Henshaw}, {Hughes}, {Kruijssen}, {Rosolowsky}, {Schruba}, {Barnes}, {Bigiel}, {Blanc}, {Chevance}, {Cao}, {Dale}, {Faesi}, {Glover}, {Grasha}, {Groves}, {Herrera}, {Klessen}, {Kreckel}, {Liu}, {Pan}, {Pety}, {Saito}, {Usero}, {Watkins}, \& {Williams}}]{2021ApJ...913..113M}
{Meidt}, S.~E., {Leroy}, A.~K., {Querejeta}, M., {et~al.} 2021, \apj, 913, 113

\bibitem[{{Meidt} {et~al.}(2018){Meidt}, {Leroy}, {Rosolowsky}, {Kruijssen}, {Schinnerer}, {Schruba}, {Pety}, {Blanc}, {Bigiel}, {Chevance}, {Hughes}, {Querejeta}, \& {Usero}}]{2018ApJ...854..100M}
{Meidt}, S.~E., {Leroy}, A.~K., {Rosolowsky}, E., {et~al.} 2018, \apj, 854, 100

\bibitem[{{Meidt} {et~al.}(2013){Meidt}, {Schinnerer}, {Garc{\'{\i}}a-Burillo}, {Hughes}, {Colombo}, {Pety}, {Dobbs}, {Schuster}, {Kramer}, {Leroy}, {Dumas}, \& {Thompson}}]{2013ApJ...779...45M}
{Meidt}, S.~E., {Schinnerer}, E., {Garc{\'{\i}}a-Burillo}, S., {et~al.} 2013, \apj, 779, 45

\bibitem[{{Meidt} {et~al.}(2012){Meidt}, {Schinnerer}, {Knapen}, {Bosma}, {Athanassoula}, {Sheth}, {Buta}, {Zaritsky}, {Laurikainen}, {Elmegreen}, {Elmegreen}, {Gadotti}, {Salo}, {Regan}, {Ho}, {Madore}, {Hinz}, {Skibba}, {Gil de Paz}, {Mu{\~n}oz-Mateos}, {Men{\'e}ndez-Delmestre}, {Seibert}, {Kim}, {Mizusawa}, {Laine}, \& {Comer{\'o}n}}]{2012ApJ...744...17M}
{Meidt}, S.~E., {Schinnerer}, E., {Knapen}, J.~H., {et~al.} 2012, \apj, 744, 17

\bibitem[{{Meidt} {et~al.}(2014){Meidt}, {Schinnerer}, {van de Ven}, {Zaritsky}, {Peletier}, {Knapen}, {Sheth}, {Regan}, {Querejeta}, {Mu{\~n}oz-Mateos}, {Kim}, {Hinz}, {Gil de Paz}, {Athanassoula}, {Bosma}, {Buta}, {Cisternas}, {Ho}, {Holwerda}, {Skibba}, {Laurikainen}, {Salo}, {Gadotti}, {Laine}, {Erroz-Ferrer}, {Comer{\'o}n}, {Men{\'e}ndez-Delmestre}, {Seibert}, \& {Mizusawa}}]{2014ApJ...788..144M}
{Meidt}, S.~E., {Schinnerer}, E., {van de Ven}, G., {et~al.} 2014, \apj, 788, 144

\bibitem[{{Moore} {et~al.}(2012){Moore}, {Urquhart}, {Morgan}, \& {Thompson}}]{2012MNRAS.426..701M}
{Moore}, T.~J.~T., {Urquhart}, J.~S., {Morgan}, L.~K., \& {Thompson}, M.~A. 2012, \mnras, 426, 701

\bibitem[{{Nair} \& {Abraham}(2010)}]{2010ApJS..186..427N}
{Nair}, P.~B. \& {Abraham}, R.~G. 2010, \apjs, 186, 427

\bibitem[{{Nakanishi} \& {Sofue}(2003)}]{2003PASJ...55..191N}
{Nakanishi}, H. \& {Sofue}, Y. 2003, \pasj, 55, 191

\bibitem[{{Noeske} {et~al.}(2007){Noeske}, {Weiner}, {Faber}, {Papovich}, {Koo}, {Somerville}, {Bundy}, {Conselice}, {Newman}, {Schiminovich}, {Le Floc'h}, {Coil}, {Rieke}, {Lotz}, {Primack}, {Barmby}, {Cooper}, {Davis}, {Ellis}, {Fazio}, {Guhathakurta}, {Huang}, {Kassin}, {Martin}, {Phillips}, {Rich}, {Small}, {Willmer}, \& {Wilson}}]{2007ApJ...660L..43N}
{Noeske}, K.~G., {Weiner}, B.~J., {Faber}, S.~M., {et~al.} 2007, \apjl, 660, L43

\bibitem[{{Oey} {et~al.}(2007){Oey}, {Meurer}, {Yelda}, {Furst}, {Caballero-Nieves}, {Hanish}, {Levesque}, {Thilker}, {Walth}, {Bland-Hawthorn}, {Dopita}, {Ferguson}, {Heckman}, {Doyle}, {Drinkwater}, {Freeman}, {Kennicutt}, {Kilborn}, {Knezek}, {Koribalski}, {Meyer}, {Putman}, {Ryan-Weber}, {Smith}, {Staveley-Smith}, {Webster}, {Werk}, \& {Zwaan}}]{2007ApJ...661..801O}
{Oey}, M.~S., {Meurer}, G.~R., {Yelda}, S., {et~al.} 2007, \apj, 661, 801

\bibitem[{{Pan} {et~al.}(2022){Pan}, {Schinnerer}, {Hughes}, {Leroy}, {Groves}, {Barnes}, {Belfiore}, {Bigiel}, {Blanc}, {Cao}, {Chevance}, {Congiu}, {Dale}, {Eibensteiner}, {Emsellem}, {Faesi}, {Glover}, {Grasha}, {Herrera}, {Ho}, {Klessen}, {Kruijssen}, {Lang}, {Liu}, {McElroy}, {Meidt}, {Murphy}, {Pety}, {Querejeta}, {Razza}, {Rosolowsky}, {Saito}, {Santoro}, {Schruba}, {Sun}, {Tomi{\v{c}}i{\'c}}, {Usero}, {Utomo}, \& {Williams}}]{2022ApJ...927....9P}
{Pan}, H.-A., {Schinnerer}, E., {Hughes}, A., {et~al.} 2022, \apj, 927, 9

\bibitem[{{Pessa} {et~al.}(2021){Pessa}, {Schinnerer}, {Belfiore}, {Emsellem}, {Leroy}, {Schruba}, {Kruijssen}, {Pan}, {Blanc}, {Sanchez-Blazquez}, {Bigiel}, {Chevance}, {Congiu}, {Dale}, {Faesi}, {Glover}, {Grasha}, {Groves}, {Ho}, {Jim{\'e}nez-Donaire}, {Klessen}, {Kreckel}, {Koch}, {Liu}, {Meidt}, {Pety}, {Querejeta}, {Rosolowsky}, {Saito}, {Santoro}, {Sun}, {Usero}, {Watkins}, \& {Williams}}]{2021A&A...650A.134P}
{Pessa}, I., {Schinnerer}, E., {Belfiore}, F., {et~al.} 2021, \aap, 650, A134

\bibitem[{{Pessa} {et~al.}(2022){Pessa}, {Schinnerer}, {Leroy}, {Koch}, {Rosolowsky}, {Williams}, {Pan}, {Schruba}, {Usero}, {Belfiore}, {Bigiel}, {Blanc}, {Chevance}, {Dale}, {Emsellem}, {Gensior}, {Glover}, {Grasha}, {Groves}, {Klessen}, {Kreckel}, {Kruijssen}, {Liu}, {Meidt}, {Pety}, {Querejeta}, {Saito}, {Sanchez-Blazquez}, \& {Watkins}}]{2022A&A...663A..61P}
{Pessa}, I., {Schinnerer}, E., {Leroy}, A.~K., {et~al.} 2022, \aap, 663, A61

\bibitem[{{Pessa} {et~al.}(2023){Pessa}, {Schinnerer}, {Sanchez-Blazquez}, {Belfiore}, {Groves}, {Emsellem}, {Neumann}, {Leroy}, {Bigiel}, {Chevance}, {Dale}, {Glover}, {Grasha}, {Klessen}, {Kreckel}, {Kruijssen}, {Pinna}, {Querejeta}, {Rosolowsky}, \& {Williams}}]{2023A&A...673A.147P}
{Pessa}, I., {Schinnerer}, E., {Sanchez-Blazquez}, P., {et~al.} 2023, \aap, 673, A147

\bibitem[{{Querejeta} {et~al.}(2015){Querejeta}, {Meidt}, {Schinnerer}, {Cisternas}, {Mu{\~n}oz-Mateos}, {Sheth}, {Knapen}, {van de Ven}, {Norris}, {Peletier}, {Laurikainen}, {Salo}, {Holwerda}, {Athanassoula}, {Bosma}, {Groves}, {Ho}, {Gadotti}, {Zaritsky}, {Regan}, {Hinz}, {Gil de Paz}, {Menendez-Delmestre}, {Seibert}, {Mizusawa}, {Kim}, {Erroz-Ferrer}, {Laine}, \& {Comer{\'o}n}}]{2015ApJS..219....5Q}
{Querejeta}, M., {Meidt}, S.~E., {Schinnerer}, E., {et~al.} 2015, \apjs, 219, 5

\bibitem[{{Querejeta} {et~al.}(2023){Querejeta}, {Pety}, {Schruba}, {Leroy}, {Herrera}, {Chiang}, {Meidt}, {Rosolowsky}, {Schinnerer}, {Schuster}, {Sun}, {Herrmann}, {Barnes}, {Be{\v{s}}li{\'c}}, {Bigiel}, {Cao}, {Chevance}, {Eibensteiner}, {Emsellem}, {Faesi}, {Hughes}, {Kim}, {Klessen}, {Kreckel}, {Kruijssen}, {Liu}, {Neumayer}, {Pan}, {Saito}, {Sandstrom}, {Teng}, {Usero}, {Williams}, \& {Zakardjian}}]{2023A&A...680A...4Q}
{Querejeta}, M., {Pety}, J., {Schruba}, A., {et~al.} 2023, \aap, 680, A4

\bibitem[{{Querejeta} {et~al.}(2021){Querejeta}, {Schinnerer}, {Meidt}, {Sun}, {Leroy}, {Emsellem}, {Klessen}, {Mu{\~n}oz-Mateos}, {Salo}, {Laurikainen}, {Be{\v{s}}li{\'c}}, {Blanc}, {Chevance}, {Dale}, {Eibensteiner}, {Faesi}, {Garc{\'\i}a-Rodr{\'\i}guez}, {Glover}, {Grasha}, {Henshaw}, {Herrera}, {Hughes}, {Kreckel}, {Kruijssen}, {Liu}, {Murphy}, {Pan}, {Pety}, {Razza}, {Rosolowsky}, {Saito}, {Schruba}, {Usero}, {Watkins}, \& {Williams}}]{2021A&A...656A.133Q}
{Querejeta}, M., {Schinnerer}, E., {Meidt}, S., {et~al.} 2021, \aap, 656, A133

\bibitem[{{Querejeta} {et~al.}(2019){Querejeta}, {Schinnerer}, {Schruba}, {Murphy}, {Meidt}, {Usero}, {Leroy}, {Pety}, {Bigiel}, {Chevance}, {Faesi}, {Gallagher}, {Garc{\'\i}a-Burillo}, {Glover}, {Hygate}, {Jim{\'e}nez-Donaire}, {Kruijssen}, {Momjian}, {Rosolowsky}, \& {Utomo}}]{2019A&A...625A..19Q}
{Querejeta}, M., {Schinnerer}, E., {Schruba}, A., {et~al.} 2019, \aap, 625, A19

\bibitem[{{Ragan} {et~al.}(2018){Ragan}, {Moore}, {Eden}, {Hoare}, {Urquhart}, {Elia}, \& {Molinari}}]{2018MNRAS.479.2361R}
{Ragan}, S.~E., {Moore}, T.~J.~T., {Eden}, D.~J., {et~al.} 2018, \mnras, 479, 2361

\bibitem[{{Rebolledo} {et~al.}(2012){Rebolledo}, {Wong}, {Leroy}, {Koda}, \& {Donovan Meyer}}]{2012ApJ...757..155R}
{Rebolledo}, D., {Wong}, T., {Leroy}, A., {Koda}, J., \& {Donovan Meyer}, J. 2012, \apj, 757, 155

\bibitem[{{Roberts} {et~al.}(1975){Roberts}, {Roberts}, \& {Shu}}]{1975ApJ...196..381R}
{Roberts}, W.~W., J., {Roberts}, M.~S., \& {Shu}, F.~H. 1975, \apj, 196, 381

\bibitem[{{Roberts}(1969)}]{1969ApJ...158..123R}
{Roberts}, W.~W. 1969, \apj, 158, 123

\bibitem[{{S{\'a}nchez-Menguiano} {et~al.}(2020){S{\'a}nchez-Menguiano}, {S{\'a}nchez}, {P{\'e}rez}, {Ruiz-Lara}, {Galbany}, {Anderson}, \& {Kuncarayakti}}]{2020MNRAS.492.4149S}
{S{\'a}nchez-Menguiano}, L., {S{\'a}nchez}, S.~F., {P{\'e}rez}, I., {et~al.} 2020, \mnras, 492, 4149

\bibitem[{{Santoro} {et~al.}(2022){Santoro}, {Kreckel}, {Belfiore}, {Groves}, {Congiu}, {Thilker}, {Blanc}, {Schinnerer}, {Ho}, {Kruijssen}, {Meidt}, {Klessen}, {Schruba}, {Querejeta}, {Pessa}, {Chevance}, {Kim}, {Emsellem}, {McElroy}, {Barnes}, {Bigiel}, {Boquien}, {Dale}, {Glover}, {Grasha}, {Lee}, {Leroy}, {Pan}, {Rosolowsky}, {Saito}, {Sanchez-Blazquez}, {Watkins}, \& {Williams}}]{2022A&A...658A.188S}
{Santoro}, F., {Kreckel}, K., {Belfiore}, F., {et~al.} 2022, \aap, 658, A188

\bibitem[{{Sarkar} {et~al.}(2023){Sarkar}, {Narayanan}, {Banerjee}, \& {Prakash}}]{2023MNRAS.518.1022S}
{Sarkar}, S., {Narayanan}, G., {Banerjee}, A., \& {Prakash}, P. 2023, \mnras, 518, 1022

\bibitem[{{Schinnerer} {et~al.}(2019){Schinnerer}, {Hughes}, {Leroy}, {Groves}, {Blanc}, {Kreckel}, {Bigiel}, {Chevance}, {Dale}, {Emsellem}, {Faesi}, {Glover}, {Grasha}, {Henshaw}, {Hygate}, {Kruijssen}, {Meidt}, {Pety}, {Querejeta}, {Rosolowsky}, {Saito}, {Schruba}, {Sun}, \& {Utomo}}]{2019ApJ...887...49S}
{Schinnerer}, E., {Hughes}, A., {Leroy}, A., {et~al.} 2019, \apj, 887, 49

\bibitem[{{Schinnerer} {et~al.}(2013){Schinnerer}, {Meidt}, {Pety}, {Hughes}, {Colombo}, {Garc{\'{\i}}a-Burillo}, {Schuster}, {Dumas}, {Dobbs}, {Leroy}, {Kramer}, {Thompson}, \& {Regan}}]{2013ApJ...779...42S}
{Schinnerer}, E., {Meidt}, S.~E., {Pety}, J., {et~al.} 2013, \apj, 779, 42

\bibitem[{{Schmidt}(1959)}]{1959ApJ...129..243S}
{Schmidt}, M. 1959, \apj, 129, 243

\bibitem[{{Schruba} {et~al.}(2010){Schruba}, {Leroy}, {Walter}, {Sandstrom}, \& {Rosolowsky}}]{2010ApJ...722.1699S}
{Schruba}, A., {Leroy}, A.~K., {Walter}, F., {Sandstrom}, K., \& {Rosolowsky}, E. 2010, \apj, 722, 1699

\bibitem[{{Semenov} {et~al.}(2017){Semenov}, {Kravtsov}, \& {Gnedin}}]{2017ApJ...845..133S}
{Semenov}, V.~A., {Kravtsov}, A.~V., \& {Gnedin}, N.~Y. 2017, \apj, 845, 133

\bibitem[{{Sheth} {et~al.}(2010){Sheth}, {Regan}, {Hinz}, {Gil de Paz}, {Men{\'e}ndez-Delmestre}, {Mu{\~n}oz-Mateos}, {Seibert}, {Kim}, {Laurikainen}, {Salo}, {Gadotti}, {Laine}, {Mizusawa}, {Armus}, {Athanassoula}, {Bosma}, {Buta}, {Capak}, {Jarrett}, {Elmegreen}, {Elmegreen}, {Knapen}, {Koda}, {Helou}, {Ho}, {Madore}, {Masters}, {Mobasher}, {Ogle}, {Peng}, {Schinnerer}, {Surace}, {Zaritsky}, {Comer{\'o}n}, {de Swardt}, {Meidt}, {Kasliwal}, \& {Aravena}}]{2010PASP..122.1397S}
{Sheth}, K., {Regan}, M., {Hinz}, J.~L., {et~al.} 2010, \pasp, 122, 1397

\bibitem[{{Smith} {et~al.}(2014){Smith}, {Glover}, {Clark}, {Klessen}, \& {Springel}}]{2014MNRAS.441.1628S}
{Smith}, R.~J., {Glover}, S. C.~O., {Clark}, P.~C., {Klessen}, R.~S., \& {Springel}, V. 2014, \mnras, 441, 1628

\bibitem[{{Stuber} {et~al.}(2023){Stuber}, {Schinnerer}, {Williams}, {Querejeta}, {Meidt}, {Emsellem}, {Barnes}, {Klessen}, {Leroy}, {Neumann}, {Sormani}, {Bigiel}, {Chevance}, {Dale}, {Faesi}, {Glover}, {Grasha}, {Kruijssen}, {Liu}, {Pan}, {Pety}, {Pinna}, {Saito}, {Usero}, \& {Watkins}}]{2023A&A...676A.113S}
{Stuber}, S.~K., {Schinnerer}, E., {Williams}, T.~G., {et~al.} 2023, \aap, 676, A113

\bibitem[{{Sun} {et~al.}(2022){Sun}, {Leroy}, {Rosolowsky}, {Hughes}, {Schinnerer}, {Schruba}, {Koch}, {Blanc}, {Chiang}, {Groves}, {Liu}, {Meidt}, {Pan}, {Pety}, {Querejeta}, {Saito}, {Sandstrom}, {Sardone}, {Usero}, {Utomo}, {Williams}, {Barnes}, {Benincasa}, {Bigiel}, {Bolatto}, {Boquien}, {Chevance}, {Dale}, {Deger}, {Emsellem}, {Glover}, {Grasha}, {Henshaw}, {Klessen}, {Kreckel}, {Kruijssen}, {Ostriker}, \& {Thilker}}]{2022AJ....164...43S}
{Sun}, J., {Leroy}, A.~K., {Rosolowsky}, E., {et~al.} 2022, \aj, 164, 43

\bibitem[{{Sun} {et~al.}(2020){Sun}, {Leroy}, {Schinnerer}, {Hughes}, {Rosolowsky}, {Querejeta}, {Schruba}, {Liu}, {Saito}, {Herrera}, {Faesi}, {Usero}, {Pety}, {Kruijssen}, {Ostriker}, {Bigiel}, {Blanc}, {Bolatto}, {Boquien}, {Chevance}, {Dale}, {Deger}, {Emsellem}, {Glover}, {Grasha}, {Groves}, {Henshaw}, {Jimenez-Donaire}, {Kim}, {Klessen}, {Kreckel}, {Lee}, {Meidt}, {Sandstrom}, {Sardone}, {Utomo}, \& {Williams}}]{2020ApJ...901L...8S}
{Sun}, J., {Leroy}, A.~K., {Schinnerer}, E., {et~al.} 2020, \apjl, 901, L8

\bibitem[{{Teng} {et~al.}(2024){Teng}, {Chiang}, {Sandstrom}, {Sun}, {Leroy}, {Bolatto}, {Usero}, {Ostriker}, {Querejeta}, {Chastenet}, {Bigiel}, {Boquien}, {den Brok}, {Cao}, {Chevance}, {Chown}, {Colombo}, {Eibensteiner}, {Glover}, {Grasha}, {Henshaw}, {Jim{\'e}nez-Donaire}, {Liu}, {Murphy}, {Pan}, {Stuber}, \& {Williams}}]{2024ApJ...961...42T}
{Teng}, Y.-H., {Chiang}, I.~D., {Sandstrom}, K.~M., {et~al.} 2024, \apj, 961, 42

\bibitem[{{Tress} {et~al.}(2020){Tress}, {Smith}, {Sormani}, {Glover}, {Klessen}, {Mac Low}, \& {Clark}}]{2020MNRAS.492.2973T}
{Tress}, R.~G., {Smith}, R.~J., {Sormani}, M.~C., {et~al.} 2020, \mnras, 492, 2973

\bibitem[{{Urquhart} {et~al.}(2021){Urquhart}, {Figura}, {Cross}, {Wells}, {Moore}, {Eden}, {Ragan}, {Pettitt}, {Duarte-Cabral}, {Colombo}, {Schuller}, {Csengeri}, {Mattern}, {Beuther}, {Menten}, {Wyrowski}, {Anderson}, {Barnes}, {Beltr{\'a}n}, {Billington}, {Bronfman}, {Giannetti}, {Kainulainen}, {Kauffmann}, {Lee}, {Leurini}, {Medina}, {Montenegro-Montes}, {Riener}, {Rigby}, {S{\'a}nchez-Monge}, {Schilke}, {Schisano}, {Traficante}, \& {Wienen}}]{2021MNRAS.500.3050U}
{Urquhart}, J.~S., {Figura}, C., {Cross}, J.~R., {et~al.} 2021, \mnras, 500, 3050

\bibitem[{{Utomo} {et~al.}(2018){Utomo}, {Sun}, {Leroy}, {Kruijssen}, {Schinnerer}, {Schruba}, {Bigiel}, {Blanc}, {Chevance}, {Emsellem}, {Herrera}, {Hygate}, {Kreckel}, {Ostriker}, {Pety}, {Querejeta}, {Rosolowsky}, {Sandstrom}, \& {Usero}}]{2018ApJ...861L..18U}
{Utomo}, D., {Sun}, J., {Leroy}, A.~K., {et~al.} 2018, \apjl, 861, L18

\bibitem[{{Vogel} {et~al.}(1988){Vogel}, {Kulkarni}, \& {Scoville}}]{1988Natur.334..402V}
{Vogel}, S.~N., {Kulkarni}, S.~R., \& {Scoville}, N.~Z. 1988, \nat, 334, 402

\bibitem[{{Walter} {et~al.}(2008){Walter}, {Brinks}, {de Blok}, {Bigiel}, {Kennicutt}, {Thornley}, \& {Leroy}}]{2008AJ....136.2563W}
{Walter}, F., {Brinks}, E., {de Blok}, W.~J.~G., {et~al.} 2008, \aj, 136, 2563

\bibitem[{{Willett} {et~al.}(2013){Willett}, {Lintott}, {Bamford}, {Masters}, {Simmons}, {Casteels}, {Edmondson}, {Fortson}, {Kaviraj}, {Keel}, {Melvin}, {Nichol}, {Raddick}, {Schawinski}, {Simpson}, {Skibba}, {Smith}, \& {Thomas}}]{2013MNRAS.435.2835W}
{Willett}, K.~W., {Lintott}, C.~J., {Bamford}, S.~P., {et~al.} 2013, \mnras, 435, 2835

\bibitem[{{Williams} {et~al.}(2022){Williams}, {Kreckel}, {Belfiore}, {Groves}, {Sandstrom}, {Santoro}, {Blanc}, {Bigiel}, {Boquien}, {Chevance}, {Congiu}, {Emsellem}, {Glover}, {Grasha}, {Klessen}, {Koch}, {Kruijssen}, {Leroy}, {Liu}, {Meidt}, {Pan}, {Querejeta}, {Rosolowsky}, {Saito}, {S{\'a}nchez-Bl{\'a}zquez}, {Schinnerer}, {Schruba}, \& {Watkins}}]{2022MNRAS.509.1303W}
{Williams}, T.~G., {Kreckel}, K., {Belfiore}, F., {et~al.} 2022, \mnras, 509, 1303

\bibitem[{{Williams} {et~al.}(2024){Williams}, {Lee}, {Larson}, {Leroy}, {Sandstrom}, {Schinnerer}, {Thilker}, {Belfiore}, {Egorov}, {Rosolowsky}, {Sutter}, {DePasquale}, {Pagan}, {Anand}, {Barnes}, {Bigiel}, {Boquien}, {Cao}, {Chastenet}, {Chevance}, {Chown}, {Dale}, {Eibensteiner}, {Emsellem}, {Faesi}, {Glover}, {Grasha}, {Hannon}, {Hassani}, {Henshaw}, {Jim{\'e}nez-Donaire}, {Kim}, {Klessen}, {Koch}, {Li}, {Liu}, {Meidt}, {M{\'e}ndez-Delgado}, {Murphy}, {Neumann}, {Neumann}, {Neumayer}, {Oakes}, {Pathak}, {Pety}, {Pinna}, {Querejeta}, {Ramambason}, {Romanelli}, {Sormani}, {Stuber}, {Sun}, {Teng}, {Usero}, {Watkins}, \& {Weinbeck}}]{2024arXiv240115142W}
{Williams}, T.~G., {Lee}, J.~C., {Larson}, K.~L., {et~al.} 2024, arXiv e-prints, arXiv:2401.15142

\bibitem[{{Yajima} {et~al.}(2021){Yajima}, {Sorai}, {Miyamoto}, {Muraoka}, {Kuno}, {Kaneko}, {Takeuchi}, {Yasuda}, {Tanaka}, {Morokuma-Matsui}, \& {Kobayashi}}]{2021PASJ...73..257Y}
{Yajima}, Y., {Sorai}, K., {Miyamoto}, Y., {et~al.} 2021, \pasj, 73, 257

\end{thebibliography}
\bibliographystyle{aa}{}

\newpage

\normalsize
\appendix

\onecolumn

\section{Spiral galaxy sample}
\label{sec:appendix_sample}

\begin{center}
\begin{longtable}{lccccc}
\caption{Spiral galaxy sample studied in this paper. \label{table:sample}}\\ 
\hline\hline 
& log($M_\star$/${\rm M}_\odot$) & log(SFR/[${\rm M}_\odot$\,yr$^{-1}$]) & spiral morphology & bar & MUSE \\ 
\hline
%\endfirsthead
%\caption{Continued.} \\
%\hline\hline
%& log($M_\star$/$M_\odot$) & log(SFR/[$M_\odot$\,yr$^{-1}$]) & $V_{\rm min}$/[km\,s$^{-1}$] & $V_{\rm max}$/[km\,s$^{-1}$]  \\ 
%\hline
%\endhead
%\hline
%\endfoot
IC\,1954     &    9.7 &  -0.44 &    M & 1 &   0  \\
NGC\,0628    &   10.3 &   0.24 &    M & 0 &    1  \\
NGC\,1097    &   10.8 &   0.68 &   G & 1 &   0  \\
NGC\,1300    &   10.6 &   0.07 &   G & 1 &   1  \\
NGC\,1365    &   11.0 &   1.23 &   G & 1 &   1  \\
NGC\,1385    &   10.0 &   0.32 &   F & 0 &   1  \\
NGC\,1512    &   10.7 &   0.11 &    G & 1 &   1  \\
NGC\,1566    &   10.8 &   0.66 &   G & 1 &   1  \\
NGC\,1637    &    9.9 &  -0.19 &    M & 1 &    0  \\
NGC\,1672    &   10.7 &   0.88 &   G & 1 &   1  \\
NGC\,2090    &   10.0 &  -0.39 &    G & 0 &   0  \\
NGC\,2283    &    9.9 &  -0.28 &    F & 1 &    0  \\
NGC\,2566    &   10.7 &   0.94 &   G & 1 &   0  \\
NGC\,2835    &   10.0 &   0.09 &    G & 1 &    1  \\
NGC\,2997    &   10.7 &   0.64 &    G & 0 &   0  \\
NGC\,3507    &   10.4 &  -0.00 &    G & 1 &   0  \\
NGC\,3627    &   10.8 &   0.58 &    G & 1 &   1  \\
NGC\,4254    &   10.4 &   0.49 &   M & 0 &   1  \\
NGC\,4303    &   10.5 &   0.73 &   M & 1 &   1  \\
NGC\,4321    &   10.7 &   0.55 &   G & 1 &   1  \\
NGC\,4535    &   10.5 &   0.33 &   M & 1 &   1  \\
NGC\,4536    &   10.4 &   0.54 &   M & 1 &   0  \\
NGC\,4548    &   10.7 &  -0.28 &    G & 1 &    0  \\
NGC\,4579    &   11.1 &   0.34 &   G & 1 &   0  \\
NGC\,4731    &    9.5 &  -0.22 &   G & 1 &   0  \\
NGC\,5248    &   10.4 &   0.36 &   G & 1 &   0  \\
NGC\,5643    &   10.3 &   0.41 &   M & 1 &   0  \\
NGC\,6744    &   10.7 &   0.38 &   M & 1 &   0  \\
\end{longtable}
\tablefoot{Stellar mass and star formation rates of the spiral galaxies studied in this paper \citep{2021ApJS..257...43L}. 
The spiral morphology lists the spiral family according to \citet{2015ApJS..217...32B} or the definitions adopted in \citet{2021ApJ...913..113M} for galaxies outside the S$^4$G survey (`G' stands for grand-design, `M' for multi-armed, and `F' for flocculent). The bar presence (1 for barred) reflects the environmental masks from \citet{2021A&A...656A.133Q}. The last column indicates if the galaxy is in the PHANGS--MUSE subsample (1) or not (0).}
\end{center}

\twocolumn

\section{Calibration of SFR arm/interarm contrasts: The effect of extinction on H$\alpha$}
\label{sec:AppendixMUSE}

\begin{figure}[t]
\begin{center}
\includegraphics[trim=0 0 0 0, clip,width=0.48\textwidth]{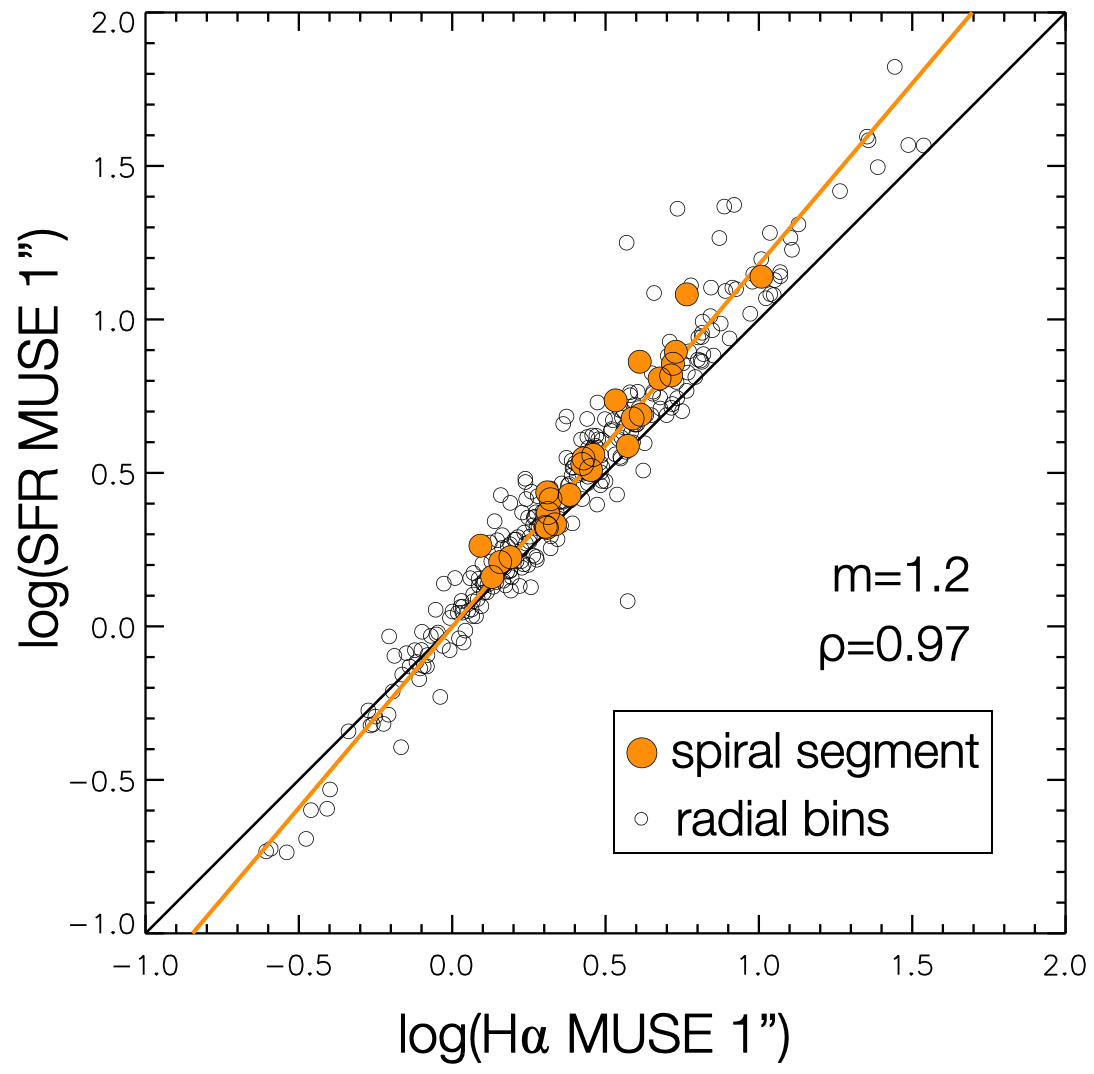}
\end{center}
\caption{arm/interarm contrast in SFR surface density based on extinction-corrected H$\alpha$ from MUSE as a function of the SFR contrast inferred directly from H$\alpha$ (without accounting for extinction). Open circles represent radial measurements (in radial bins of 500\,pc) across all PHANGS--MUSE galaxies, while orange circles show the mean contrast for each spiral segment (typically contributing two points per galaxy).}
\label{fig:SFR_calibration1}
\end{figure}

Here, we use the PHANGS-MUSE dataset to calibrate the effect of neglecting extinction on SFR arm/interarm contrasts based on H$\alpha$. For 13 galaxies, we have both narrow-band and MUSE data, and we consider the arm/interarm contrasts in matched radial bins for the field of view in common. Figure~\ref{fig:SFR_calibration1} shows this relation, in which the degree of correlation is remarkably high (Spearman rank coefficient 0.97). Yet, the relation is not one-to-one, and amplifies the contrast, in such a way that the largest contrasts get a proportionally larger boost when extinction is accounted for. This amplification effect is well described by the power-law relation shown in the figure, $\log\left({\rm SFR_{contrast}^{ext-corr}}\right)=-0.0013 + 1.1799\, \log\left({\rm H\alpha_{contrast}^{no-ext}}\right)$, where ${\rm SFR_{contrast}^{ext-corr}}$ is the arm/interarm contrast of SFR from MUSE, including the extinction correction based on the Balmer decrement, and ${\rm H\alpha_{contrast}^{no-ext}}$ is the SFR arm/interarm contrast consistently from MUSE H$\alpha$, but switching off the extinction correction. This means that we can use H$\alpha$ contrasts from our narrow-band imaging as a surrogate for the real SFR contrasts, without the need to degrade to lower resolution to apply a hybrid recipe that accounts for obscured star formation.

The scatter around the relation shown in Figure~\ref{fig:SFR_calibration1} is only 0.1\,dex (standard deviation in the $y$-axis relative to the power-law fit). This informs us about the uncertainty when neglecting extinction in our SFR contrast measurements. However, there are other sources of error that can also affect the contrasts based on narrow-band H$\alpha$. Indeed, the narrow-band H$\alpha$ maps are not perfectly identical to the MUSE-based H$\alpha$ maps for a number of reasons, including the [NII] contamination removal and filter transmission, neglecting H$\alpha$ absorption when subtracting the continuum broad-band from the narrow-band images, and from an imperfect calibration of narrow-band against broad-band.

\begin{figure}[t]
\begin{center}
\includegraphics[trim=0 0 0 0, clip,width=0.48\textwidth]{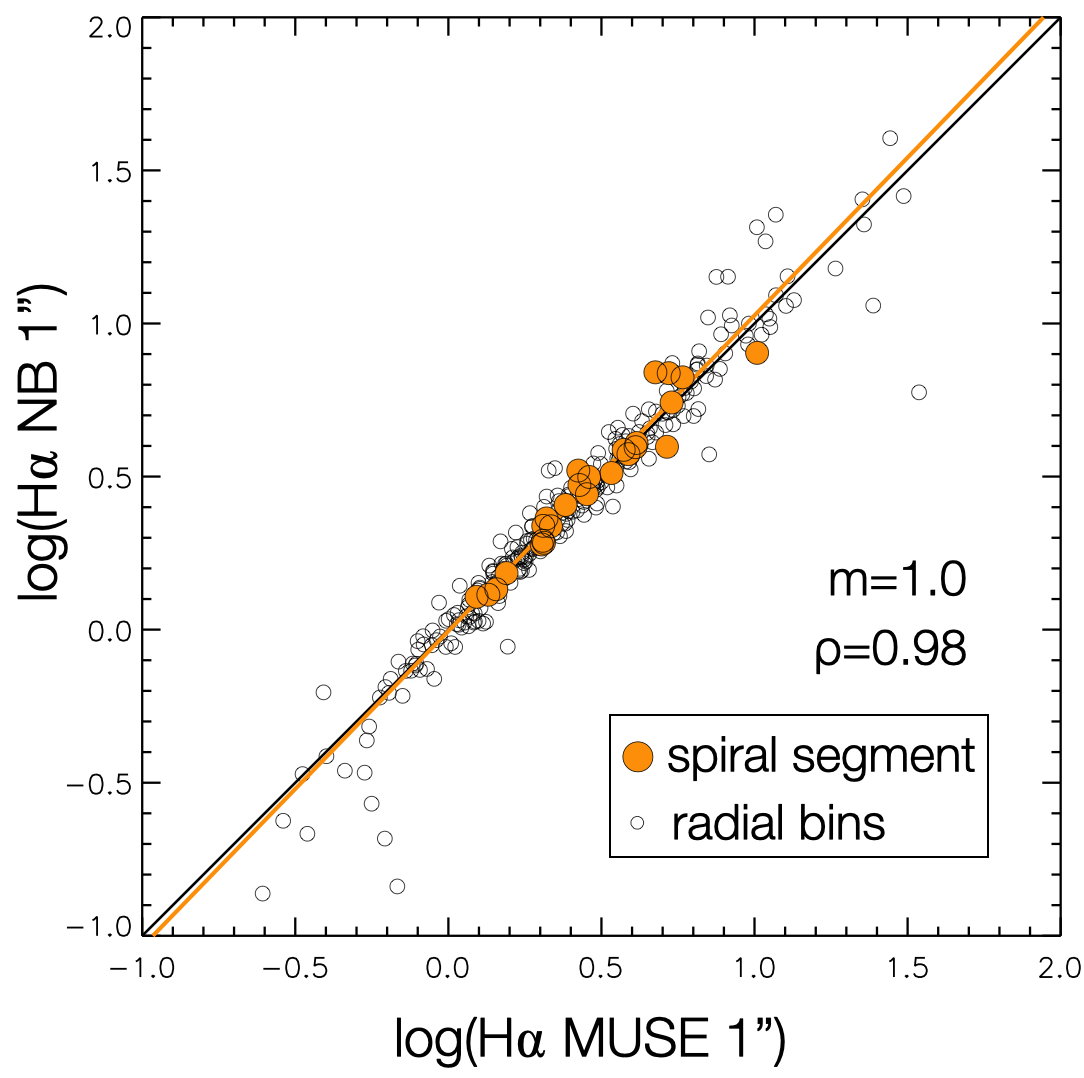}
\end{center}
\caption{arm/interarm contrast of narrow-band H$\alpha$ surface density as a fuction of the MUSE-based H$\alpha$ surface density contrast (without accounting for extinction). Open circles represent radial measurements, and orange circles, means for each spiral segment, as in Fig.\,\ref{fig:SFR_calibration1}.}
\label{fig:SFR_calibration2}
\end{figure}

To gauge the uncertainty associated with narrow-band H$\alpha$ calibration effects, we quantify the scatter when plotting the narrow-band H$\alpha$ contrasts as a function of the MUSE-based H$\alpha$ contrasts, for the same field of view and radial bins in galaxies where both are available. As shown in Figure~\ref{fig:SFR_calibration2}, there is a very good correlation between both variables (Spearman rank coefficient 0.98), with a vertical scatter of 0.059\,dex around the 1:1 relation; we consider this representative of the uncertainty associated with SFR contrasts due to calibration limitations. We add it in quadrature to the uncertainty in our empirical calibration for the effect of extinction. Like this, we obtain a final uncertainty of 0.122\,dex on the SFR surface density contrasts. 

\begin{figure}[t]
\begin{center}
\includegraphics[trim=0 0 0 0, clip,width=0.48\textwidth]{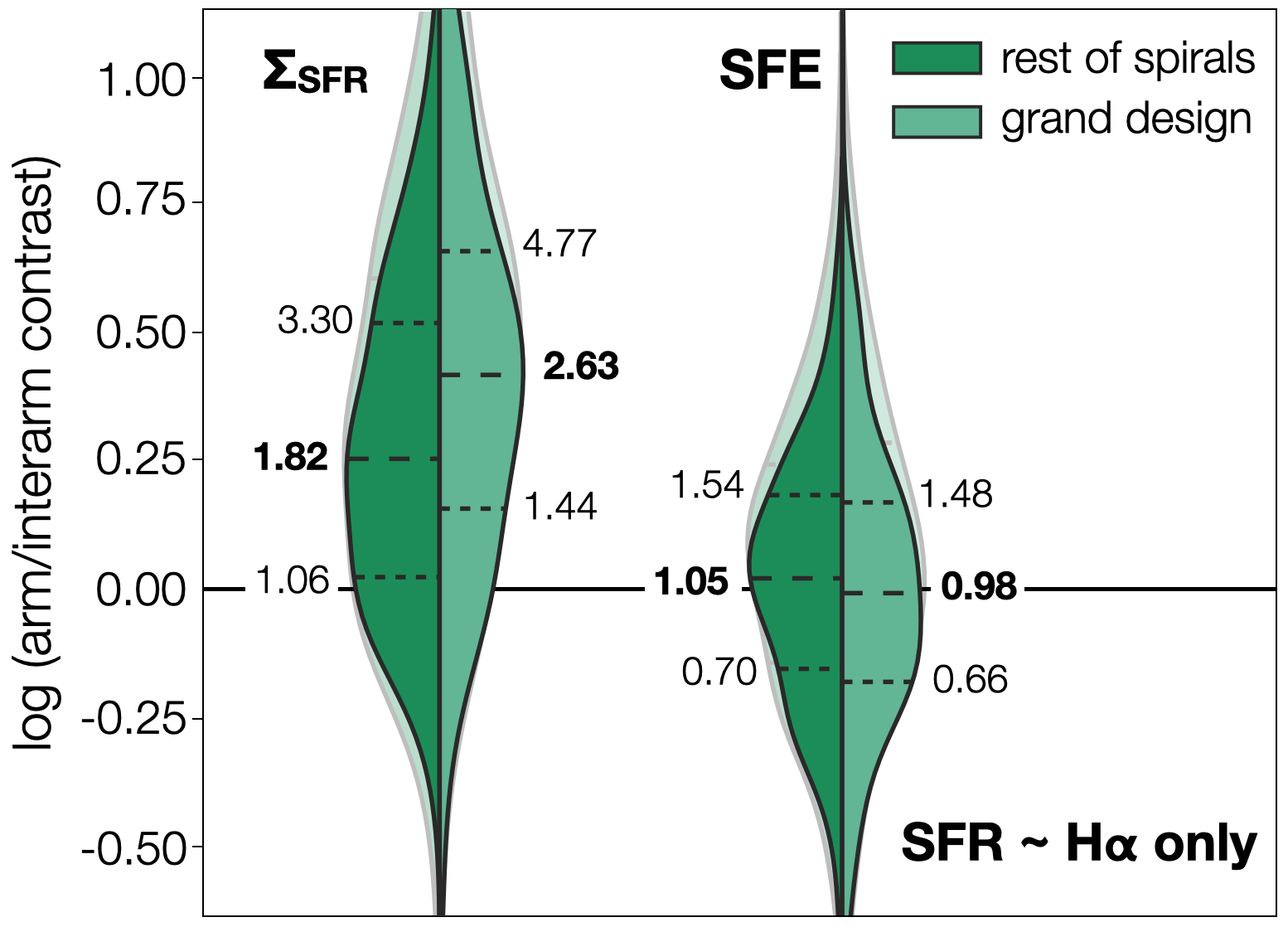}
\end{center}
\caption{Violin plot showing the distribution of SFR and SFE ($1/\tau_{\rm dep}$) arm/interarm contrasts in logarithmic scale for all radial bins across the PHANGS targets. As opposed to Fig.\,\ref{fig:violin_plot_GD-vs-rest}, here the SFR contrast is calculated directly as the ratio of H$\alpha$, without applying any corrections to account for extinction. The equivalent violins following our nominal approach (those from Fig.\,\ref{fig:violin_plot_GD-vs-rest}) are shown as translucent in the background for reference; these include the calibration introduced in Sect.\,\ref{sec:AppendixMUSE} to consider extinction effects. The long dashed line shows the median of the distribution in each case, while the short dashed lines display the 25$^{\rm th}$ and 75$^{\rm th}$ percentiles of the data, with labels indicating for reference the corresponding values on a linear scale.
}
\label{fig:violin_plot_GD-vs-rest-Ha-only}
\end{figure}

With this empirical calibration we circumvent the limitation of obscured star formation and assign realistic uncertainties to the SFR contrasts. These can be propagated to the measurement of SFE contrasts, so that we can conclusively tell in how many cases SFE is statistically enhanced in spiral arms relative to the interarm. For reference, Fig.~\ref{fig:violin_plot_GD-vs-rest-Ha-only} shows the distribution of SFR contrasts (split into grand-design spirals and the rest) if we totally neglect extinction and take the ratio of arm-to-interarm H$\alpha$ surface densities.

\section{Stellar arm/interarm contrasts using different tracers}
\label{Sec:stellarcon}

\begin{figure*}[t]
\begin{center}
\includegraphics[trim=0 0 0 0, clip,width=0.95\textwidth]{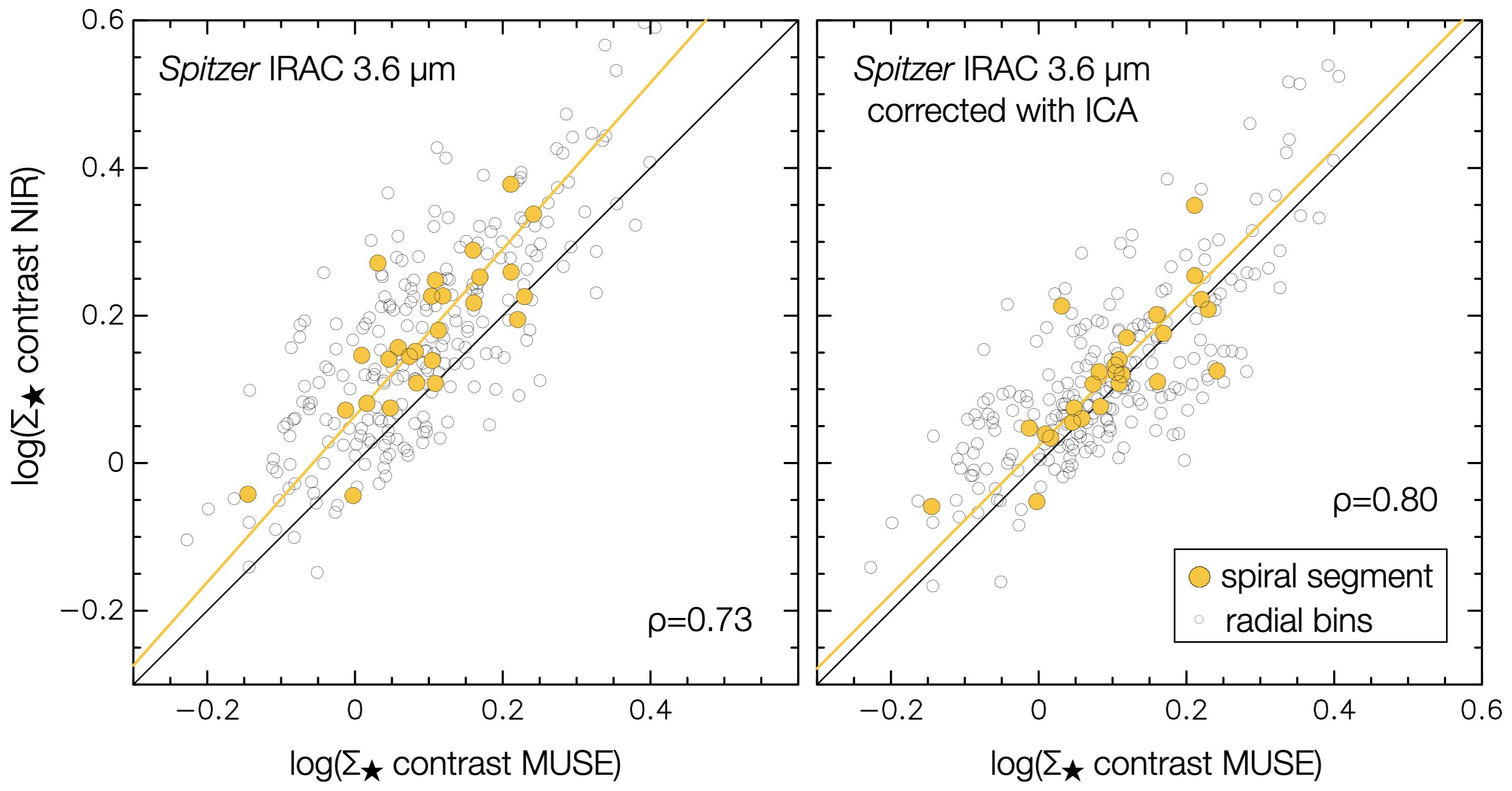}
\end{center}
\caption{Stellar mass surface density (\SigStar) arm/interarm contrast measured from NIR imaging as a function of the equivalent contrast from MUSE \citep{2023A&A...673A.147P}. These sanity checks are possible for a subset of 13 galaxies and for a more limited field of view. The right panel demonstrates that the agremeent improves when applying the ICA correction (Sect.~\ref{Sec:stellarmass}) to the  {\it Spitzer} $3.6$\,$\mu$m maps. The median absolute deviation drops from 0.08 to 0.03\,dex.
}
\label{fig:MUSE_SigStar_vs_IRAC-ICA}
\end{figure*}

Ground- and space-based NIR imaging is a common tracer of stellar mass, but the presence of dust emission can locally bias our view of the distribution of stellar surface density. This is particularly true when it comes to estimating arm/interarm contrasts. In this paper, we rely on stellar mass maps based on {\it Spitzer}/IRAC imaging corrected for non-stellar emission using an independent component analysis (ICA) method \citep{2012ApJ...744...17M,2015ApJS..219....5Q}.

Figure~\ref{fig:MUSE_SigStar_vs_IRAC-ICA} compares the stellar arm/interarm contrast from NIR imaging against fully independent estimates based on PHANGS--MUSE. These stellar mass maps rely on stellar population fitting (via pPXF) to Voronoi bins on the MUSE cubes, as implemented in the PHANGS--MUSE data analysis pipeline \citep{2022A&A...659A.191E} and presented in \citet{2021A&A...650A.134P,2022A&A...663A..61P,2023A&A...673A.147P}. 
The change from the left to the right panel in Fig.~\ref{fig:MUSE_SigStar_vs_IRAC-ICA} shows that the stellar contrast drops with the ICA correction, and comes much closer to the MUSE estimate. This decrease is expected, as the dust contribution that ICA corrects for is stronger around star-forming regions, and, thus, in spiral arms. The median stellar contrast drops from 1.71 to 1.56 after applying the ICA correction (while the mean contrast drops from 1.98 to 1.71).

With the ICA correction, \SigStar\ contrasts show a better agreement with MUSE. In particular, the best-fit power law to means along spiral segments changes from $0.064 + 1.128\, \log\left({\rm \Sigma_{\star ~ contrast}^{MUSE}}\right)$ to $0.023 + 1.005\, \log\left({\rm \Sigma_{\star ~ contrast}^{MUSE}}\right)$. Thus, the best-fit power-law becomes more linear and the vertical offset with respect to the 1:1 line drops by a factor of three. The correlation also becomes slightly stronger (from $\rho = 0.73$ to $\rho = 0.80$) and the median absolute deviation drops from 0.08 to 0.03\,dex. This sanity check is possible for a subset of 13 galaxies. For M51, \citet{2019A&A...625A..19Q} also found that the ICA-corrected stellar mass map agrees well (better than the original IRAC) with an independent stellar mass map of M51 obtained through Bayesian marginalisation analysis by \citet{2017ApJ...835...93M}.

\begin{figure}[t]
\begin{center}
\includegraphics[trim=0 0 0 0, clip,width=0.48\textwidth]{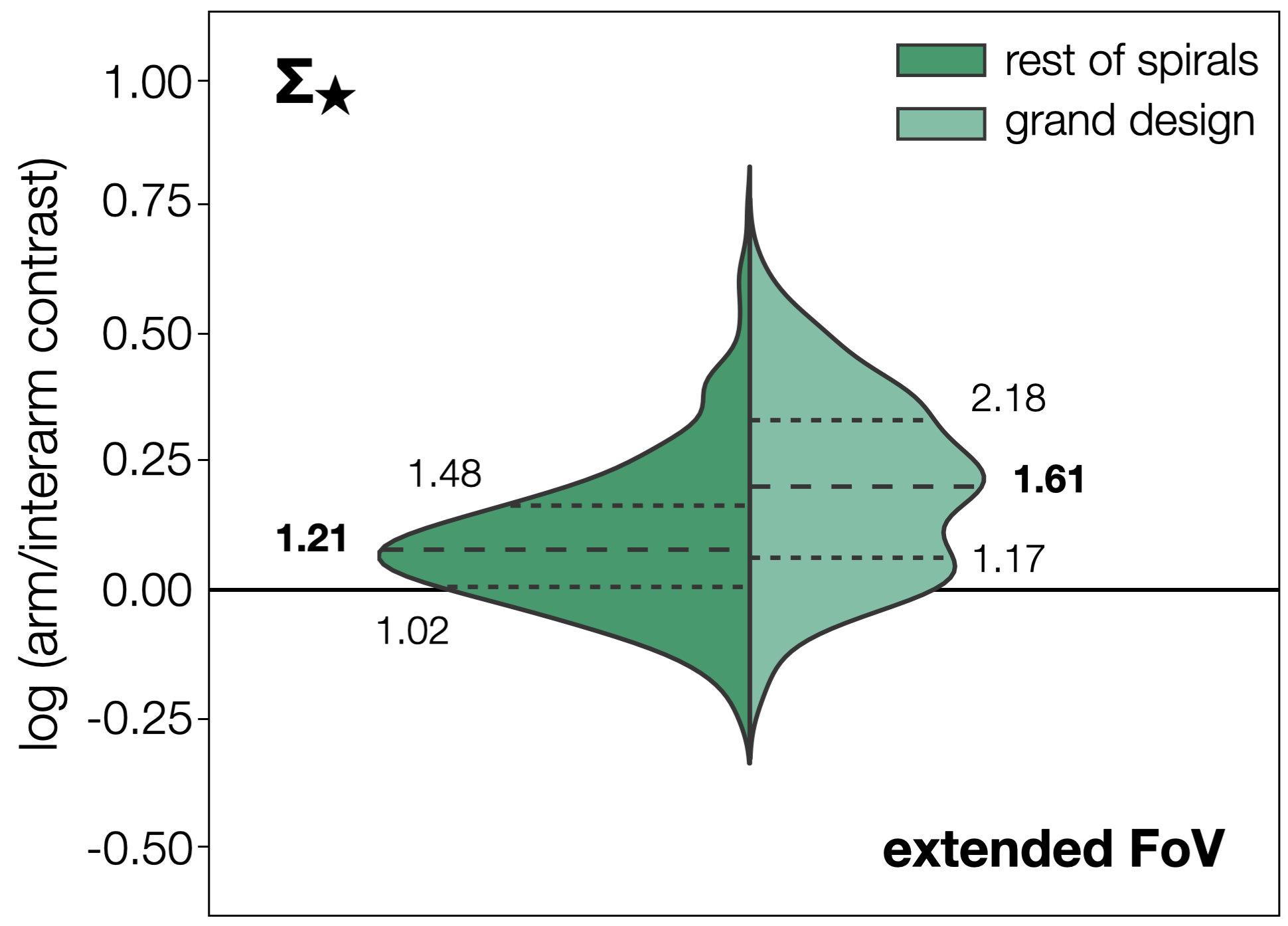}
\end{center}
\caption{Violin plot showing the distribution of stellar mass arm/interarm contrasts in logarithmic scale for all radial bins across the PHANGS targets, considering an extended field of view (all \SigStar\ measurements across spiral arms, not limited to the PHANGS--ALMA field of view). The violin plots are split into grand-design galaxies (right) and the rest of spirals (left). The long dashed line shows the median of the distribution in each case, while the short dashed lines display the 25$^{\rm th}$ and 75$^{\rm th}$ percentiles of the data, with labels indicating for reference the corresponding values on a linear scale.
}
\label{fig:violin_plot_GD-vs-rest-extended}
\end{figure}

For completeness, Fig.~\ref{fig:violin_plot_GD-vs-rest-extended} shows our nominal stellar contrasts based on ICA-corrected $3.6$\,$\mu$m for an extended field of view (beyond the end of PHANGS--ALMA CO coverage). This confirms the significantly larger stellar contrasts associated with grand-design spirals.

\section{The implications of DIG for SFR arm/interarm contrast}
\label{Sec:DIG}

While H$\alpha$ is a traditional tracer of star formation, as much as $\sim$50\% of H$\alpha$ emission does not originate directly from \hii{} regions, but from a more diffuse and extended component, often known as diffuse ionised gas \citep[DIG;][]{2007ApJ...661..801O,2009RvMP...81..969H,2016ApJ...827..103K,2020MNRAS.493.2872C}.
As shown by \citet{2022A&A...659A..26B}, this component originates predominantly from leaking ionising photons from \hii{} regions, which can travel $\sim$kpc distances before they contribute to the DIG emission. Therefore, there could be a fair concern that the DIG component introduces some dilution in the arm/interarm contrast of \SigSFR. In this section we examine the issue and conclude that DIG is unlikely to have a major impact in SFR contrast dilution.

While photons can travel significant distances away from \hii{} regions before powering DIG (mean free path for the ionising radiation of 1.9 kpc according to \citealt{2022A&A...659A..26B}), this leakage is expected to happen mostly towards higher disc scale-heights, where densities are lower, away from the thin cold gas disc. This means that, since our targets have moderate inclinations (maximum inclination in the spiral sample considered here is 66$^\circ$), if photons mostly escape away from the galaxy plane, thanks to projection we will end up finding those photons relatively close to the \hii{} region that they originated from. Since our spiral masks are reasonably broad (1-2\,kpc), based on this argument we would not expect a large leakage of ionising photons from arm to interarm (we would expect most DIG due to spiral star formation within the spiral masks).

The maps from PHANGS--MUSE show that the most intense DIG seems to accumulate immediately around \hii{} regions, mostly within distances of a few hundred parsecs. 
To confirm this visual impression, we performed a test using the PHANGS--MUSE data, for which \hii{} region masks are available \citep{2022A&A...658A.188S,2023MNRAS.520.4902G}. Our main concern is whether a significant fraction of the H$\alpha$ emission outside \hii{} regions in the interarm arises from DIG due to leaked photons from spiral arms, which could bias the arm/interarm SFR (and SFE) contrasts. 
If this were the dominant contribution to DIG in the interarm, we would expect this component to be distributed relatively smoothly across the interarm and not preferentially close to \hii{} regions in the interarm. We find that, on average, $\sim$70\% of the (Balmer-corrected) H$\alpha$ flux in the interarm is inside the \hii{} region masks, covering an area of $\sim$15\% of the interarm region. However, the remaining $\sim$30\% of H$\alpha$ flux is not evenly distributed across the remaining pixels. On the contrary, most of the remaining flux is close to interarm \hii{} regions. If we dilate the \hii{} region masks to cover $\sim$25\% of the remaining area, we do not recover 25\% of the remaining flux, but as much as 60\% of the remaining flux. Therefore, most DIG in the interarm regions seems quite clearly associated with interarm \hii{} regions, and not due to leaked photons from the spiral arms.

Based on the arguments we\ examine above, we believe that DIG should be included in our arm/interarm contrast measurements, as we have done across this paper. Previous studies tried to correct for DIG, under the premise that it is not directly tracing star formation \citep[e.g.][]{2022ApJ...927....9P}. If we consider the arm/interarm contrast purely arising from flux found within \hii{} regions (setting to zero all pixels outside the \hii{} region masks for the MUSE targets), we find that the SFR contrasts tend to go up (by 45\% on average, or 34\% median). However, such measurements suffer from an important bias, because the filling factor of \hii{} regions in spiral arms is much higher than in the interarm, and this means that a leaked photon is more likely to be found within the footprint of a nearby \hii{} region in spiral arms than in the interarm. As \hii{} regions are more sparse in the interarm, leaked photons are less likely to be found within the footprint of another \hii{} region.

\section{Effect of mask width} 
\label{Sec:narrow}

\begin{figure*}[t]
\begin{center}
\includegraphics[trim=0 0 0 0, clip,width=0.9\textwidth]{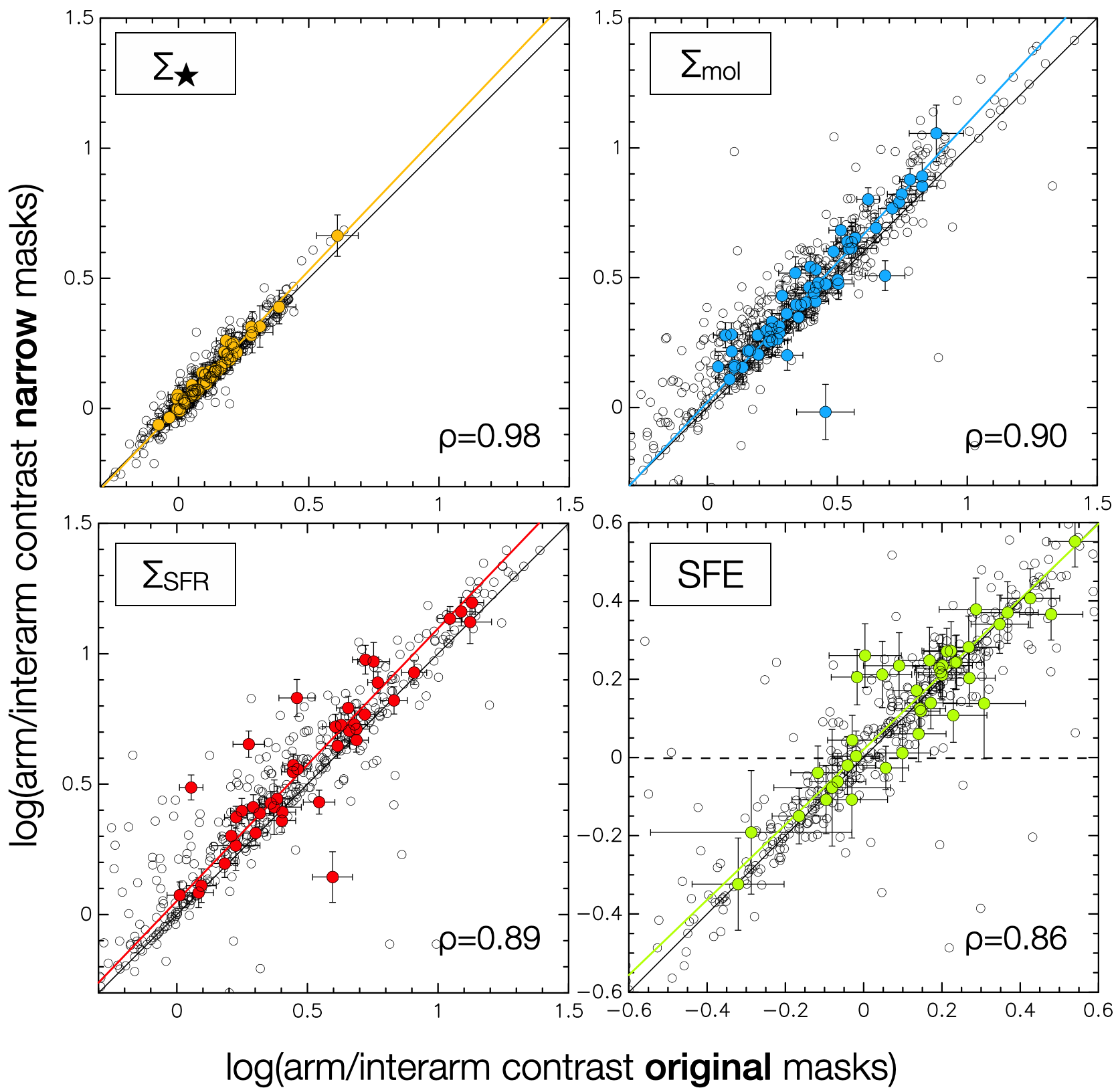}
\end{center}
\caption{Comparison of original versus narrow spiral masks (as defined in Sect.\,\ref{Sec:thinnersp}).
}
\label{fig:orig_vs_narrow}
\end{figure*}

The spiral masks released in \citet{2021A&A...656A.133Q}, the nominal option adopted in this paper, are relatively broad, with a typical width of $1{-}2$\,kpc. This is in order to accommodate for local departures from a perfect log-spiral function and to encompass the different tracers that we consider (NIR, CO, H$\alpha$). In this Section, we explicitly measure the effect of spiral mask width by comparing our nominal results against the contrast using narrower spiral masks, which closely follow the ridge of molecular gas and star formation, as introduced in Sect.~\ref{Sec:thinnersp}. These masks have typical widths between 500 and 1000\,pc, so about half the width of the original masks (still keeping a few resolution elements across the spiral mask).

Figure~\ref{fig:orig_vs_narrow} shows that contrasts based on the original and narrow masks track each other very well (Spearman rank correlation coefficients $\rho \gtrsim 0.9$). If we fit a power-law for each tracer separately, we find that the relation is essentially linear with a small offset towards higher contrasts for the narrow masks in \SigMol\ and \SigSFR. Indeed, the median increase in \SigStar\ contrasts when measured using the narrow masks is just 1\%, whereas this number is 10\% and 11\% for \SigMol\ and \SigSFR, respectively. This is much smaller than the ${\sim} 40\%$ increase that we measured when shifting from low (1.5\,kpc) to high (${\sim} 100$\,pc) resolution. The change in \SigMol\ and \SigSFR\ when adopting the narrower masks is such that the resulting SFE remains on average the same.
Therefore, the contrasts are not strongly affected by the width of the spiral mask, and the effect is far more limited than spatial resolution. Yet, there is some scatter in the plots that makes it clear that, even though it should not affect global trends, the details of the spiral masks do affect individual measurements in a non-trivial way.

\section{Sanity checks and alternatives to measure contrasts}
\label{sec:sanity_checks}

\begin{figure*}[t]
\begin{center}
\includegraphics[trim=0 0 0 0, clip,width=0.95\textwidth]{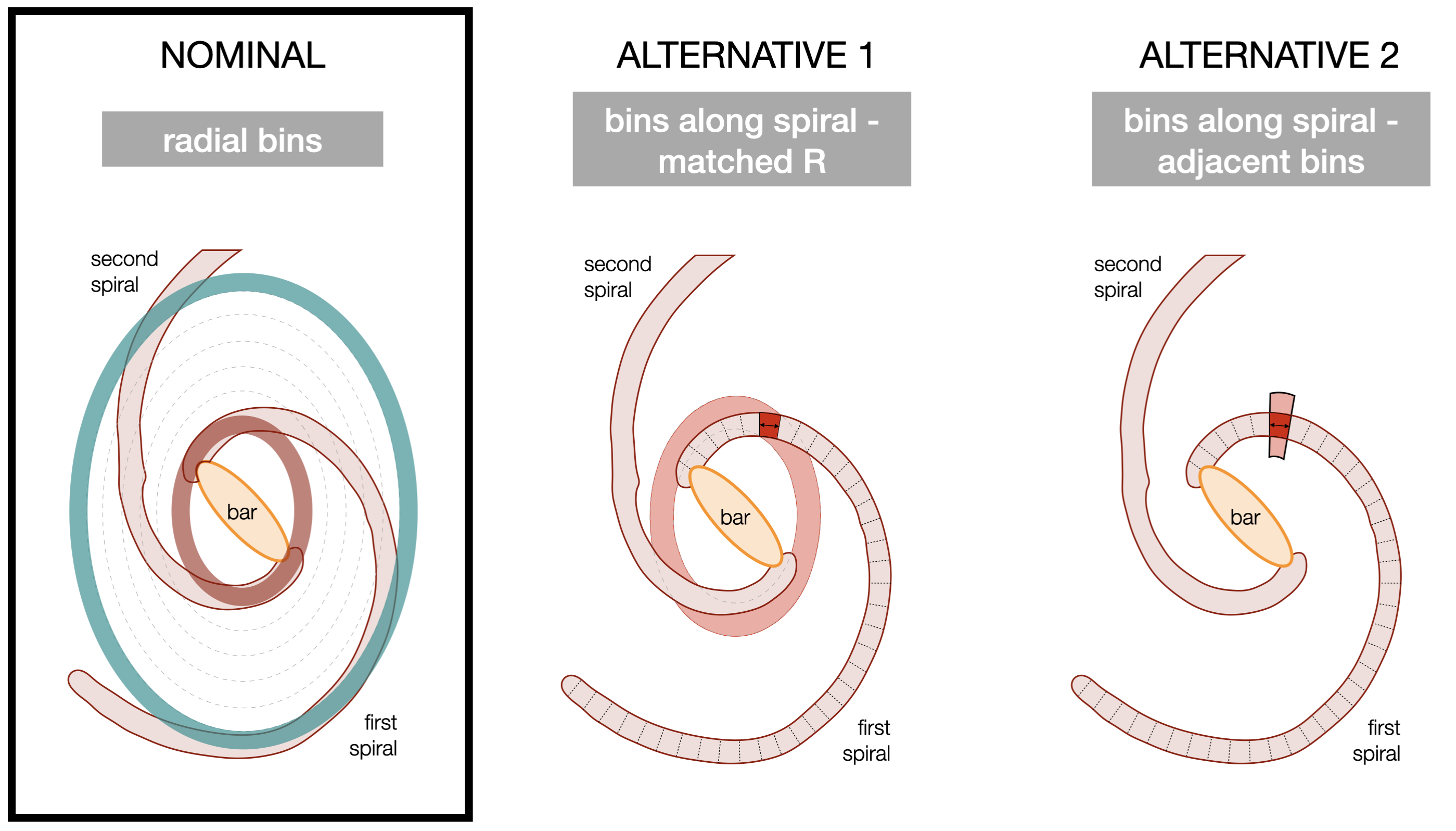}
\end{center}
\caption{Cartoon illustrating the two different binning alternatives introduced in this Appendix. Both alternatives involve defining bins along each spiral arm (`boxes' delimited by regularly spaced cuts perpendicular the spiral spine). In the first alternative, the corresponding interarm region is defined as an elliptical annulus covering all pixels with the range of galactocentric distances present in the spiral bin, and excluding the spiral mask. In the second alternative, the interarm value is obtained as the average surface density within two adjacent `boxes' adjacent to the spiral bin and immediately outside the spiral mask.}
\label{fig:binning_alternatives}
\end{figure*}

Here we consider as sanity checks several alternatives to measure contrasts. Table\,\ref{table:stats_appendix} shows the resulting rank correlation coefficients (Spearman $\rho$) and the slope and intercept of bisector fits for different relations examined in this paper: the contrast in molecular gas (\SigMol), star formation rate (\SigSFR), and star formation efficiency (SFE) as a function of stellar contrast (\SigStar), and the star formation rate contrast as a function of the molecular gas contrast. For each of these relations, we show the correlation coefficients and fits based on the individual spiral segments (one average value per spiral segment, so typically two datapoints per galaxy), and also for all radial bins (hundreds of datapoints across all galaxies). The various blocks of rows show different alternatives, either changing tracers or methodology. The former includes using all tracers at lower resolution ({Sect.~\ref{Sec:effect_res}}), measuring SFR directly from H$\alpha$ (without applying the empirical correction for extinction introduced in Sect.\,\ref{sec:AppendixMUSE}), estimating SFR using extinction-corrected H$\alpha$ (based on the Balmer decrement) from MUSE, which is only available for a subset of galaxies and a smaller field of fiew, or, also for MUSE galaxies, calculating the \SigSFR\ contrast only within \hii{} regions (blanking the SFR map outside the \hii{} region mask, \citealt{2022A&A...659A..26B}). The methodological alternatives include using narrower spiral masks (Appendix~\ref{Sec:narrow}), using bins that follow each spiral segment at uniform length steps compared to interarm regions at matched galactocentric radii (Fig.\,\ref{fig:binning_alternatives}), or the same bins along spiral segments with interarm regions defined as adjacent `boxes' (immediately on top and below each spiral bin).

We see that rank coefficients become smaller when considering radial bins instead of entire spiral segments. This is expected since, on the smaller radial bins, we are more sensitive to local, stochastic variations in surface density and we are generally more affected by the noise of the individual measurements; by averaging over entire spiral segments, the contrast measurements become more robust.

In agreement with what we present in {Sect.~\ref{Sec:effect_res}} and {Appendix}~\ref{Sec:narrow}, the effect of using narrower masks or employing a lower resolution can affect the actual contrast measurements, but does not affect dramatically the relations between contrasts. The slopes that we find are typically compatible within a few $\sigma$.

As shown by Fig.\,\ref{fig:violin_plot_GD-vs-rest-Ha-only}, the application of the empirical calibration that we introduced in Sect.\,\ref{sec:AppendixMUSE} to account for extinction in the \SigSFR\ contrasts does not affect qualitatively the distribution of SFR or SFE contrasts. Table\,\ref{table:stats_appendix} confirms that this choice does not have a strong impact on the relations between contrasts either. If we instead use MUSE extinction-corrected H$\alpha$ for \SigSFR, we do find slopes which are quite different, but also the error bars become much larger. Since MUSE is only available for a subset of galaxies and a smaller field of fiew, the noise increases considerably. The same applies to SFR calculated for MUSE within \hii{} regions.

If we employ either wider (1000\,pc) or narrower (250\,pc) radial bins, the trends remain qualitatively the same, with similar correlation coefficients and slopes that are well compatible within the uncertainties. The alternative sampling schemes introduce in practice many more bins and therefore noise increases significantly. This is particularly true if we consider all radial bins together for interarm regions defined as immediately adjacent to each spiral bin, which increases the stochasticity; averaging over an entire ring at a given range of galactocentric radii provides a more stable reference value than measuring the average surface densities on small boxes next to the arm, where local fluctuations can lead to very different arm/interarm surface density ratios.

In conclusion, the precise values of the contrasts are sensitive to the various choices in methodology and tracers, but our qualitative conclusions remain valid independently {from} these details.

\begin{table*}[t!]
\small
\begin{center}
\caption[h!]{Sanity checks showing how contrast measurements are affected by different choices of tracer and methodology.}
\begin{tabular}{llcccccc}
\hline\hline
   \noalign{\smallskip}
 & & \multicolumn{3}{c}{individual spiral segments} & \multicolumn{3}{c}{all radial bins}  \\
      \cmidrule(lr){3-5} \cmidrule(lr){6-8}
 & & Spearman $\rho$ & Slope & Intercept & Spearman $\rho$ & Slope & Intercept  \\
   \hline
   \hline
   \noalign{\smallskip}
\multirow{4}{*}{Nominal} &  \SigStar -- \SigMol & $0.73$ & $1.83 \pm  0.46$ & $0.16 \pm  0.03$ & $0.46$ & $2.23 \pm  0.15$ & $0.08 \pm  0.02$ \\
 &  \SigStar -- \SigSFR & $0.61$ & $2.89 \pm  1.29$ & $0.15 \pm  0.04$ & $0.55$ & $3.36 \pm  0.37$ & $0.01 \pm  0.02$ \\
 &  \SigStar -- SFE & $0.34$ & $1.68 \pm  0.30$ & $-0.09 \pm  0.04$ & $0.27$ & --- & --- \\
 &  \SigMol -- \SigSFR  & $0.78$ & $1.61 \pm  0.25$ & $-0.13 \pm  0.08$ & $0.69$ & $1.35 \pm  0.07$ & $-0.06 \pm  0.03$ \\
   \noalign{\smallskip}
   \hline
   \noalign{\smallskip}
\multirow{4}{*}{Narrow masks} &  \SigStar -- \SigMol & $0.65$ & $1.83 \pm  0.48$ & $0.19 \pm  0.03$ & $0.48$ & $2.19 \pm  0.17$ & $0.13 \pm  0.02$ \\
 &  \SigStar -- \SigSFR & $0.47$ & $2.77 \pm  1.45$ & $0.21 \pm  0.05$ & $0.51$ & $3.20 \pm  0.41$ & $0.08 \pm  0.03$ \\
 &  \SigStar -- SFE & $0.19$ & $1.50 \pm  0.29$ & $-0.07 \pm  0.05$ & $0.24$ & --- & --- \\
 &  \SigMol -- \SigSFR  & $0.77$ & $1.56 \pm  0.23$ & $-0.10 \pm  0.08$ & $0.68$ & $1.36 \pm  0.07$ & $-0.07 \pm  0.04$ \\
   \noalign{\smallskip}
   \hline
   \noalign{\smallskip}
\multirow{4}{*}{1.5 kpc resolution} &  \SigStar -- \SigMol & $0.83$ & $1.91 \pm  0.25$ & $0.05 \pm  0.02$ & $0.68$ & $1.86 \pm  0.09$ & $0.05 \pm  0.01$ \\
 &  \SigStar -- \SigSFR & $0.75$ & $2.65 \pm  0.43$ & $0.02 \pm  0.04$ & $0.64$ & $2.31 \pm  0.11$ & $0.04 \pm  0.01$ \\
 &  \SigStar -- SFE & $0.36$ & $1.10 \pm  0.12$ & $-0.05 \pm  0.02$ & $0.14$ & --- & --- \\
 &  \SigMol -- \SigSFR  & $0.82$ & $1.28 \pm  0.10$ & $-0.02 \pm  0.02$ & $0.78$ & $1.23 \pm  0.05$ & $-0.02 \pm  0.01$ \\
   \noalign{\smallskip}
   \hline
   \noalign{\smallskip}
\multirow{4}{*}{\parbox{1.7cm}{SFR directly from H$\alpha$}} &  \SigStar -- \SigMol & $0.73$ & $1.83 \pm  0.46$ & $0.16 \pm  0.03$ & $0.46$ & $2.23 \pm  0.15$ & $0.08 \pm  0.02$ \\
 &  \SigStar -- \SigSFR & $0.61$ & $2.43 \pm  0.82$ & $0.11 \pm  0.03$ & $0.55$ & $2.87 \pm  0.26$ & $0.01 \pm  0.02$ \\
 &  \SigStar -- SFE & $0.27$ & $1.42 \pm  0.27$ & $-0.14 \pm  0.04$ & $0.19$ & --- & --- \\
 &  \SigMol -- \SigSFR  & $0.78$ & $1.35 \pm  0.18$ & $-0.12 \pm  0.07$ & $0.69$ & $1.16 \pm  0.06$ & $-0.05 \pm  0.03$ \\
   \noalign{\smallskip}
   \hline
   \noalign{\smallskip}
\multirow{4}{*}{\parbox{2.4cm}{SFR from MUSE (ext.-corr. H$\alpha$)}} &  \SigStar -- \SigMol & $0.55$ & $2.47 \pm  2.06$ & $0.19 \pm  0.06$ & $0.40$ & $2.66 \pm  0.51$ & $0.12 \pm  0.03$ \\
 &  \SigStar -- \SigSFR & $0.37$ & $3.05 \pm  3.80$ & $0.32 \pm  0.12$ & $0.46$ & $4.12 \pm  1.10$ & $0.12 \pm  0.04$ \\
 &  \SigStar -- SFE & $0.17$ & $1.08 \pm  0.85$ & $0.10 \pm  0.06$ & $0.23$ & --- & --- \\
 &  \SigMol -- \SigSFR  & $0.71$ & $1.69 \pm  0.33$ & $-0.12 \pm  0.14$ & $0.69$ & $1.57 \pm  0.08$ & $-0.08 \pm  0.04$ \\
   \noalign{\smallskip}
   \hline
   \noalign{\smallskip}
\multirow{4}{*}{\parbox{2.4cm}{SFR from MUSE (only \hii{} regions)}} &  \SigStar -- \SigMol & $0.55$ & $2.47 \pm  2.06$ & $0.19 \pm  0.06$ & $0.40$ & $2.66 \pm  0.51$ & $0.12 \pm  0.03$ \\
 &  \SigStar -- \SigSFR & $0.45$ & $2.14 \pm  1.41$ & $0.16 \pm  0.04$ & $0.49$ & $2.47 \pm  0.37$ & $0.09 \pm  0.02$ \\
 &  \SigStar -- SFE & $-0.12$ & $-1.14 \pm  0.31$ & $0.12 \pm  0.04$ & $-0.01$ & --- & --- \\
 &  \SigMol -- \SigSFR  & $0.69$ & $0.87 \pm  0.08$ & $-0.01 \pm  0.04$ & $0.59$ & $0.92 \pm  0.05$ & $-0.02 \pm  0.03$ \\
   \noalign{\smallskip}
   \hline
   \noalign{\smallskip}
\multirow{4}{*}{\parbox{2.0cm}{Narrower radial bins (250 pc)}} &  \SigStar -- \SigMol & $0.74$ & $1.86 \pm  0.48$ & $0.16 \pm  0.03$ & $0.45$ & --- & --- \\
 &  \SigStar -- \SigSFR & $0.58$ & $2.99 \pm  1.44$ & $0.14 \pm  0.04$ & $0.53$ & $3.43 \pm  0.28$ & $0.00 \pm  0.02$ \\
 &  \SigStar -- SFE & $0.25$ & $1.58 \pm  0.41$ & $-0.06 \pm  0.05$ & $0.26$ & --- & --- \\
 &  \SigMol -- \SigSFR  & $0.79$ & $1.65 \pm  0.26$ & $-0.14 \pm  0.08$ & $0.67$ & --- & --- \\
   \noalign{\smallskip}
   \hline
   \noalign{\smallskip}
\multirow{4}{*}{\parbox{2.0cm}{Wider radial bins (1000 pc)}} &  \SigStar -- \SigMol & $0.71$ & $1.64 \pm  0.38$ & $0.16 \pm  0.03$ & $0.45$ & $2.04 \pm  0.20$ & $0.10 \pm  0.02$ \\
 &  \SigStar -- \SigSFR & $0.55$ & $2.65 \pm  1.00$ & $0.13 \pm  0.04$ & $0.56$ & $3.11 \pm  0.45$ & $0.04 \pm  0.03$ \\
 &  \SigStar -- SFE & $0.38$ & $1.60 \pm  0.26$ & $-0.10 \pm  0.03$ & $0.29$ & --- & --- \\
 &  \SigMol -- \SigSFR  & $0.77$ & $1.72 \pm  0.23$ & $-0.18 \pm  0.08$ & $0.72$ & $1.26 \pm  0.11$ & $-0.02 \pm  0.05$ \\
   \noalign{\smallskip}
   \hline
   \noalign{\smallskip}
\multirow{4}{*}{Bins along spiral} &  \SigStar -- \SigMol & $0.74$ & $1.76 \pm  0.34$ & $0.16 \pm  0.03$ & $0.50$ & $1.92 \pm  0.07$ & $0.08 \pm  0.01$ \\
 &  \SigStar -- \SigSFR & $0.63$ & $2.83 \pm  1.14$ & $0.23 \pm  0.11$ & $0.58$ & $3.08 \pm  0.15$ & $-0.02 \pm  0.02$ \\
 &  \SigStar -- SFE & $0.41$ & $1.43 \pm  0.26$ & $0.05 \pm  0.05$ & $0.35$ & --- & --- \\
 &  \SigMol -- \SigSFR  & $0.71$ & $1.37 \pm  0.22$ & $0.10 \pm  0.07$ & $0.72$ & $1.49 \pm  0.05$ & $-0.12 \pm  0.02$ \\
   \noalign{\smallskip}
   \hline
   \noalign{\smallskip}
\multirow{4}{*}{\parbox{2.0cm}{Bins with adjacent interarm}} &  \SigStar -- \SigMol & $0.57$ & $1.37 \pm  0.37$ & $0.41 \pm  0.04$ & $0.29$ & --- & --- \\
 &  \SigStar -- \SigSFR & $0.60$ & $1.54 \pm  0.55$ & $0.59 \pm  0.04$ & $0.35$ & $3.19 \pm  0.38$ & $0.24 \pm  0.02$ \\
 &  \SigStar -- SFE & $0.14$ & $1.02 \pm  0.19$ & $0.28 \pm  0.04$ & $0.15$ & $1.60 \pm  0.20$ & $-0.03 \pm  0.02$ \\
 &  \SigMol -- \SigSFR  & $0.78$ & $1.12 \pm  0.09$ & $0.10 \pm  0.04$ & $0.52$ & --- & --- \\
   \noalign{\smallskip}
   \hline
   \hline
\end{tabular}
\label{table:stats_appendix}
\end{center}
\end{table*}

\end{document}